\documentclass[11pt,a4paper]{article}
\usepackage{multicol}
\usepackage{comment}
\usepackage[dvipsnames]{xcolor}
\usepackage{jheppub}
\usepackage{caption, subcaption}  
\usepackage{color,latexsym, array,multirow,  verbatim, enumerate,cancel,graphicx}
\usepackage{slashed}
\usepackage{bm}

\def\beq {\begin{equation}}
\def\eeq {\end{equation}}
\def\bi {\begin{itemize}}
\def\ei {\end{itemize}}
\def\bea {\begin{eqnarray}}
\def\eea {\end{eqnarray}}

\numberwithin{equation}{section} 

\title{Dedicated Triggers for Displaced Jets using Timing Information from Electromagnetic Calorimeter at HL-LHC}

\author{Biplob Bhattacherjee$^1$, Tapasi Ghosh$^2$, Rhitaja Sengupta$^1$, Prabhat Solanki$^1$}

\affiliation{\vspace*{0.1in}$^1$ Centre for High Energy Physics, Indian Institute of Science, Bengaluru 560012, India}
\affiliation{\vspace*{0.1in}$^2$ M. S. Ramaiah University of Applied Sciences, Bengaluru 560054, India}

\emailAdd{biplob@iisc.ac.in}
\emailAdd{tapasighosh.ss@msruas.ac.in}
\emailAdd{rhitaja@iisc.ac.in}
\emailAdd{prabhats@iisc.ac.in}

\abstract{In this paper, we study the prospect of ECAL barrel timing to develop triggers dedicated to long-lived particles decaying to jets, at the level-1 of HL-LHC. We construct over 20 timing based variables, and identify three of them which have better performances and are robust against increasing PU. We estimate the QCD prompt jet background rates accurately using the ``stitching'' procedure for varying thresholds defining our triggers, and compute the signal efficiencies for different LLP scenarios for a permissible background rate. The trigger efficiencies can go up to $\mathcal{O}(80\%)$ for the most optimal trigger for pair-produced heavy LLPs having high decay lengths, which degrades with decreasing mass and decay length of the LLP. We also discuss the prospect of including the information of displaced L1 tracks to our triggers, which further improves the results, especially for LLPs characterised by lower decay lengths.
}
 
\begin{document}
\maketitle
\flushbottom


\section{Introduction}
\label{sec:intro}

In 2024, LHC will witness the last run of Phase-I before closing down for the Phase-II upgrade. Till the completion of Run-II in 2018, most of the results were consistent with the Standard Model of particle physics (SM), and we have not come across any clear indication of new physics with consequential statistical significance. Novel methods to search for physics beyond the Standard Model (BSM) are being studied and employed to leave no stones unturned. Still, several avenues in BSM searches have not yet been explored to their full potential. In recent years, the need to extend our searches to include such exotic particles have gained quite some traction due to both exclusions of large regions of new physics parameter space with particles having conventional signatures at current experiments as well as the development of High Luminosity LHC (HL-LHC) with several hardware and software level improvements. Till now, most of the experimental searches focused on BSM particles decaying promptly. However, recently, the focus has shifted to look for particles with longer lifetimes or long-lived particles (LLPs).

The presence of LLPs is well motivated in many BSM theories. Minimal super-symmetric extension of SM (MSSM)\,\cite{Fayet:1976et,Fayet:1977yc} is one such laboriously studied BSM theory in the context of LLPs. A number of new particles with a longer lifetime can arise in MSSM. R-parity violating SUSY model\,\cite{Barbier:2004ez} is one of the riveting SUSY models which predict the presence of long-lived particles. LLPs also arise in many other SUSY scenarios like gauge-mediated SUSY (GMSB)\,\cite{Dimopoulos:1996vz,Giudice:1998bp}, anomaly mediated SUSY (AMSB)\,\cite{Feng:1999fu}, split SUSY\,\cite{Arkani-Hamed:2004ymt,Arvanitaki:2012ps}, and stealth SUSY\,\cite{Fan:2011yu}. Particles with longer lifetimes can also arise in many non-SUSY theories, like, heavy neutrino theories, dark matter theories, gauge and Higgs portal theories. For a detailed review of many such LLP models, see Ref.\,\cite{Alimena:2019zri} and the references therein.

Several phenomenological studies explore the possibilities of LLP searches covering a wide range of models and signatures, such as those in Refs.\,\cite{Ilten:2015hya,Gago:2015vma,Banerjee:2017hmw,KumarBarman:2018hla,Bhattacherjee:2019fpt,CidVidal:2019urm,Banerjee:2019ktv,Jones-Perez:2019plk,Bhattacherjee:2020nno,Gershtein:2020mwi,Fuchs:2020cmm,Chakraborty:2020cpa,Cheung:2020ndx,Liu:2020vur,Evans:2020aqs,Linthorne:2021oiz,Alimena:2021mdu,Bhattacherjee:2021rml,Sakurai:2021ipp,Du:2021cmt}.
Collaborations of two major general purpose detectors at LHC, CMS and ATLAS, have also been actively performing several BSM studies involving various LLP signatures. Displaced object searches, mainly focused on looking for displaced jets, displaced vertices and displaced leptons, have been extensively performed by both CMS\,\cite{CMS:2020iwv,CMS-PAS-EXO-16-022,CMS:2017kku,CMS:2014hka,CMS-PAS-FTR-18-002,CMS:2018lab,CMS-PAS-EXO-19-013,CMS:2021kdm,CMS:2021yhb} and ATLAS\,\cite{ATLAS:2015xit,ATLAS:2015itk,ATLAS:2019tkk,ATLAS:2018rjc,ATLAS:2019fwx,ATLAS:2020wjh,ATLAS:2017tny,ATLAS:2019kpx,ATLAS:2020xyo,ATLAS:2021jig} collaborations.
Along with these, CMS and ATLAS have also performed several non-vertex based LLP searches, involving disappearing tracks, trackless jets, jets with low electromagnetic energy fraction, non-pointing photons and emerging jets\,\cite{CMS-PAS-EXO-16-036,CMS:2020atg,CMS:2019zxa,CMS:2018bvr,ATLAS:2015wsk,ATLAS:2018lob,ATLAS-CONF-2021-015,ATLAS:2014kbb,ATLAS:2019qrr,ATLAS:2018niw}. While ATLAS has been performing several LLP searches using the muon spectrometer\,\cite{ATLAS:2018tup,ATLAS:2019jcm,ATL-PHYS-PUB-2019-002,ATLAS-CONF-2021-032}, LLP signatures are now being probed at the CMS for the first time in the muon spectrometer\,\cite{CMS:2021juv}.
At LHCb, several LLP searches have been performed involving dark photons, displaced leptons and displaced jets\,\cite{LHCb:2016buh,LHCb:2016inz,LHCb:2016awg,LHCb:2017xxn,LHCb:2019vmc,LHCb:2020akw}.
There are already strong constraints present on the production cross section of LLPs and their decay to displaced jets. For example, the CMS experiment has put stringent limits till date on LLPs decaying to jets using 132\,fb$^{-1}$ of data collected at CMS till the completion of Run-II in Phase-I of LHC\,\cite{CMS:2020iwv}. For LLPs of a particular mass, most restrictive limits arise in regions with mean proper decay length, c$\tau$, between 3\,mm and 300\,mm. For smaller and higher decay lengths, limits are less restrictive. Most confining limits are put on LLPs with masses greater than 500\,GeV, while they are much less stringent for LLPs with smaller masses. For LLP of mass 50\,GeV, pair production cross-sections above $\approx$ 10\,fb  and 100\,fb are excluded for mean proper decay lengths of 1\,cm and 30\,cm respectively.

Since large cross-sections of LLP processes are ruled out from experiments, it becomes very crucial to select LLP events having even lower cross sections.
One of the important aspects of any analysis is to select events efficiently at the online trigger levels, namely level-1 (L1) and High Level Trigger (HLT). Therefore, we need to ensure that we efficiently select events with LLPs at the first level trigger itself, otherwise these events are lost forever.
With L1 triggers having a smaller latency period, selecting LLP events become much more challenging because of the innate property of LLPs to decay away from the beamline with significant time delay. Till Run-II of Phase-I of LHC, several standard L1 triggers like single and multi-jet triggers, lepton triggers, missing-$E_T$ (MET), and $H_T$ triggers were employed to select LLP events at the first level, which was possible due to the relatively low amount of pile-up (PU). At HL-LHC, the high amount of PU will pose a huge challenge, and L1 triggers will be largely affected by the increased number of PU interactions per bunch crossing. Lighter LLPs, for example, those coming from the decay of the 125\,GeV Higgs boson, will deposit a small amount of energy in the calorimeters and will be very hard to trigger on amid the huge PU background. 

To deal with PU at HL-LHC, hardware, as well as software upgrades are proposed for detectors. The outer tracker will be upgraded to accommodate track reconstruction at L1\,\cite{CERN-LHCC-2020-004}. Due to the availability of tracks at L1, tracklessness property of LLP can be efficiently used to construct dedicated L1 triggers for LLPs as shown in Ref.\,\cite{Bhattacherjee:2020nno}. A whole new timing detector, MIP Timing Detector (MTD), will be placed between the outer tracker and the electromagnetic calorimeter (ECAL) to get precise timing information of minimum-ionising particles (MIPs), which will help to mitigate in-time as well as out of time PU\,\cite{CMS:2667167}. Endcaps of both the electromagnetic and hadronic calorimeters (HCAL) will also undergo a massive upgrade, and the current detector elements will be replaced by the high granularity calorimeter (HGCAL)\,\cite{CERN-LHCC-2017-023}. As a result of the upgrade to faster and more efficient electronic systems on-board detectors, particle flow (PF) candidates can be reconstructed at L1. Availability of tracks and PF candidates at L1 allows implementation of Pileup Per Particle Identification (PUPPI)\,\cite{Bertolini:2014bba} algorithm at L1, which would help to control the background rate for the triggers, which will be most affected by the huge amount of PU at HL-LHC.

Possibility of using timing information in LLP searches has been extensively studied in many experimental and phenomenological studies\,\cite{ATLAS:2014kbb,CMS:2016kce,CMS:2017kku,Liu:2018wte,ATLAS:2019gqq,CMS:2019qjk,CMS:2019zxa,Bhattacherjee:2020nno,ATLAS:2021mdj,Chiu:2021sgs}. The possibility of using MTD based timing information at L1 to trigger on LLPs is studied in detail in Ref.\,\cite{Bhattacherjee:2020nno}. The CMS Phase-II L1 TDR\,\cite{CERN-LHCC-2020-004} mentions a potential scope of using inputs from the MTD in an ``External Trigger'', which can be fed into the Global Trigger at L1.
Upgraded calorimeters will also have improved timing capabilities. While timing information of calorimeter energy deposits from ECAL and HCAL will be available at L1, HGCAL will not be able to provide timing information at L1 due to bandwidth constraints\,\cite{CERN-LHCC-2020-004}. Availability of timing information at L1 from ECAL barrel calorimeter is expected to play an important role in triggering long-lived particles, especially the ones where mean proper decay length\,\footnote{For simplicity, mean proper decay length of the LLP will be referred to as decay length hereafter, unless stated otherwise.} of the LLP is large. 

To trigger on LLP events, CMS has specifically designed two dedicated L1 triggers harnessing the tracking capabilities, which include extended tracking till a transverse impact parameter ($d_0$) of 8\,cm, and ECAL barrel timing\,\cite{CERN-LHCC-2020-004}. Apart from the CMS Phase-II TDR\,\cite{CERN-LHCC-2020-004}, and Ref.\,\cite{Bhattacherjee:2020nno}, there are not many detailed realistic studies for developing dedicated LLP triggers for HL-LHC that properly includes the effect of increased PU and the detector resolution, which are essential parts of HL-LHC. The CMS TDR contains results for limited LLP benchmark points for a fixed amount of PU, which cannot be directly used for other LLP models.
Extrapolating these results to other benchmark points within the same model
with different PU is also difficult due to the absence of exact simulation details $-$ like the ECAL barrel timing resolution used in their results and the dependence of this resolution with pseudorapidity ($\eta$).
This motivates us to study in detail the prospect of triggering LLPs in scenarios where the LLPs decay to displaced jets. We focus on the following major questions in the present work $-$
\begin{itemize}
\item Are PUPPI based jet triggers which are optimised for prompt jets efficient to trigger on displaced jets coming from LLP decays?
\item What are the most optimal timing variables from ECAL barrel information to be used in L1 triggers? What factors affect them? Are these variables robust against the increasing PU and degrading timing resolution?
\item Are these timing variables equally efficient for various LLP scenarios decaying to displaced jets?
\item How can we make use of timing-based triggers to their full potential to trigger on various LLP scenarios? How can the trigger efficiencies be improved further?
\item Can the displaced track collection at L1 be combined with timing information efficiently to improve further trigger efficiency of timing-based jet triggers for various LLP scenarios, and how do they complement each other?  
\end{itemize}
Accurate estimation of background rates of a trigger without double counting when combining different regions of the phase-space of QCD multijet events is very crucial in understanding the feasibility of implementing the trigger. One of the major goals of the present work is to compute the background rates precisely and to study how the rates of the triggers developed by us in this work vary in different contexts.
Following a thorough discussion of trigger prospects at L1 with ECAL barrel timing and L1 tracking, we present a brief discussion of trigger prospects at the HLT $-$ inspecting whether the events that we selected at L1 using triggers developed by us in this work can pass the HLT with background rates reduced further by an order.


The rest of the paper is outlined as follows: in Section\,\ref{sec:signal}, we define the signal models and backgrounds for our study along with the simulation details. In Section\,\ref{sec:hllhcups}, we discuss about HL-LHC upgrades at CMS and their updated L1 trigger menu for jets. We also talk about the possibility of using PUPPI based triggers for triggering events with displaced jets. In Section\,\ref{sec:ecaltiming}, we discuss the adverse effects of high PU along with the effect of ECAL timing resolution on jet timing as the PU increases and timing resolution degrades with time. We also examine why some prompt QCD jets might have high timing values. In Section\,\ref{sec:ecaltrigger}, we define several timing variables constructed using different statistical measures and calculate the signal efficiency and background rate for various benchmark points of different LLP scenarios using triggers based on jet timing. In section\,\ref{sec:improve}, we present some discussions regarding the effect of the resolution, addition of displaced L1 tracks, narrower jets, high PU, and triggering at HLT, some of which enhance the signal acceptance keeping the rate at the acceptable level.
Finally, we conclude in Section\,\ref{sec:concl}.

\section{LLP signal models, current bounds and computational setup}
\label{sec:signal}

In this section, we briefly discuss various production processes of LLPs and their decay to jets as considered in this work, along with examples of concrete models that can give rise to such LLP scenarios. We present the current experimental bounds from the CMS and ATLAS studies of long-lived particles decaying to jets relevant for these scenarios. Broadly, LLPs can have two possible production modes in the colliders $-$ direct production or from the decay of some on-shell SM or BSM particle. We study three different scenarios in the present work, where we cover both these kinds of LLP production.
After introducing the LLP scenarios, we briefly discuss the background process considered in this work and our computational setup.


\subsection{(A) LLPs coming from the decay of the SM Higgs boson}
\label{ssec:scenarioA}

$$\mathbf{  p p \rightarrow h (125 GeV) \rightarrow X X, X \rightarrow j j}$$
In the first scenario, we have considered production of LLPs through the decay of an on-shell 125 GeV Higgs boson, and the subsequent decay of each LLP to a pair of jets. 
The Higgs portal is one of the leading renormalizable portal connecting new gauge-singlet particles to the SM. It is also experimentally motivated due to the scope to include couplings of Higgs boson to new physics particles as well as different production modes of the Higgs boson.  

An example of a model where the Higgs boson decays to exotic particles which are long-lived is the Dark Matter model with light fermionic WIMPs and light scalar mediator\,\cite{Matsumoto:2018acr}. Such a model can generate large velocity-dependent scattering cross-sections between the WIMPs, solving the small scale crisis of structure formation in the Universe.
The scalar mediator mixes with the SM Higgs boson and therefore, its couplings to SM fermions are all suppressed by this mixing. For very small values of mixing, the mediator can have long lifetimes, given the WIMPs are heavier than the mediator. Since the mediators couple to SM fermions through their Yukawa couplings, mediators having masses greater than 10\,GeV predominantly decay to $b$-quarks, giving rise to displaced jets. 

Search of displaced vertices in the ATLAS inner detector have excluded branching ratios of Higgs to long-lived mediators above 10\% at 95\% confidence level for decay lengths between 4\,mm and 100\,mm\,\cite{ATLAS:2021jig} for an LLP of mass $\sim$40\,GeV with 139\,fb$^{-1}$ of data using the $Zh$ production mode. In comparison, the CMS displaced jets search uses the gluon-gluon fusion (ggF) production mode and excludes branching fractions larger than 1\% at 95\% CL for decay lengths as low as 1\,mm and as large as 340\,mm with an integrated luminosity of 132\,fb$^{-1}$\,\cite{CMS:2020iwv}. For a recent phenomenological study computing the sensitivity of long-lived scalar mediators coming from Higgs boson decay combining all the dominant production modes of Higgs boson using the CMS muon spectrometer, see Ref.\,\cite{Bhattacherjee:2021rml}. In the present work, we focus on the dominant production mode of SM Higgs boson, i.e., the ggF, which has a cross-section of 50.32\,pb at center of mass energy, $\sqrt{s}=14$\,TeV. 

In this scenario, masses of LLPs will be relatively smaller ($m_{X}\lesssim \frac{m_{h}}{2}\approx62.5$\,GeV).
We study various mass points ranging between 10\,GeV to 50\,GeV with mean decay lengths ranging between 1\,cm to 500\,cm. Decay lengths of LLP in the lab frame depend on the boost of the LLP as well as on the mean decay length of the LLP in its rest frame.
Since the LLP comes from the decay of the on-shell Higgs boson with mass 125 GeV, the former's boost depends on its mass difference with the Higgs boson. As a result, 50\,GeV LLPs will be pair produced with very small boost as compared to 10\,GeV LLPs. Overall, LLPs in this scenario have smaller hadronic activities in the detectors due to small boosts and therefore, will be harder to trigger on. 


\subsection{(B) LLPs directly pair-produced and each decaying to a pair of jets}
\label{ssec:scenarioB}

$$\mathbf{pp  \rightarrow XX, X \rightarrow jj}$$


In the second scenario, we have considered direct pair-production of LLPs in the colliders from a quark-initiated process mediated by $\gamma/Z$. The long-lived particles are assumed to further decay into a pair of quarks each, giving rise to two displaced jets. This kind of scenario can arise in $R$-parity violating (RPV) supersymmetric models. For instance, a pair of sneutrinos ($\tilde{\nu}$) are produced in the colliders which eventually decay to two quarks each, using the RPV LQD coupling\,\cite{Barbier:2004ez}. Constraints from experiments restrict these kinds of couplings to very small values, which can make the sneutrinos have longer lifetimes. 
For this scenario, we have considered the decay of sneutrinos to a pair of light quarks ($u/d/s$).

The CMS displaced jets search has excluded pair-production cross-sections greater than 0.07\,fb\,\cite{CMS:2020iwv} for LLPs heavier than 500\,GeV which decay to a quark-antiquark pair with 100\% branching fraction and $c\tau$ between 2\,mm and 250\,mm with 132\,fb$^{-1}$ of data. Limits are less restrictive for lower masses, like cross-sections till 1\,fb are allowed even for the most sensitive decay length of 10\,mm for a 100\,GeV LLP. The CMS collaboration have also translated the results of their displaced jets search to models with specific RPV-LQD coupling in stop decays ($\tilde{t}\rightarrow b \ell$) and RPV-UDD coupling in gluino decays ($\tilde{g}\rightarrow tbs$) in Ref.\,\cite{CMS:2020iwv}.

\subsection{(C) LLPs directly pair-produced and each decaying to three jets}
\label{ssec:scenarioC}

 $$\mathbf{pp  \rightarrow XX, X \rightarrow jjj}$$
 
In the third scenario, we have considered a similar production process as in the previous scenario (B), however, each of the LLPs, $X$, decays to three quarks in this scenario, leading to six displaced jets in the final state if both the LLPs decay before the calorimeters. A scenario like this can also arise in RPV SUSY models, where a pair of lightest neutralinos ($\chi_{1}^{0}$) are pair-produced directly from a quark-initiated process, and further decay to three quarks due to a UDD type RPV coupling\,\cite{Barbier:2004ez}. As discussed earlier, RPV couplings are usually constrained to small values which can make the lightest supersymmetric particle (LSP), here $\chi_{1}^{0}$, long-lived. 
In this scenario, we have considered the decay of neutralinos to three light quarks ($u/d/s$).
This scenario has an increased jet multiplicity in comparison to the previous two scenarios and we will study whether this enhances its sensitivity or reduces it due to more softer jets.

For both scenarios (B) and (C), we study LLPs with masses ranging between 100\,GeV to 500\,GeV having $c\tau$ between 1\,cm to 500\,cm. Since the LLPs are directly produced in these two scenarios, we can consider much heavier LLPs than the first scenario. Jets coming from the decay of such LLPs will, therefore, have higher transverse momenta and hence, more hadronic activity.
Thus, they will be relatively easier to trigger on as compared to scenario (A). For rest of the paper, we have denoted benchmark points in each scenario with the notation (M\underline{\hspace{0.3cm}}, c$\tau$\underline{\hspace{0.3cm}}) where M\underline{\hspace{0.3cm}} denotes the mass of the LLP in GeV and c$\tau$\underline{\hspace{0.3cm}} denotes its mean proper decay length in cm. 

\begin{figure}[hbt!]
    \centering
    \includegraphics[width=0.5\textwidth]{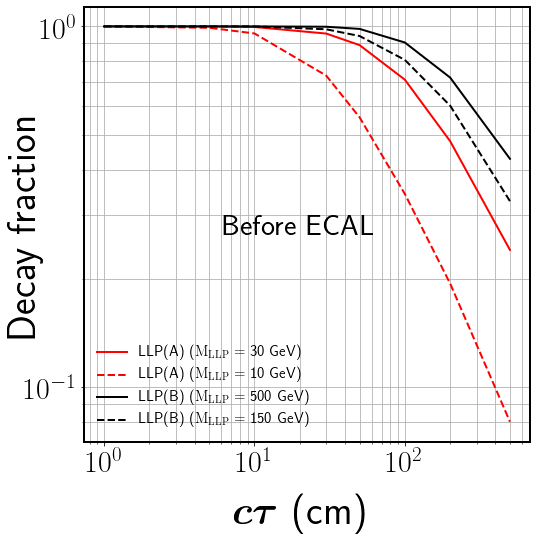}
    \caption{Fraction of events decaying before the ECAL, i.e., within a radial distance of 1.29\,m, and half-length of 3\,m, as a function of the mean proper decay length ($c\tau$) of the LLP having a mass of 10\,GeV and 30\,GeV in scenario (A), and 150\,GeV and 500\,GeV in scenario (B).}
    \label{fig:decay_frac}
\end{figure}

In the present work, we focus mostly on the ECAL barrel timing, and therefore, it is important to have an idea of how much fraction of the LLPs having various masses and decay lengths from the scenarios considered here decay before the ECAL, i.e., within a radial distance of 1.29\,m, and half-length of 3\,m.
Fig.\,\ref{fig:decay_frac} shows the variation of the fraction of LLPs decaying before the ECAL with the decay length of the LLP for LLPs coming from Higgs boson decay (scenario (A)) with masses 10\,GeV and 30\,GeV, and pair-produced LLPs (scenario (B)) with masses 150\,GeV and 500\,GeV. We observe that with decreasing mass and increasing decay length of the LLP this fraction decreases, and it falls off to very low values ($\sim 8$\%) close to $c\tau=500$\,cm for our lightest benchmark of 10\,GeV. For the 30\,GeV LLP coming from Higgs boson decay, around 100\% of the LLPs decay within the ECAL for $c\tau=10$\,cm which decreases to about 70\% for $c\tau=100$\,cm, whereas for the 500\,GeV benchmark, which has direct production, even at a $c\tau=100$\,cm, the fraction of decays within the ECAL is 90\%. Thus, we have significant decays of LLPs within the ECAL spanning a range of masses and lifetimes in all our scenarios. We now need to study how many of these pass L1 triggers when their decay products are required to leave energy deposits in the ECAL barrel with significant time delays.

\subsection{Background}
\label{ssec:bkg}

Most of the SM processes are prompt and therefore, can be easily separated from the displaced and time delayed signatures of long-lived particles. Still LLP searches have backgrounds from sources like beam halo, or interactions with the detector materials, or spurious detector noise, simulation of which are outside the scope of the present phenomenological study. However, we simulate the backgrounds that will dominantly affect the analysis using jet timing from the ECAL. These backgrounds are briefly described below and their simulation details are outlined in Section\,\ref{ssec:simu}.

\begin{itemize}
    \item \textbf{Jets from QCD dijet background:} For LLP scenarios considered in our study, QCD dijet events will constitute a major background in our analysis using jet timing. Although these jets consist of mostly prompt particles which are relativistic, some of the lower $p_T$ jets can have higher values of jet timing. The cross-section of QCD dijet process is around 0.21\,mb with jets having a transverse momentum, $p_T>30$\,GeV.
    This is overwhelmingly large compared to the cross-section of our LLP signal processes, which are already constrained to be below a few fb. Therefore, even a small number of QCD dijet events having jets with large timing can contribute as a major background in our analysis.
    \item \textbf{Jets from pile-up:} Due to the increased luminosity, the HL-LHC runs suffer from high PU, as we will further discuss in Section\,\ref{sec:hllhcups}. There will be around 140 (200) vertices per bunch crossing of protons in the beginning (end) of the HL-LHC runs.
    Jets from these vertices can lie in the tail of the jet timing distribution due to low $p_T$ as well as spatial and temporal spread of the PU vertices. 
    In addition, many particles from different PU vertices can fall within the jets coming from the hard process, which can bias the timing of the jet to lower or higher values.
    It is, therefore, important to properly simulate the high PU environment of HL-LHC for this study.
    
\end{itemize}

\subsection{Simulation details and computational setup}
\label{ssec:simu}

We simulate all our LLP processes as well as the QCD background using \texttt{PYTHIA8}\,\cite{Sjostrand:2014zea} at a center of mass energy of $\sqrt{s}=14$\,TeV.
For LLP signals, we generate the processes described in Sections\,\ref{ssec:scenarioA}, \ref{ssec:scenarioB} and \ref{ssec:scenarioC} with varying masses and decay lengths as specified earlier in each of these scenarios.
For the current study, we have simulated QCD dijet events divided in several parton level $p_T$ bins for a proper simulation of the tail of the $p_T$ distribution. We have generated events in eight $p_T^{gen}$ bins $-$ \{30,50\}\,GeV, \{50,75\}\,GeV, \{75,100\}\,GeV, \{100,125\}\,GeV, \{125,150\}\,GeV, \{150,175\}\,GeV, \{175,200\}\,GeV, and $>200$\,GeV. For the PU events, we simulate 1,\,000,\,000 minbias events with the inelastic soft QCD using the \texttt{SoftQCD:inelastic} option in \texttt{PYTHIA8}. For all the processes we use \texttt{LHAPDF6}\,\cite{Buckley:2014ana} with the \texttt{cteq6l1} PDF set and the corresponding CMS UE tune.

We have used \texttt{Delphes-3.5.0} \cite{deFavereau:2013fsa} along with the Phase-II detector card available with the $\texttt{Delphes}$ package for the fast detector simulation. We have reconstructed jets using anti-$k_T$ jet clustering algorithm with some fixed cone-size which we motivate later. 
We merge the minbias events with the hard collision events of various LLP signals and QCD dijet background using the \texttt{PileUpMerger} module of \texttt{Delphes}. In addition to a hard $pp$ collision event, we add an average number of PU vertices, similar to Ref.\,\cite{Bhattacherjee:2020nno}\footnote{\texttt{PileUpMerger} module of the \texttt{Delphes} is adapted to correctly place pileup vertices on the beamline.}. \texttt{Delphes} is primarily developed to simulate processes involving prompt particles, and therefore, \texttt{Delphes} is modified to accommodate displaced particles. All particles are propagated to the beginning of the ECAL since we will use its timing information, and their pseudorapidity ($\eta$) and azimuthal angle ($\phi$) are taken to be the one at the ECAL for jet formation. Particles produced within the ECAL and till the end of HCAL are also used for jet formation, since in the experiment these particles will have energy deposits in the calorimeters and hence, contribute to the jets, however might not contribute to the jet timing. In the next section, we briefly discuss the Phase-II upgrades of the HL-LHC runs along with the PUPPI algorithm and how the latter affects displaced jets from LLPs.

\section{High luminosity LHC: Phase-II upgrades and PUPPI algorithm}
\label{sec:hllhcups}

The Phase-II upgrade of the LHC or High Luminosity LHC (HL-LHC) will witness increased $pp$ collisions during each bunch crossing with instantaneous luminosity increasing to $\approx 5.0 \times 10^{34}$\,cm$^{-2}$s$^{-1}$ and peaking at $\approx 7.5 \times 10^{34} \,{\rm cm^{-2}s^{-1}}$ for the ultimate scenario. Amount of PU will increase to 140 interactions per bunch crossing and will peak at 200 compared to a peak instantaneous luminosity of $1.0\times 10^{34}\,{\rm cm^{-2}s^{-1}}$ for Run-1 and 2 of Phase-I LHC design with around 30-50 PU vertices. Several hardware upgrades related to electronic systems and various sub-detectors are proposed to improve the physics performance and to mitigate the adverse effects of the high PU on physics analyses. Also, PUPPI algorithm can be used at L1 of HL-LHC. In this section we discuss these in some detail.

\subsection{Phase-II upgrades}
\label{ssec:upgrades}

Major hardware upgrades for HL-LHC are motivated by the requirement to maintain the Phase-I physics performance at higher luminosity and increased PU interactions. We start this section with a brief discussion of the Phase-II upgrades available at L1, most of which we have already introduced in Section\,\ref{sec:intro}: 
\begin{itemize}
    \item One of the major hardware upgrades at HL-LHC will be the upgrade of electronics of barrel calorimeters for both front-end and back-end systems to accommodate extended level-1 (L1) latency period of 12.5\,$\mu$s, increased L1 trigger rate up to 750\,kHz and high data transfer rate \cite{CERN-LHCC-2017-011}. 
    \item Upgrade of electronic systems on-board the ECAL detector will also provide precise timing information and will help in reducing the noise coming due to increased luminosity. After the hardware upgrade, ECAL is expected to deliver 30\,ps timing resolution for 20\,GeV energy deposition at the beginning of HL-LHC when integrated luminosity will be around 300\,fb$^{-1}$\,\cite{CERN-LHCC-2017-011}. However, this timing resolution is expected to degrade with increasing luminosity.
    \item Outer tracker of CMS at HL-LHC will be upgraded which will pave way for the availability of the tracking information at L1 trigger system\,\cite{CERN-LHCC-2020-004}. The Phase-II outer tracker will be upgraded to stacked silicon strip modules which will enable the reconstruction of tracks at L1. Tracks will be reconstructed at L1 using hits correlations between two closely stacked silicon strip sensors. Tracks with $p_T>2$\,GeV, with a $d_0=0$ constraint, will be selected based on the stubs formed in the tracker layers. 
    \item Extended tracking for displaced tracks is being studied at L1 for HL-LHC\,\cite{CERN-LHCC-2020-004}. Decay products of LLPs come from a displaced vertex and if charged, will lead to displaced tracks in the tracker. Availability of tracking information for these displaced particles will enables us to have more handle on triggering events for e.g., with LLPs decaying to jets. Extended tracking is being studied for displaced tracks with $d_0$ up to 8\,cm. 
    \item Another important upgrade at the CMS for HL-LHC will be the implementation of the particle flow (PF) algorithm at L1. The PF algorithm which has already been implemented at HLT and offline analyses at CMS for Phase-I of LHC helps in reconstructing and identifying each particle individually after combining information from various sub-detectors of CMS. Availability of both track and calorimeter information at L1 has made possible the integration of PF algorithm at L1 in the ``Correlator Trigger'' where information from various standalone detector segments is optimally combined to precisely identify and reconstruct all particles\,\cite{CERN-LHCC-2020-004}.
    \item Finally, availability of tracking and PF information at L1 opens the door for implementing PU mitigation techniques at L1.
    One of the most important PU mitigation algorithm already being used during Phase-I of LHC in HLT and offline analyses, known as PUPPI, is being studied to be implemented at L1.
\end{itemize}

As outlined in Ref.\,\cite{CERN-LHCC-2020-004}, the L1 trigger menu is proposed to use jets after applying the PUPPI algorithm to perform selections robust under the high PU environment of HL-LHC. Therefore, to understand the efficiency of standard jet triggers at L1 of HL-LHC, it is important to check the performance of PUPPI algorithm on displaced jets, which is otherwise optimised for prompt jets. In the subsequent sections, we first explain the PUPPI algorithm and how it mitigates PU, and then study the effect of applying it to LLP scenarios where we have displaced jets in the final state.

\subsection{PUPPI algorithm at HL-LHC: prompt vs displaced jets}
\label{ssec:L1trigger}

At HL-LHC, presence of track collection and particle flow information at L1 has opened the possibility of reconstructing jets, combining both tracks and calorimeter deposits along with the application of PUPPI algorithm, for jet based triggers. PUPPI algroithm\,\cite{Bertolini:2014bba} uses the PF information collected and combined from various sub-detectors to mitigate the neutral PU contribution on particle-by-particle basis. Availability of track information at L1 also aids in separately identifying the charged PU contribution with the knowledge of the Primary Vertex (PV) position. 
For limited usage at L1, we only need position of the hard scattering vertex. PF tracks are assigned different bins along the $z$-axis after sorting them according to their $z$-position. Bin corresponding to the maximum scalar sum $p_T$ is taken as the position of the PV. This method of vertexing is known as ``FastHisto'' method and can be implemented at L1 well within the latency limit.

Tracks which are not originating from the PV are regarded to be coming from PU vertices and are discarded. For each neutral particle, PUPPI algorithm defines a local shape variable to distinguish between a hard collision versus a soft component coming from PU. 
ECAL and HCAL energy deposits with no reconstructed track in the tracker fall in the category of neutral particles.
PUPPI algorithm uses the PV position to calculate the local shape variable for each neutral particle in an event by taking into account the charged particles in the neighbourhood of the neutral particle within a cone of certain radius. Finally, every neutral particle is re-weighted by the weight calculated using the local shape variable, $p_T$ of the particle, and amount of PU in each event. A modified version of PUPPI algorithm is implemented at L1\,\cite{CERN-LHCC-2020-004}, keeping in mind small latency period available at L1. Collection of PF candidates is used as the input to the L1 PUPPI algorithm, which will be available at L1 as a result of presence of tracking at L1. For each neutral particle, the local shape variable, $\alpha$, is calculated using Eq.\,\ref{eq:alpha_shape}.
\begin{equation}
\alpha = \text{log} \sum_{i\in \text{PV}, \Delta \text{R} < 0.3}\frac{\text{min}(p_T^i,p_T^{\text{max}})^2}{\text{max}(\Delta R, \Delta R^{\text{min}})^2}
\label{eq:alpha_shape}
\end{equation}
The sum in Eq.\,\ref{eq:alpha_shape} runs over all the charged particles with $p_T>2$\,GeV coming from the PV which lie within a cone of radius $0.07 (\Delta R^{\text{min}})$ $< \Delta R < 0.3$ from the neutral particle in the barrel region. To avoid disproportionate contribution from a single charged particle, an upper limit is set on the $p_T$ of the charged particle ($p_T^{\text{max}}$), taken to be 50\,GeV here.

Particle flow algorithm is already emulated in \texttt{Delphes-3.5.0} with good consistency\,\cite{deFavereau:2013fsa}. There are three collections required as inputs to the PUPPI algorithm at L1, namely, PF photons, neutral hadrons and tracks, where the tracks collection also include displaced tracks from ``Extended Tracking''. The displaced tracks at L1 follow an efficiency curve as a function of their transverse impact parameter from the beamline as given in the L1 TDR\,\cite{CERN-LHCC-2020-004}. An important thing to note here is that the displaced tracks which are not reconstructed at L1 will appear in the collection of neutral particles since they have energy deposits in the calorimeter with no corresponding tracks.

\begin{figure}[t]
\centering
\includegraphics[width=0.5\textwidth]{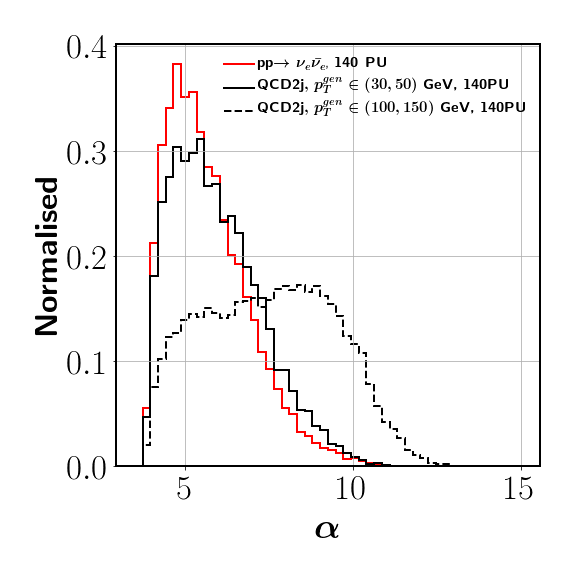}
\caption{Local shape variable, $\alpha$, for neutral particles coming solely from soft QCD which dominates the PU, and QCD dijet events in two $p_T^{gen}$ bins $-$ $p_T^{gen}\in \{30,50\}$\,GeV and $p_T^{gen}\in \{100,150\}$\,GeV, merged with an average of 140 PU vertices.}
\label{fig:alpha_weight}
\end{figure}

To compare the distribution of the local shape variable of neutral particles coming from PU interactions, which are mostly soft, versus those coming from the hard scattering, we have calculated $\alpha$ for these two cases following Eq.\,\ref{eq:alpha_shape}.
As outlined in Section\,\ref{ssec:simu}, PU is generated using \texttt{PYTHIA8} soft QCD process and then merged with the hard process using \texttt{PileUpMerger} module of \texttt{Delphes}. To study the sole effect of PU, we merge the PU interactions with the $pp \rightarrow \nu\bar{\nu}$ process since the neutrinos will go undetected throughout the detector and will not deposit any energy in the calorimeters. For the hard interaction, we use QCD dijet events, generated in $p_T^{gen}$ bins of \{30,50\}\,GeV and \{100, 150\}\,GeV, also merged with PU. We have used an average of 140 PU vertices for each collision event in both the cases to simulate the conditions at the beginning of the HL-LHC runs. Fig.\,\ref{fig:alpha_weight} shows the distributions of $\alpha$ calculated for neutral particles coming solely from PU as well as QCD dijet events in two $p_T^{gen}$ bins of \{30,50\}\,GeV and \{100,150\}\,GeV, in the presence of PU.

We observe from Fig.\,\ref{fig:alpha_weight} that neutral particles coming from the hard QCD dijet process with $p_T^{gen}\in$\,\{100,150\}\,GeV have higher values of $\alpha$ compared to particles coming purely from PU which peak at smaller $\alpha$ values. The distribution for the former is broader, extending even to smaller values of $\alpha$, which is a result of PU particles contaminating the hard jets from the QCD dijet process.
Neutral particles from PU have smaller values of $\alpha$ compared to the hard process since these particles mostly have softer charged PU particles in their vicinity and are also spread over larger $\Delta$R, in contrast to hard scattered process where each neutral particle has increased probability of having more hard tracks in its neighbourhood within smaller $\Delta$R. 
Distribution of $\alpha$ for the low $p_T^{gen}$ bin of \{30,50\}\,GeV is similar to the PU distribution, which is expected since jets coming from PU are also low $p_T$ QCD jets, although it still has a slightly longer tail compared to the distribution of PU. 
The resulting local shape variable is then translated to some weight, where particles from PU are mostly assigned smaller weights while particles coming from the hard QCD process, most probably associated to the vertex identified as the PV, have higher weights. The distributions of weights are shown in Fig.\,\ref{fig:puppi_weights} of Appendix\,\ref{app:puppiwt} .

PUPPI jets are reconstructed by binning the PUPPI candidates in pseudo-trigger towers in the $\eta$-$\phi$ plane divided in 120$\times$72 bins corresponding to bin size of 0.083$\times$0.087. Trigger towers are then clustered in a 7$\times$7 window centered around a local maxima to form jets having a size of 0.6$\times$0.6 in the $\eta$-$\phi$ plane. The transverse momentum of jets is calculated as the scalar sum of $p_T$ of all the trigger towers in the window and the $\eta$-$\phi$ position of the central tower (seed) is taken as the jet position. Jet clustering done at L1 using sliding window method as described above gives similar efficiency as the clustering done using anti-$k_T$ algorithm with a cone size of $R=0.4$. Transverse momentum of the PUPPI candidates is re-weighted according to the weights obtained from PUPPI algorithm to correct for the PU contamination. 

Now that the PUPPI algorithm is discussed in some detail, and we understand how it mitigates PU for prompt processes, like QCD dijet events, we study this algorithm in light of displaced scenarios arising from LLPs. We will study the feasibility of triggering LLP events by triggers constructed using PUPPI jets. As described previously, PUPPI algorithm calculates a local shape variable for each neutral particle in an event by running a sum over ratio of $p_T$ and $\Delta R$ of all the charged particles coming from the primary vertex within $\Delta R<$ 0.3. From Fig.\,\ref{fig:alpha_weight} we observed that the L1 PUPPI algorithm performs reasonably good in discriminating a particle originating from the PV and a particle from a PU vertex. Thus, PUPPI jets can be effectively used to trigger the events where particles are promptly produced. However, for LLPs we face a two-fold problem $-$
\begin{itemize}
    \item The PUPPI algorithm depends on the identification of the PV. For LLPs, due to lack of enough prompt tracks at L1, probability of reconstructing the PV incorrectly increases, as also pointed out in Ref.\,\cite{Bhattacherjee:2020nno}. Therefore, in a fair amount of cases some PU vertex can satisfy the maximum $\sum p_T^2$ condition and be identified as the PV, whereas the vertex actually corresponding to the LLP production might get misidentified as a PU vertex.
    \item Decay products of the LLPs can have higher displacements such that they evade even the ``Extended Tracking'', and as mentioned earlier, these will mostly contribute to the collection of neutral particles at L1. As a result, the number of L1 tracks that can be fed into the PUPPI algorithm to calculate the local shape variable will decrease, which will further reduce the efficacy of using PUPPI jets to trigger on displaced jets. 
\end{itemize}

Performance of the PUPPI algorithm degrades in going from heavier to lighter LLPs since the heavier LLPs will comparatively have more number of associated charged tracks at their production vertex which will help in correctly identifying the PV. Moreover, lighter LLPs have more boost and longer decay lengths in the lab frame which will lead to lesser number of L1 tracks associated with them, even with the extension to displaced tracking at L1. Similarly, for LLPs having longer decay lengths, less and less number of tracks can be reconstructed at L1, thus, decreasing the efficacy of the PUPPI algorithm further with increasing decay length. As a demonstration of how these problems affect the distribution of the local shape variable, we have calculated $\alpha$ for neutral particles in LLP scenarios (A) and (B), as described in Sections\,\ref{ssec:scenarioA} and \ref{ssec:scenarioB} and compare the distributions of $\alpha$ in some benchmark points from these scenarios, merged with an average of 140 PU vertices per collision, with that obtained from merging $pp\rightarrow\nu\bar{\nu}$ process with PU, where the latter captures the sole effect of PU.

\begin{figure}[hbt!]
\centering
\includegraphics[width=0.49\textwidth]{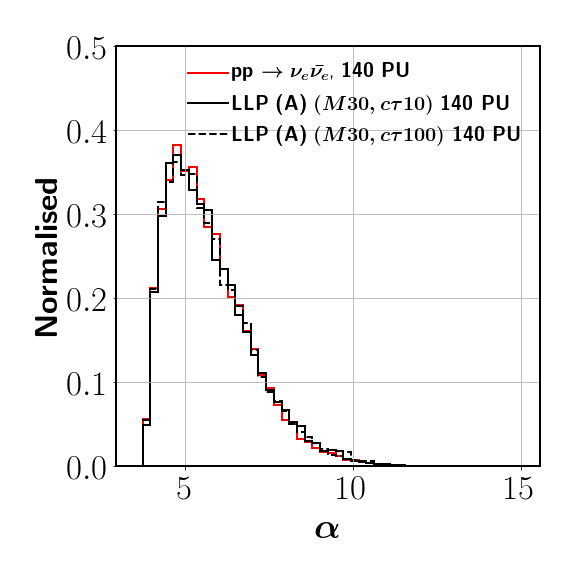}~~
\includegraphics[width=0.49\textwidth]{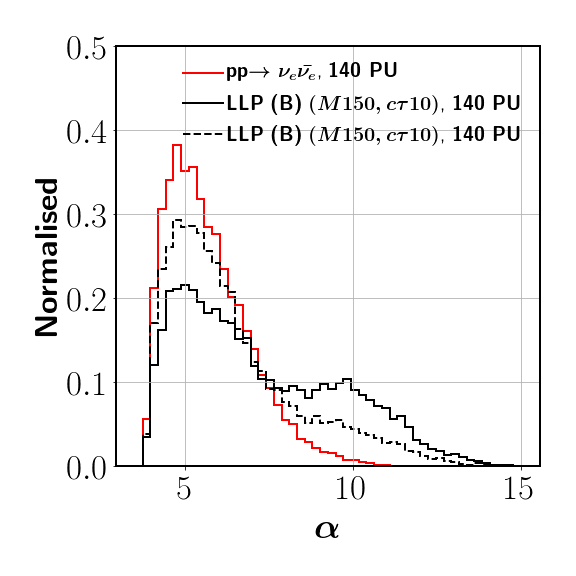}
\caption{Local shape variable, $\alpha$, for LLPs from scenario (A) produced from the decay of 125\,GeV Higgs boson with a mass of 30\,GeV (\textit{left}) and for pair produced LLPs from scenario (B) with a mass of 150\,GeV (\textit{right})$-$ both with decay lengths of 10\,cm and 100\,cm, all merged with an average of 140 PU vertices. Distribution coming from solely the merged 140 PU vertices is also shown for comparison.}
\label{fig:alpha_weight_LLP}
\end{figure} 

The {\it left} panel of Fig.\,\ref{fig:alpha_weight_LLP} shows the distributions of the local shape variable for LLPs having a mass of 30\,GeV with decay lengths 10\,cm and 100\,cm decaying to jets from scenario (A) in the 140 PU environment of HL-LHC compared to the distribution obtained only with PU. As we can clearly see, PUPPI algorithm fails to distinguish neutral particles coming from PU and the LLP signal since there is almost no distinction between the two distributions.
The {\it right} panel of Fig.\,\ref{fig:alpha_weight_LLP} shows similar distributions for LLPs from scenario (B) with a mass of 150\,GeV and decay lengths of 10\,cm and 100\,cm. In this case, there is some distinction between neutral particles from just PU and LLP merged with PU, the latter having a longer tail at higher $\alpha$ values, as compared to scenario (A), where the LLP is lighter.
Also, we observe that with increasing decay length more and more particles start looking like particles from the PU vertex, for reasons as described earlier.

To sum up our observations, PU mitigation using the PUPPI algorithm is not optimal for LLP scenarios, especially for lighter LLPs and those with higher decay lengths. For heavier LLPs, triggers constructed using PUPPI objects can perform better than lighter LLPs; however, even for heavier LLPs, the performance of the PUPPI algorithm in displaced searches is going to be sub-optimal. 

In addition to PUPPI based jet triggers, two separate triggers are designed specifically to explore physics involving displaced jets. As discussed earlier, ``Extended Tracking'' available at L1 enables reconstruction of displaced tracks at L1 with the tracking efficiency varying with $d_0$. Trigger based on jets constructed using L1 tracks, using both extended and prompt track-finding approaches, is included in the trigger menu dedicated for the displaced physics searches. Another displaced jet trigger is constructed based on the ECAL timing information which will also be available at L1. Timing is an important way to select out LLP events since heavier LLPs with large decay lengths travel slower in the colliders due to lesser boost before decaying, and as a result, their decay products will have a time delay compared to prompt relativistic particles from SM processes. For this time based calo-jet trigger, ECAL timing for jets is calculated by averaging the energy-weighted ECAL time-stamp of ECAL towers ($T_{raw}$)\,\cite{CMS:2012bbi}. 

Information regarding displaced triggers as given in L1 TDR \cite{CERN-LHCC-2020-004} is not sufficient enough to understand the peculiarities of time based jet triggers. We now advance our discussion and study closely the timing of a jet $-$ its different measures, how efficiently these can distinguish between displaced and prompt jets, and how these are affected by the various experimental conditions, like detector resolutions and PU.

\section{ECAL barrel timing of a jet: ramifications of the HL-LHC conditions}
\label{sec:ecaltiming}

In the previous section, we discussed how the standard PUPPI jet trigger might not be the best choice for selecting LLP events, especially in the scenarios where LLPs are lighter with larger decay lengths. Dedicated triggers developed by the CMS for displaced searches give an extra handle on selecting LLP events. We now focus on the timing of a jet obtained from ECAL at L1, and discuss the possibility of exploring it for further development of dedicated triggers for LLPs in the light of physics scenarios mentioned in Section\,\ref{sec:signal}.
However, before advancing, there are several effects which are needed to be studied in detail which will have a major influence on how dedicated timing triggers perform in various conditions. The initial HL-LHC runs will have 140 PU interactions which are expected to increase to 200 interactions per event by the end of HL-LHC. The timing resolution of the detectors will also degrade over time as luminosity increases. In the following sections, we will discuss the effect of increased PU interactions on jet distributions, especially on the timing of a jet as well as how we can deal with PU efficiently. We will also discuss how the timing of the jet will get modified with degrading resolution as we move from an integrated luminosity of 300\,fb$^{-1}$ to 4500\,fb$^{-1}$.

\subsection{Effect of PU on the timing of a jet}
\label{ssec:effect-of-PU}

Mitigation of PU for displaced jets is going to be a big challenge at L1. As seen in Section\,\ref{ssec:L1trigger}, jet triggers constructed using PUPPI objects will be suitable for triggering events with prompt jets at L1 without the high PU being much of an issue, however, standard PUPPI jet triggers will not be efficient in selecting events which contain jets originating from the decay of an LLP. We would require different strategies to get a handle on PU for searches involving LLPs decaying to jets. 

\begin{figure}[hbt!]
\centering
\includegraphics[width=0.6\textwidth]{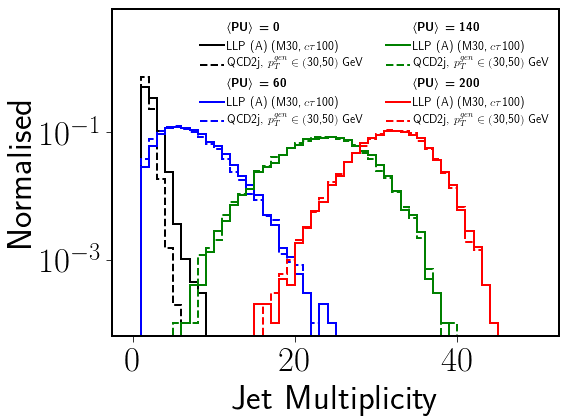}
\caption{Multiplicity of jets having $p_T>20$\,GeV and $|\eta|<1.44$ in an event for QCD dijet process ($p_T^{gen}\in$\,\{30,50\}\,GeV) and LLP benchmark (M30, c$\tau$100) from scenario (A) for 0 PU, 60 PU, 140 PU, and 200 PU.}
\label{fig:jet_multi}
\end{figure}

Let us begin by studying the effect of PU on jet distributions, like jet multiplicity, for LLP signals and QCD background. For this, we have chosen the benchmark with a mass of the LLP being 30\,GeV and decay length 100\,cm from LLP scenario (A). As discussed earlier, LLPs coming from the decay of 125\,GeV Higgs boson are comparatively lighter, and the heavier LLPs have lesser boost, and therefore, further decay to jets with relatively smaller energy deposition in the calorimeters (ECAL and HCAL), making them harder to trigger on. 
For comparison, we have considered QCD dijet events in the bin with $p_T^{gen}\in$\,\{30,50\}\,GeV as the background process. We have considered four PU scenarios with 0, 60, 140, and 200 PU interactions per bunch crossing, where the first one corresponds to the ideal situation, and the other three correspond to PU conditions at Run-II, and beginning and end of HL-LHC, respectively.
The spread of vertices in the time and $z$-direction is taken as given in the \texttt{Delphes} card for Phase-II CMS.
We have clustered the ECAL, and HCAL towers into jets with the anti-$k_T$ algorithm with fixed cone size, which is required to have at least one tower with hadronic energy ($E_{had}$) greater than 2\,GeV in the jet. We have also applied a scale factor of 1.2 in the \texttt{Delphes} card for the jets.
Fig.\,\ref{fig:jet_multi} shows the multiplicity of jets, constructed using anti-$k_T$ jet clustering algorithm with $R=0.4$, having $p_T>20$\,GeV and $|\eta|<1.44$, in an event for signal and background in the four different PU scenarios. 

As we can see from Fig.\,\ref{fig:jet_multi}, mean jet multiplicity is around 2 for both signal and background at 0 PU, which increases to $\approx$ 6 for 60 PU, $\approx$ 20 for 140 PU, and $\approx$ 30 for 200 PU. There is a dramatic increase in the jet multiplicity as we increase the PU to 200, which clearly indicates that at high PU, jet distributions will be dominated by the jets coming from PU interactions. Since PU will have the same effect on both QCD and LLP distributions, it will be impossible to distinguish between signal and background at high PU. Also, since PU will be uniformly distributed throughout the detector, signal jets will get highly contaminated with energy deposits coming from PU as we move towards high PU conditions. This contamination will have an  adverse effect on physics variables constructed for jets.

\begin{figure}[hbt!]
\centering
\includegraphics[width=\textwidth]{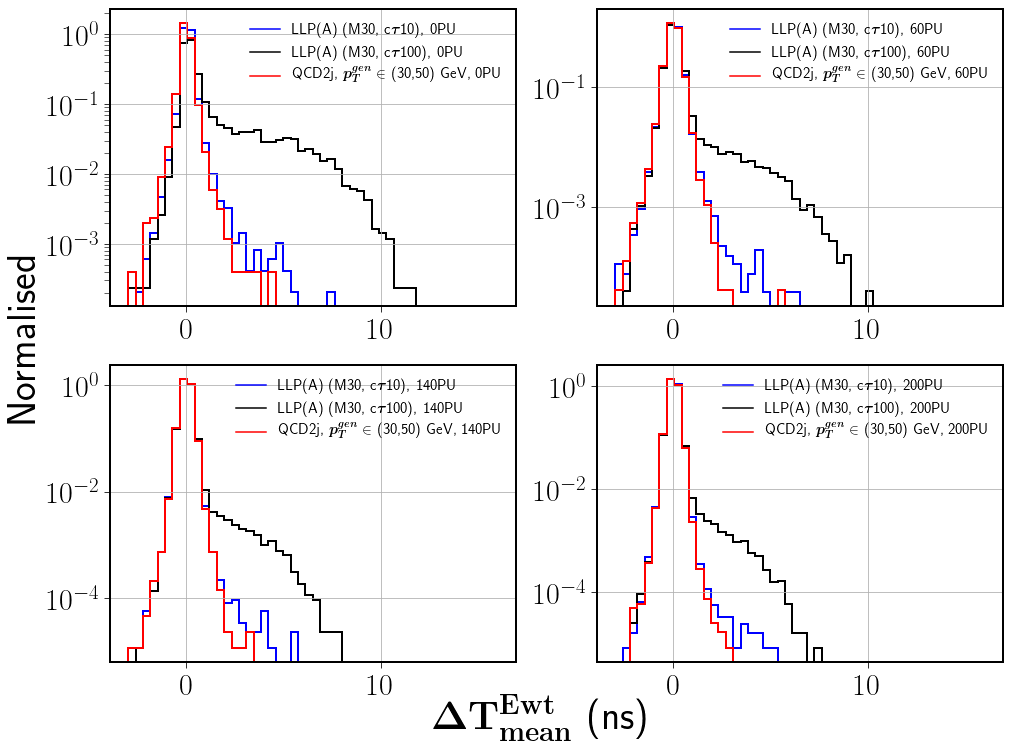}
\caption{Energy-weighted timing of the jets ($\Delta T_{mean}^{Ewt}$) with $R = 0.4$ for two LLP benchmark points in LLP scenario (A) ($M_X$ = 30 GeV and $c\tau=10$\,cm, $M_X$ = 30 GeV and $c\tau=100$\,cm) and QCD dijet process ($p_T^{gen}\in\{30,50\}$\,GeV) for 0 PU, 60 PU, 140 PU, and 200 PU.}
\label{fig:ewt_time_0_4}
\end{figure}

Let us now discuss the effect of PU on ECAL timing of a jet.
ECAL timing for jets is calculated by averaging the energy-weighted ECAL time-stamp of ECAL towers ($T_{raw}$)\,\cite{CMS:2012bbi}. The timing for each tower is calibrated using the definition $\Delta t$ = $T_{raw} - T_{prompt}$ where $T_{prompt}$ is the time taken by a relativistic prompt particle produced at the interaction point to reach at the same ($\eta$-$\phi$) position as the tower.
In order to remove out-of-time PU, we reject ECAL crystals with the time difference, $|\Delta t|>20$\,ns.
We have studied four PU scenarios like before $-$ 0, 60, 140, and 200 average number of vertices per bunch crossing. In Fig.\,\ref{fig:ewt_time_0_4}, we have plotted the energy-weighted timing of the jet ($\Delta T_{mean}^{Ewt}$) with jet cone radius $R = 0.4$ for two benchmark points in LLP scenario (A) where mass of LLP is 30\,GeV with $c\tau=$\,10\,cm and 100\,cm. We have also shown the timing distribution of jets coming from QCD dijet background with parton level $p_T$ restricted in range \{30,50\}\,GeV, for comparison. We select the lowest $p_T^{gen}$ bin since it is expected to have large values of $\Delta T_{mean}^{Ewt}$ compared to other high $p_T$ bins. Timing for each jet is calculated using ECAL crystals with energy deposits greater than 0.5\,GeV as a first-level removal of some of the detector noise and energy deposits from PU, which can also have a high time delay, without affecting the jet $p_T$.

As we can see from Fig.\,\ref{fig:ewt_time_0_4}, the timing of the jet peaks at 0\,ns with a tail reaching up to 15\,ns for the LLP process characterised by the larger decay length (100\,cm). At 0 PU, we are just capturing the jets from the LLP process with no PU contribution, hence, we see a longer tail from the signal jets for the 0 PU scenario. At high PU, the jets will dominantly have energy deposits coming from PU inside them, contributing to the timing of the jet. 
We find that the tail of the distribution reduces with increasing PU, the effect being more pronounced for 140 and 200 PU. A large number of particles from PU inside the signal jet smear the jet timing, and jets that had larger timing, to begin with, now have smaller timing values due to  huge PU contribution. 

\begin{figure}[hbt!]
\centering
\includegraphics[width=\textwidth]{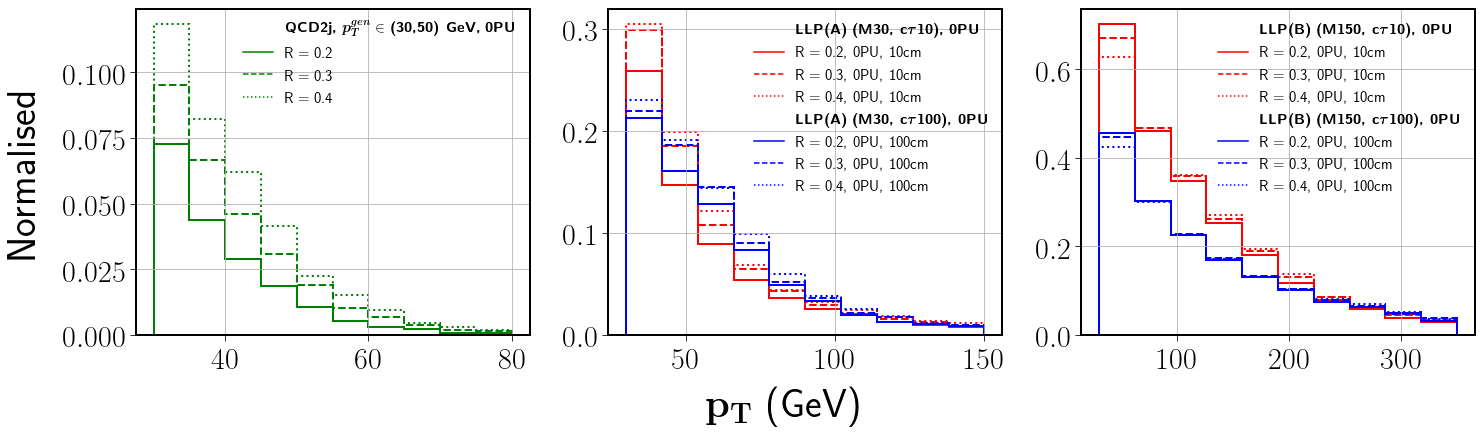}
\caption{The $p_T$ distribution of the jets with jet cone radius $R =$\,0.2, 0.3, and 0.4 for QCD dijet process ($p_T^{gen}\in\{30,50\}$\,GeV) ({\it left}), LLP from scenario (A) $M_X$= 30 GeV with $c\tau$ = 10\,cm and 100\,cm) ({\it center}) and LLP from scenario (B) ($M_X$= 150\,GeV, $c\tau$ = 10\,cm and 100\,cm) ({\it right}) for 0 PU.}
\label{fig:cone_size_QCD_LLP_3050_30100_3010_PT}
\end{figure}

Since PU is uniformly distributed inside the detector, PU contribution inside the jet will depend on the jet area. The smaller the jet area,  the smaller will be the PU contribution. Reducing the jet cone size can effectively mitigate the PU contribution given that jets from the signal process are not affected, and a smaller cone radius captures most of the hadronic activity of the signal. This condition works in our favour, as it was shown in Refs.\,\cite{Banerjee:2017hmw,Bhattacherjee:2019fpt,Bhattacherjee:2020nno}, that displaced jets coming from LLP decays deposit energy in a smaller area of the $\eta$-$\phi$ plane. To further study the feasibility of using smaller cone sizes for our analysis, we show the effect of different cone sizes on jet $p_T$ distribution for LLP and QCD in Fig.\,\ref{fig:cone_size_QCD_LLP_3050_30100_3010_PT}. We have plotted the jet $p_T$ distributions for LLP scenarios (A) and (B) with 0 PU interactions along with QCD dijet events ($p_T\in\{30,50\}$\,GeV). We have considered two benchmark points for each scenario with decay lengths 10\,cm and 100\,cm $-$ for scenario (A), the mass of the LLP is taken to be 30\,GeV and for scenario (B), it is chosen as 150\,GeV. 

As we can clearly observe from Fig.\,\ref{fig:cone_size_QCD_LLP_3050_30100_3010_PT}, $p_T$ of QCD jets are most affected by the reduction in jet cone size which indicates that QCD jets are broader and their energy deposit cannot be entirely contained in smaller jet cone radius than 0.4 for low $p_T$ ($p_T^{gen}\in\{30,50\}$\,GeV). On the other hand, $p_T$ of displaced jets from LLP decay is less affected as we decrease the cone radius from 0.4 to 0.2 since they are narrower compared to QCD jets depending on their displacement and can be contained in smaller cone sizes. While most of the hadronic activity of LLP jets in scenario (A) for benchmark point M30,c$\tau$10 can be effectively contained in $R = 0.3$, for LLP with 100\,cm decay length, even narrower cone radius of $R = 0.2$ is sufficient for containing most of the energy deposit. For pair produced LLPs in scenario (B), $R = 0.2$ jets will suffice to capture most of the hadronic activity of the LLP jets, since these are more boosted compared to scenario (A).

\begin{figure}[hbt!]
\centering
\includegraphics[width=0.9\textwidth]{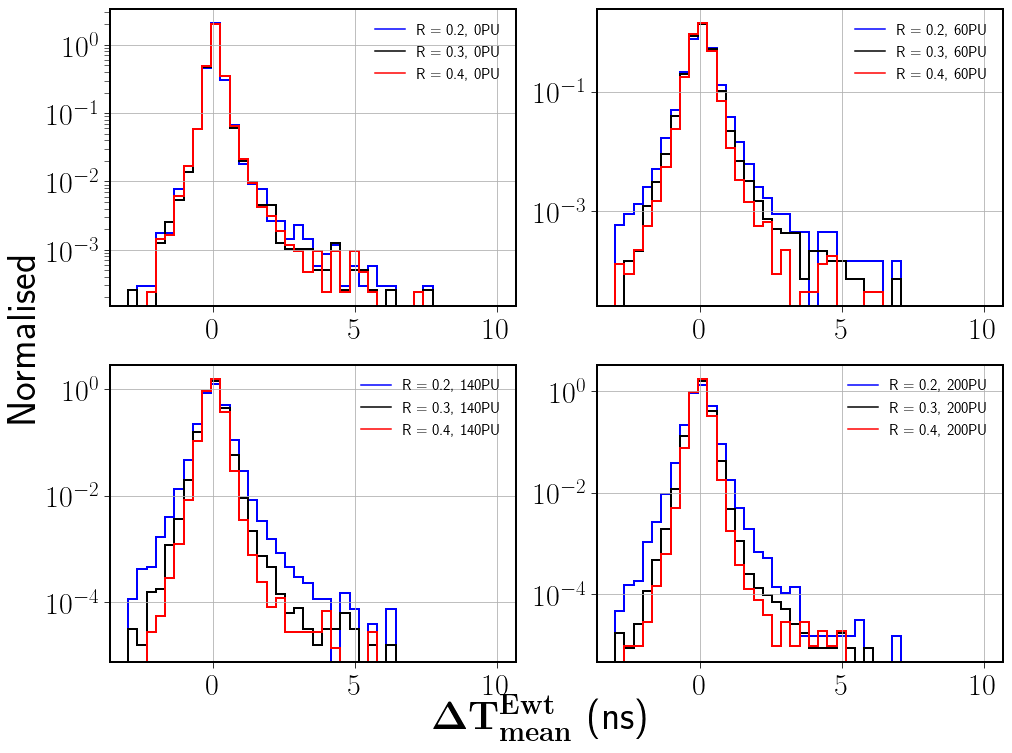}
\caption{The timing of the jets ($\Delta T_{mean}^{Ewt}$) with jet cone radius $R =$\,0.2, 0.3, and 0.4 for LLP scenario (A) with benchmark point ($M_X$= 30\,GeV, $c\tau$ = 10\,cm) for 0 PU (\textit{top left}), 60 PU (\textit{top right}), 140 PU (\textit{bottom left}), and 200 PU (\textit{bottom right}).}
\label{fig:ewt_time_difR_3010}
\end{figure}

\begin{figure}[hbt!]
\centering
\includegraphics[width=0.9\textwidth]{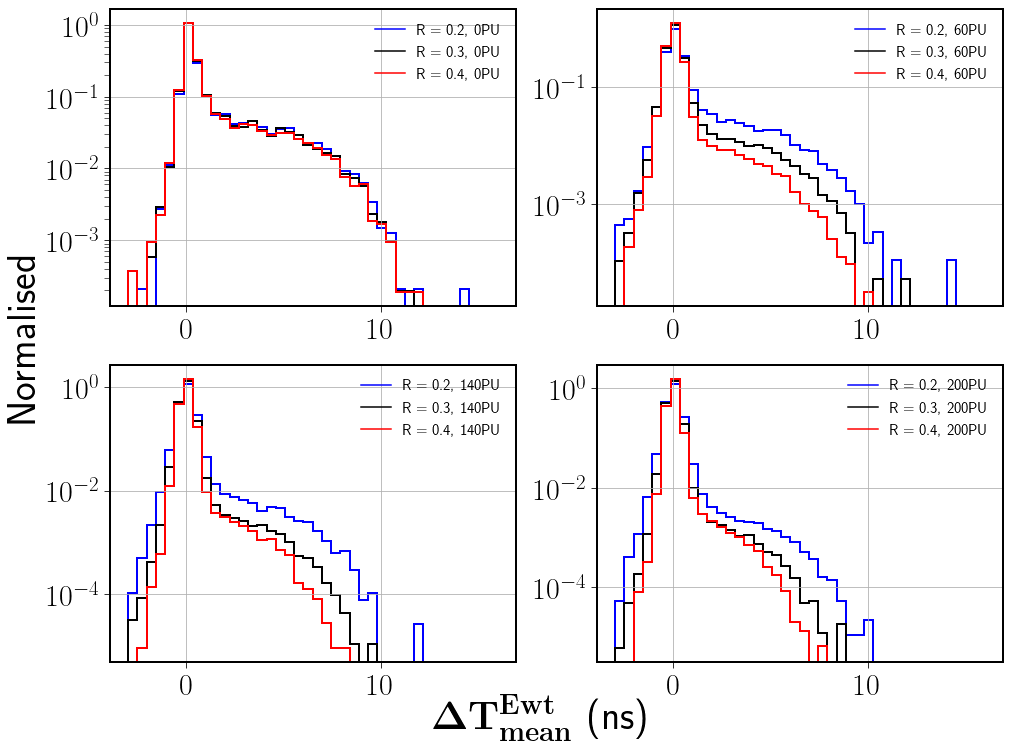}
\caption{The timing of the jets ($\Delta T_{mean}^{Ewt}$) with jet cone radius $R =$\,0.2, 0.3, and 0.4 for LLP scenario (A) with benchmark point ($M_X$= 30\,GeV, $c\tau$ = 100\,cm) for 0 PU (\textit{top left}), 60 PU (\textit{top right}), 140 PU (\textit{bottom left}), and 200 PU (\textit{bottom right}).}
\label{fig:ewt_time_difR_30100}
\end{figure}

We now study the effect of varying the jet cone size on the timing distributions of jets for scenario (A). Figs.\,\ref{fig:ewt_time_difR_3010} and \ref{fig:ewt_time_difR_30100} show the timing distribution of jets from LLP scenario (A) for benchmark ($M_X$= 30\,GeV, $c\tau$ = 10\,cm) and ($M_X$= 30\,GeV, $c\tau$ = 100\,cm), respectively, for jet cone radii of $R =$\,0.2, 0.3, and 0.4, each in the 0 PU, 60 PU, 140PU, and 200 PU conditions.
We can clearly see from the \textit{top left} panel of  Figs.\,\ref{fig:ewt_time_difR_3010} and \ref{fig:ewt_time_difR_30100}, that for 0 PU, when jets are clustered purely from the energy deposits coming from the decay of the LLP signal with no PU contribution, the timing of the jet does not change with a decrease in the jet cone radius from $R=0.4$ to $R=0.2$, which again inspires the use of smaller cone radius when dealing with displaced jets from LLP processes.
Fig.\,\ref{fig:cone} in Appendix\,\ref{app:conesize} shows similar plots where we fix the PU scenario and draw the $\Delta T_{mean}^{Ewt}$ distributions for varying cone-sizes for two LLP benchmarks from scenario (A), which implies that for all values of cone-size, increasing the PU reduces the tail towards high values of $\Delta T_{mean}^{Ewt}$ variable.

As the amount of average number of PU per event is increased to 200, $R = 0.4$ jets being most contaminated by the PU have a shorter tail in the timing distribution, while for $R = 0.2$ and 0.3 jets, we can see more jets with larger jet timing with longer tail in the $\Delta T_{mean}^{Ewt}$ distribution which clearly indicates that jets clustered with $R =$\,0.2 and 0.3 have much less PU contamination even in the ultimate PU scenario. As discussed above, $R = 0.3$ jets will be able to mostly contain the energy deposit of displaced jets in LLP scenarios (A) as well as (B), and therefore, it won't affect the efficiency of $p_T$ threshold in triggering of events. The distributions in 0 PU condition show that timing of the signal jets is not affected by the reduction of jet cone size from $R = 0.4$ to $R = 0.3$, however, the longer tails in the timing distributions are revived at higher PU scenarios for $R=0.3$.
This aids in discriminating displaced jets from prompt ones in a better way at HL-LHC. 
For the rest of the paper, we have used jets clustered using anti-$k_T$ algorithm with jet cone radius fixed at $R = 0.3$, unless stated otherwise, since using narrower jets effectively decreases the PU contamination inside the jet. 

\subsection{Effect of the ECAL timing resolution on jet timing}
\label{ssec:effect-of-reso}

In time-sensitive analyses, ECAL timing resolution will be a key factor. From the preliminary studies done at CMS, it is expected that the ECAL detector will be able to achieve a timing resolution of 30\,ps for 20-25\,GeV ECAL towers at the beginning of HL-LHC. The timing resolution of the ECAL detector will degrade over time. As given in TDR\,\cite{CERN-LHCC-2017-011}, the major contribution to timing resolution will be from noise coming from readout electronics which will increase as luminosity increases over time. By the end of the HL-LHC, ECAL is expected to accommodate target timing resolution of 30\,ps for 50-90\,GeV energy towers\,\cite{CERN-LHCC-2017-011}.

\begin{figure}[hbt!]
\centering
\includegraphics[width=0.55\textwidth]{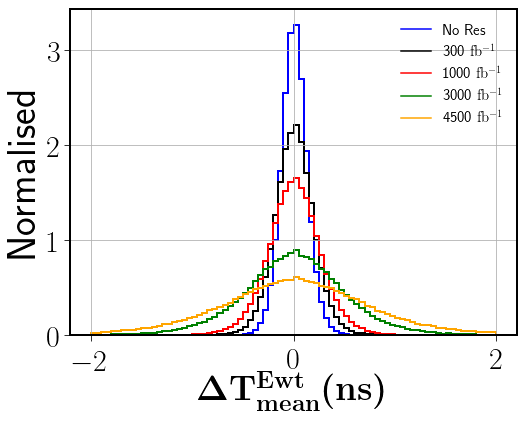}
\caption{The timing distribution ($\Delta T_{mean}^{Ewt}$) of jets for LLP benchmark from scenario (A), $M_X$ = 30\,GeV, c$\tau=$ 10\,cm, with $R = 0.3$ cone radius with no timing resolution, and resolution after 300 ${\rm fb^{-1}}$, 1000 ${\rm fb^{-1}}$, 3000 ${\rm fb^{-1}}$ and 4500 ${\rm fb^{-1}}$ of integrated luminosity collected in the 140 PU scenario.}
\label{fig:mean_ewt_diffRes_3010}
\end{figure}

ECAL TDR has shown 
the time resolution discretely for $\eta = 0.0$ and $\eta = 1.45$ as a function of energy. We have assumed that timing resolution will vary linearly between $\eta = 0.0$ and $\eta = 1.45$, and thus, we have linearly extrapolated the resolution function for all $\eta$ values between 0 and 1.45.  
In Fig.\,\ref{fig:mean_ewt_diffRes_3010}, we have shown the effect of different timing resolutions on the timing of the jet for LLP scenario (A). Timing of each ECAL crystal is smeared following the Gaussian distribution for each scenario where we have applied a $p_T$ and $\eta$ dependent timing resolution corresponding to integrated luminosity of 300\,fb$^{-1}$, 1000\,fb$^{-1}$, 3000\,fb$^{-1}$, and 4500\,fb$^{-1}$.

\begin{figure}[hbt!]
\centering
\includegraphics[width=0.474\textwidth]{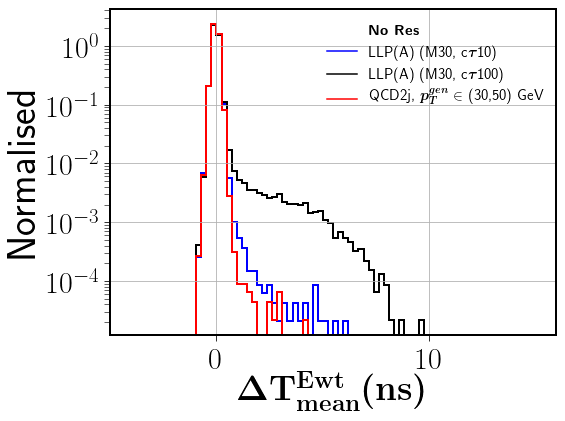}\qquad
\includegraphics[width=0.474\textwidth]{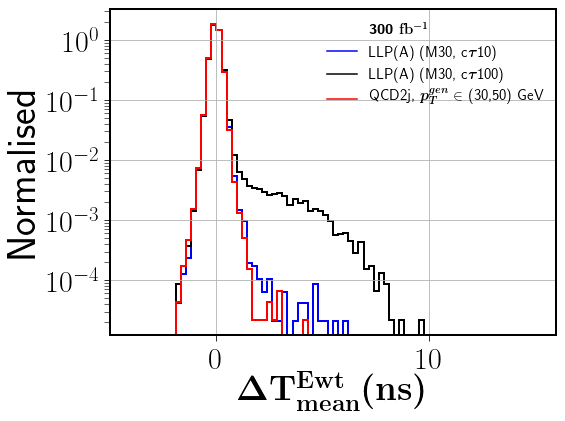}\\
\includegraphics[width=0.474\textwidth]{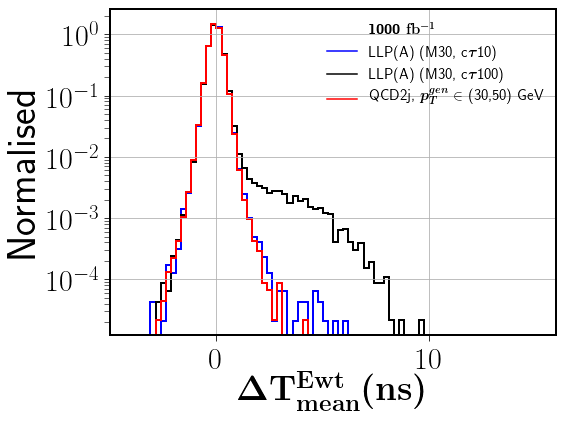}\qquad
\includegraphics[width=0.474\textwidth]{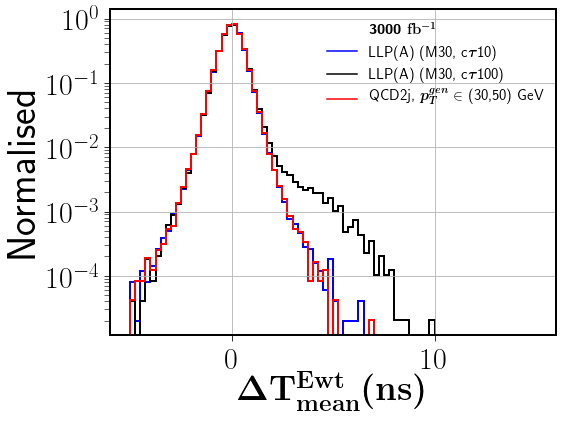}\\
\includegraphics[width=0.474\textwidth]{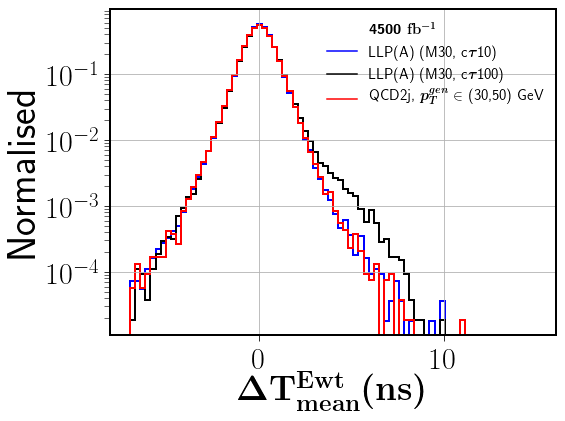}
\caption{The timing distribution ($\Delta T_{mean}^{Ewt}$) of the jets for LLP (A) benchmark ($M_X$ = 30 GeV, c$\tau=$10cm) and QCD dijet background ($p_T^{gen} =$\{30,50\}\,GeV) with $R = 0.3$ cone radius without any timing resolution (\textit{top left}), and with timing resolution after 300\,fb$^{-1}$ (\textit{top right}), 1000\,fb$^{-1}$ (\textit{center left}), 3000\,fb$^{-1}$ (\textit{center right}) and 4500\,fb$^{-1}$ (\textit{bottom}) of integrated luminosity collected in the 140 PU scenario.}
\label{fig:mean_ewt_diffRes}
\end{figure}

Fig.\,\ref{fig:mean_ewt_diffRes} shows timing distribution for two LLP benchmarks ($M_X$=30 GeV, c$\tau$ = 10\,cm and 100\,cm) from scenario (A) and QCD dijet events ($p_T^{gen} \in \{30,50\}$\,GeV). LLP with shorter decay length is affected more with degrading timing resolution. We can see a slightly longer tail in the timing distribution for LLP with c$\tau=$ 10\,cm compared to the QCD dijet events at the beginning of HL-LHC when timing resolution is much better, and this distinction degrades in going towards the end of HL-LHC. Tail of distribution for LLP characterised with a higher decay length of 100\,cm is affected comparatively less with degrading timing resolution because the timing of the jets coming from highly displaced LLPs is relatively quite large compared to the noise. With degrading timing resolution, tail in the timing distribution of QCD jets also broadens as we move from 300\,fb$^{-1}$ to 4500\,fb$^{-1}$. 
Thereby, we can conclude that timing resolution will greatly affect how timing distributions will look like for jets coming from QCD dijet and LLP processes and there will be more spread in the timing distribution of the jets as we go towards the end of the HL-LHC due to degradation in the timing resolution.
Hence it will become more and more difficult to distinguish LLP events from QCD ones towards the end of HL-LHC.

\subsection{Why do prompt QCD jets having high time delays?}
\label{ssec:qcd-time}

In the previous sections, we have observed that even some prompt jets from QCD dijet process have high timing values. In this section, we try to understand the origin of such jets from QCD. We have seen that resolution of the ECAL timing is one of the reasons for smearing the jet timing and hence giving larger values even when the actual jet is not delayed. There are two other factors that might add to the tail of the prompt jet timing distribution to higher values. First being the intrinsic spread of the beamspot in both the temporal and longitudinal direction, which can make even prompt jets appear delayed with high $\Delta T_{mean}^{Ewt}$.
Second is the presence of some long-lived SM particles whose decay products can be delayed, and hence shift the jet timing to higher values. We now discuss each of these effects briefly.

\subsubsection{Spread of the beamspot}
\label{sssec:beamspot}

\begin{figure}[hbt!]
\centering
\includegraphics[height=6.1cm]{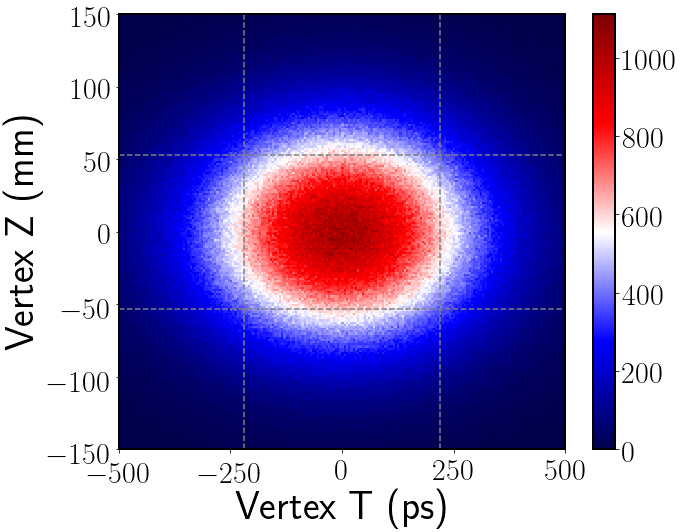}\qquad
\includegraphics[height=6.1cm]{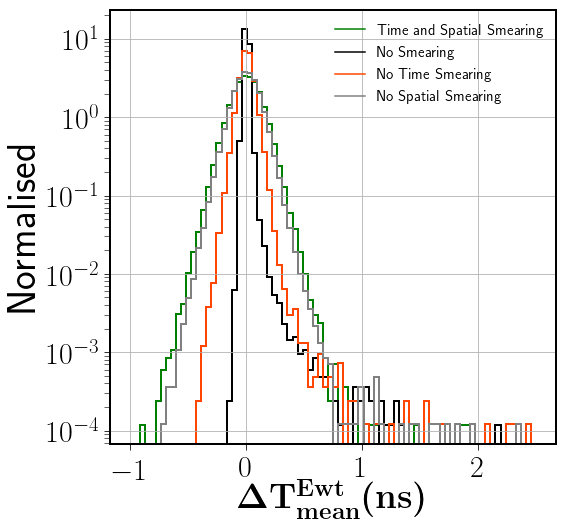}
\caption{{\it Left:} Distribution of vertices on the beamline in the temporal and longitudinal direction at HL-LHC for 1 million soft QCD events with an average of 140 vertices in each event. {\it Right:} Distribution of energy-weighted mean timing ($\Delta T_{mean}^{Ewt}$) of QCD prompt jets from dijet process in the $p_T^{gen}\in \text\{30,50\}$\,GeV bin with and without both the time and spatial smearing, along with their individual effects.}
\label{fig:qcd_time1}
\end{figure}

For HL-LHC, the position of vertices on the beamline will follow a Gaussian distribution with a  standard deviation of approximately 220\,ps in time and 55\,mm in longitudinal ($z$) direction around the center as shown in the {\it left} panel of Fig.\,\ref{fig:qcd_time1}.
We already know that spread in the timing distribution of QCD jets is determined by the ECAL timing resolution (see Fig.\,\ref{fig:mean_ewt_diffRes}). It can also be due to this intrinsic spread of the beamspot in both the temporal and longitudinal direction. We have already studied the effect of ECAL timing resolution on the shape of timing distribution in the previous section. In Fig.\,\ref{fig:qcd_time1} ({\it right}), we show energy-weighted mean timing of QCD jets without applying any ECAL timing resolution, with and without the spread in both the temporal and longitudinal directions, and also with only one of them kept on. As we can see from Fig.\,\ref{fig:qcd_time1}, while the longitudinal spread of vertices contributes to the time delay, the spread of vertices in time plays the dominant role in increasing the timing of QCD jets. The effect of longitudinal spread is less since the spread of vertices is only 55\,mm in $z$-direction compared to the radial distance of the ECAL from the beamline, which is around 1290\,mm.
The effect of this beamspot spread can also be seen in Fig.\,\ref{fig:ewt_time_0_4} where even in the 0 PU scenario, the QCD and LLP benchmark with the shorter decay length of 10\,cm have very similar distributions of $\Delta T_{mean}^{Ewt}$.

\subsubsection{Long-lived particles in SM}
\label{sssec:SM-LLP}

\begin{figure}[hbt!]
\centering
\includegraphics[width=0.5\textwidth]{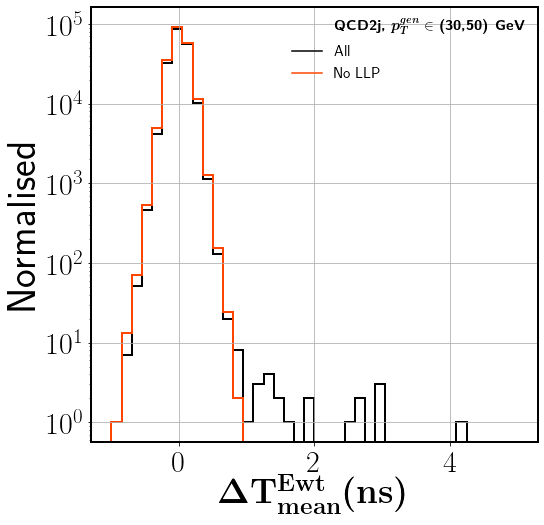}~~
\caption{Distribution of energy-weighted mean timing ($\Delta T_{mean}^{Ewt}$) of QCD jets in the $p_T^{gen}\in\text\{30,50\}$\,GeV bin with and without veto on the decay of long-lived particles in SM as discussed in the text.}
\label{fig:qcd_time2}
\end{figure}

Besides the overall change in the shape of timing distribution because of the intrinsic beamspot spread, we can also see the presence of some spurious jets with suspiciously significant time delay in the tail of the distribution. The 
high timing of these jets might be a result of the presence of SM particles with a relatively longer lifetime in the jet. 
Particles like $K_S$, $\Lambda$, $\Omega$ etc. are long lived in the detector and can decay within the tracker volume after traversing some distance depending on their boost and proper mean decay length, where the latter is few centimeters. To study the effect of presence of these long lived particles on the timing of the jets, these particles are made stable at the generator level. 
In the default \texttt{Delphes} setting, $K_S$ and $\Lambda$ are made to deposit 30\% of their energy in ECAL and the rest in HCAL.
Since we want to study the effect of what happens when these particles are removed from the jet timing measure, we make these particles deposit 100\% of their energy in the HCAL, and as a result, they do not contribute to the ECAL timing of the jet.
In Fig.\,\ref{fig:qcd_time2}, we have shown the timing distribution of prompt jets from QCD dijet events with and without the veto on the decay of above mentioned long-lived SM particles. We observe that the longer tail of high jet timing reduces when these SM long-lived particles are made stable, and therefore, we can infer that these jets in the tail of the distribution owed their high values of $\Delta T_{mean}^{Ewt}$ to the presence of SM long-lived particles.
\\

All these effects $-$ the increased amount of PU at HL-LHC, the timing resolution of the ECAL detector, which degrades with data taking, the spread of vertices in the temporal and longitudinal directions as well as the presence of SM long-lived particles, are important factors affecting the timing of a jet. We now discuss how the ECAL timing can be used for dedicated triggers of long-lived particles decaying to jets.

\section{Developing triggers for displaced jets based on ECAL timing}
\label{sec:ecaltrigger}

The previous section discusses various factors affecting the timing of a jet, and how we can control PU for LLP signal by constructing narrow jets. In this section, we study how the ECAL timing information of a jet can be exploited in various ways to identify displaced jets from the large background of prompt jets from SM processes, dominated mainly by QCD dijet events. Furthermore, we discuss how this can be used to construct L1 triggers for the HL-LHC.

\subsection{Timing variables}
\label{ssec:timing-vars}

Jets are objects consisting of many particles, and as a result, the timing of a jet has to be some statistical measure using the individual timing of the jet components. Since we are focusing on the CMS ECAL timing here, we need to define the jet timing in terms of the timing of each ECAL crystal associated with a jet.
In previous sections, we have always talked about the energy-weighted mean timing of the jet, $\Delta T_{mean}^{Ewt}$, which has also been used by the CMS collaboration in designing their trigger based on ECAL timing\,\cite{CERN-LHCC-2020-004}. We can construct many such variables defining the timing of a jet. In the present work, we have constructed several timing variables using the ECAL timing information available at L1 to understand which of them might be useful in differentiating prompt jets from displaced jets. 

The following list describes the various measures used by us to define jet timing: 
\begin{itemize}
\item $ \mathbf{\Delta T_{mean}}$: mean of the timing of all the ECAL crystals contained within the jet. 
\begin{equation}
    \Delta T_{mean} = \frac{\sum \Delta T_{i}}{N}
    \label{eq:delTmean}
\end{equation}
where $i$ runs over all the ECAL crystals inside the jet and $N$ is the total number of crystals associated with the jet.
\item $\mathbf{\Delta T_{median}}$: median of the timing of all the ECAL crystals in a jet. 
\item $\mathbf{\Delta T_{RMS}}$: RMS (root mean square) of the timing of all the ECAL crystals in a jet. 
\begin{equation}
    \Delta T_{RMS} = \sqrt{\frac{\sum \Delta T_{i}^2}{N}}
    \label{eq:delTmean}
\end{equation}
where $i$ runs over all the ECAL crystals inside the jet and $N$ is the total number of crystals associated with the jet.
\item $\mathbf{\sum \Delta T}$: sum of the timing of all the ECAL crystals in a jet. 
\item $\mathbf{\Delta T_{mean}^{Ewt}}$: energy-weighted mean of the timing of all the ECAL crystals in a jet. 
\begin{equation}
    \Delta T_{mean}^{Ewt} = \frac{\sum \Delta T_{i}\times E_{i}}{\sum E_{i}},\,\,i\equiv\text{crystals inside the jet}
    \label{eq:delTmean-ewt}
\end{equation}
\item $\mathbf{\Delta T_{mean}^{ETwt}}$: transverse energy-weighted mean of the timing of all the ECAL crystals in a jet. 
\begin{equation}
    \Delta T_{mean}^{ETwt} = \frac{\sum \Delta T_{i}\times E_{T,i}}{\sum E_{T,i}},\,\,i\equiv\text{crystals inside the jet}
    \label{eq:delTmean-eTwt}
\end{equation}
\end{itemize}
The timing of each crystal is calibrated with respect to the origin as discussed earlier, and for the energy and transverse energy-weighted measures, calibration is applied before re-weighting with energy.
We have also constructed all of the above timing variables using two more timing calibration techniques where the timing of each crystal in the jet is calibrated with respect to the PV and the jet vertex (JV) from which that particular jet is originating. Note that PV is reconstructed using prompt track collection available at L1 by selecting the vertex with largest $\sum p_T^2$. The JV is found out in a similar way by using all the prompt tracks associated with the jet, i.e., lying within $\Delta R<0.3$ of the jet axis at the ECAL, and selecting the vertex with the maximum $\sum p_T^2$ value.

\begin{figure}[t]
\centering
    \includegraphics[width=0.44\textwidth]{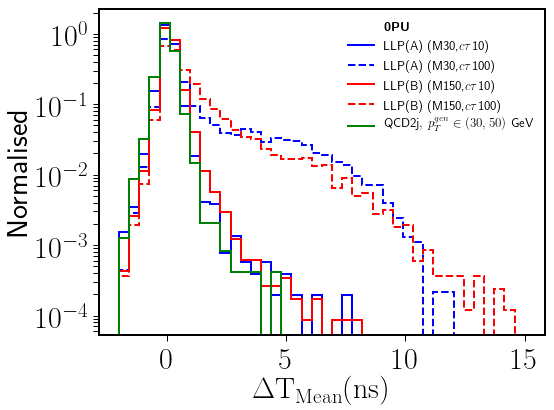}\qquad
    \includegraphics[width=0.44\textwidth]{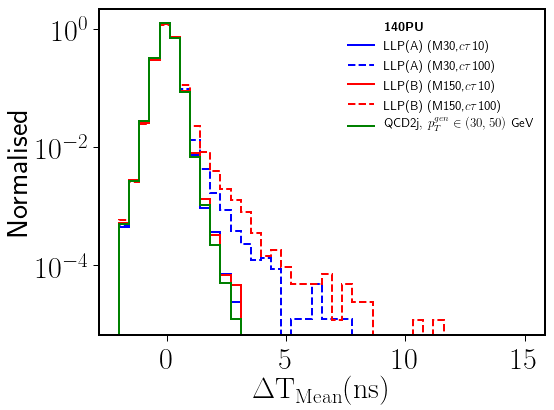}\\ 
    \includegraphics[width=0.44\textwidth]{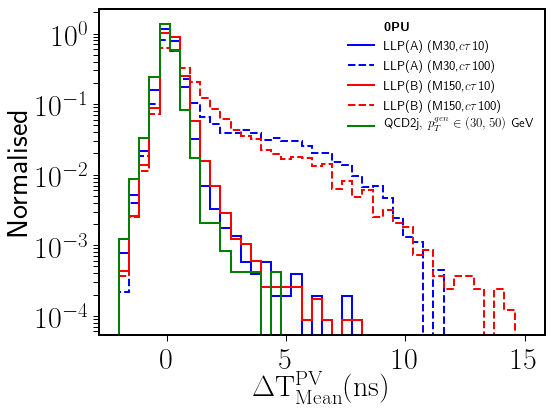}\qquad
    \includegraphics[width=0.44\textwidth]{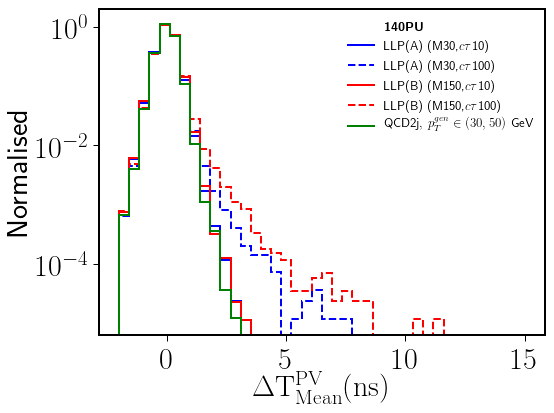}\\
    \includegraphics[width=0.44\textwidth]{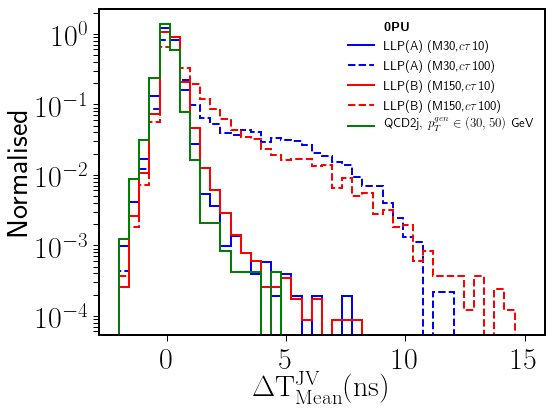}\qquad
    \includegraphics[width=0.44\textwidth]{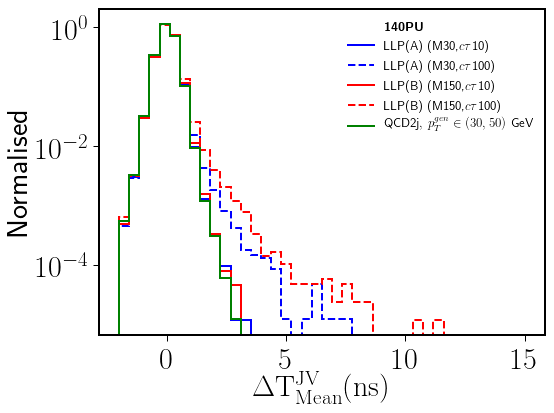}
    \caption{Distributions of mean timing ($\Delta T_{mean}$) of a jet when timing of each crystal is calibrated with respect to the origin ({\it top}), primary vertex, PV ({\it center}), and jet vertex, JV ({\it bottom}) in the ideal situation with 0 PU ({\it left}) and beginning of HL-LHC with 140 PU ({\it right}) for two benchmarks each from LLP scenarios (A) and (B), and QCD dijet events in the bin with $p_T^{gen}\in\{30,50\}$\,GeV.}
    \label{fig:mean-time}
\end{figure}
    
\begin{figure}[t]
\centering    
    \includegraphics[width=0.45\textwidth]{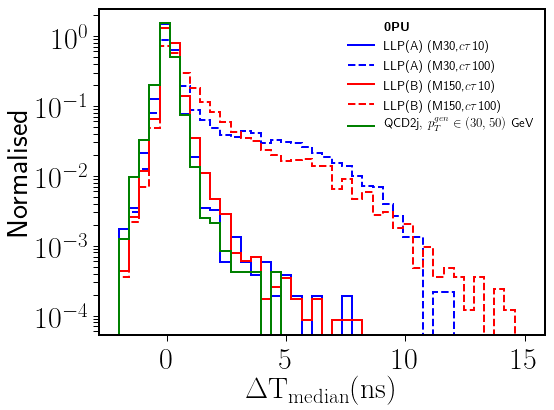}\qquad
    \includegraphics[width=0.45\textwidth]{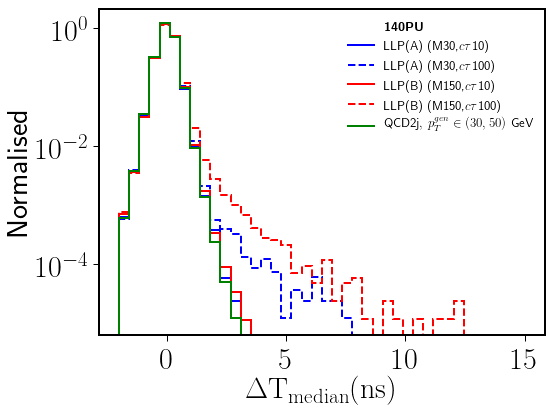}\\
    \includegraphics[width=0.45\textwidth]{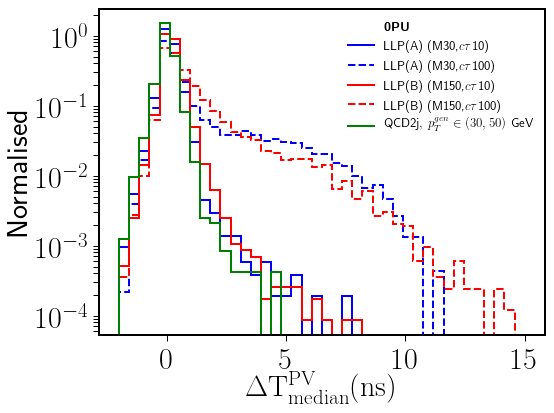}\qquad 
    \includegraphics[width=0.45\textwidth]{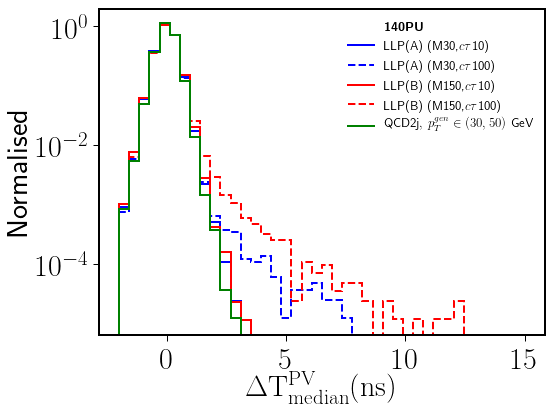}\\
    \includegraphics[width=0.45\textwidth]{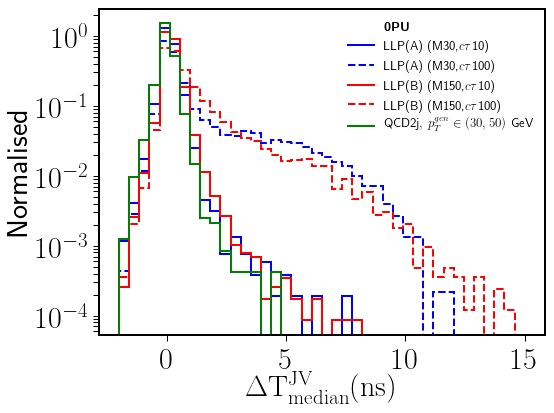}\qquad
    \includegraphics[width=0.45\textwidth]{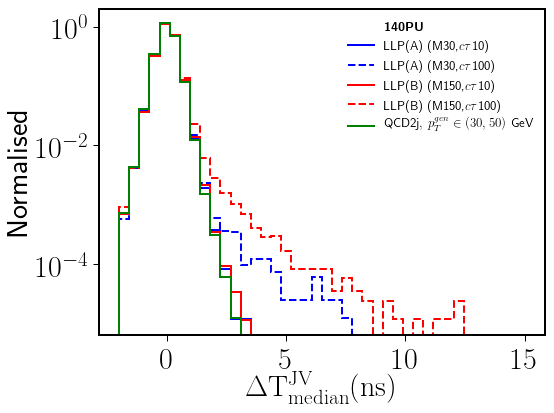}    
    \caption{Distributions of median timing ($\Delta T_{median}$) of a jet when timing of each crystal is calibrated with respect to the origin ({\it top}), primary vertex, PV ({\it center}), and jet vertex, JV ({\it bottom}) in the ideal situation with 0 PU ({\it left}) and beginning of HL-LHC with 140 PU ({\it right}) for two benchmarks each from LLP scenarios (A) and (B), and QCD dijet events in the bin with $p_T^{gen}\in\{30,50\}$\,GeV.}
    \label{fig:median-time}
\end{figure}


\begin{figure}[t]
\centering

    \includegraphics[width=0.45\textwidth]{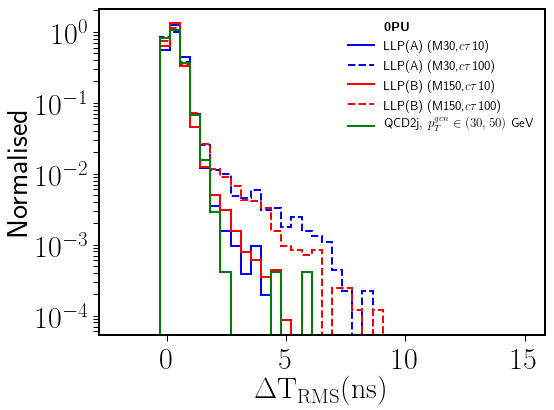}\qquad
    \includegraphics[width=0.45\textwidth]{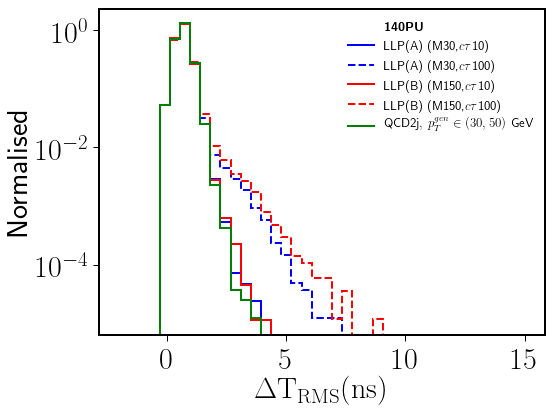}\\
    \includegraphics[width=0.45\textwidth]{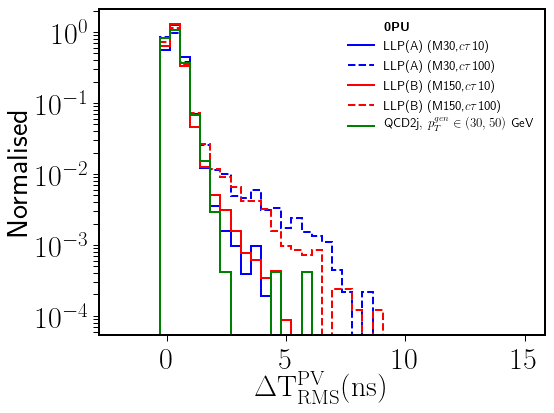}\qquad 
    \includegraphics[width=0.45\textwidth]{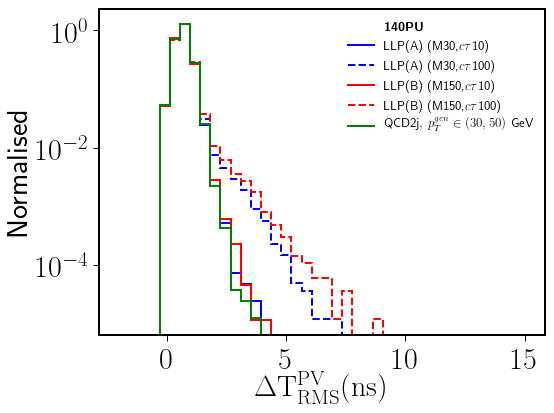}\\
    \includegraphics[width=0.45\textwidth]{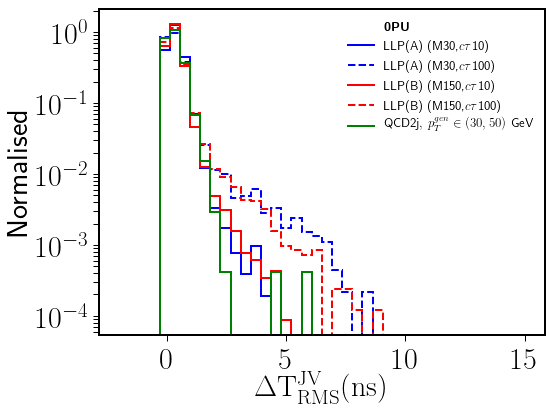}\qquad
    \includegraphics[width=0.45\textwidth]{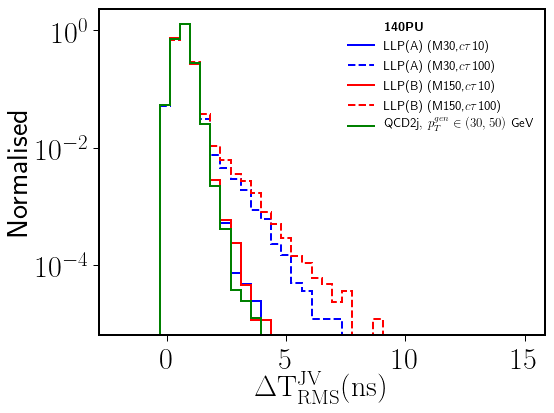}
    \caption{Distributions of RMS timing ($\Delta T_{RMS}$) of a jet when timing of each crystal is calibrated with respect to the origin ({\it top}), primary vertex, PV ({\it center}), and jet vertex, JV ({\it bottom}) in the ideal situation with 0 PU ({\it left}) and beginning of HL-LHC with 140 PU ({\it right}) for two benchmarks each from LLP scenarios (A) and (B), and QCD dijet events in the bin with $p_T^{gen}\in\{30,50\}$\,GeV.}
    \label{fig:rms-time}
\end{figure}    
    
\begin{figure}[t]
\centering    
    \includegraphics[width=0.45\textwidth]{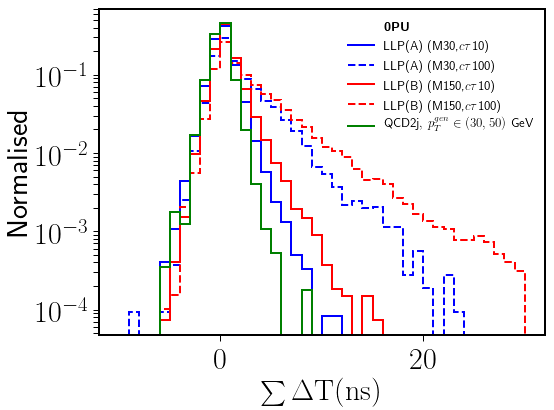}\qquad
    \includegraphics[width=0.45\textwidth]{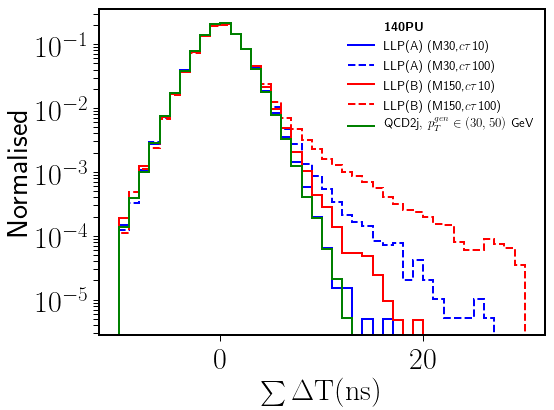}\\ 
    \includegraphics[width=0.45\textwidth]{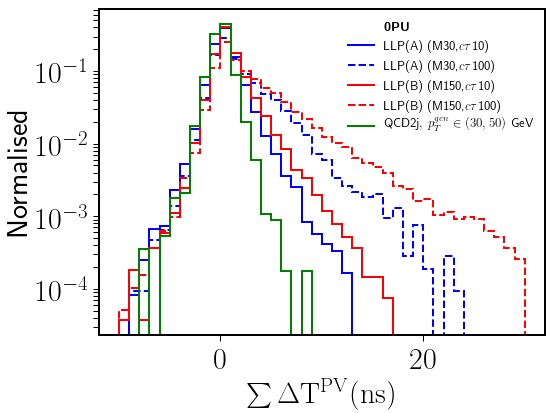}\qquad 
    \includegraphics[width=0.45\textwidth]{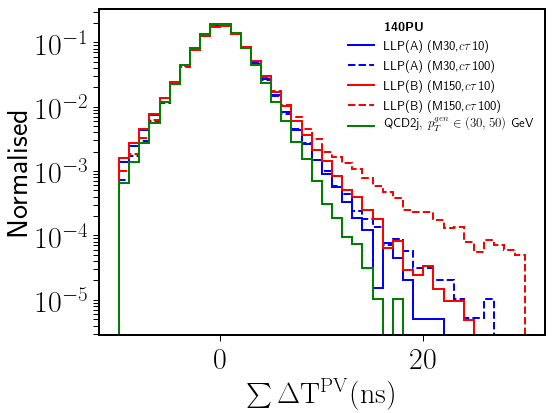}\\
    \includegraphics[width=0.45\textwidth]{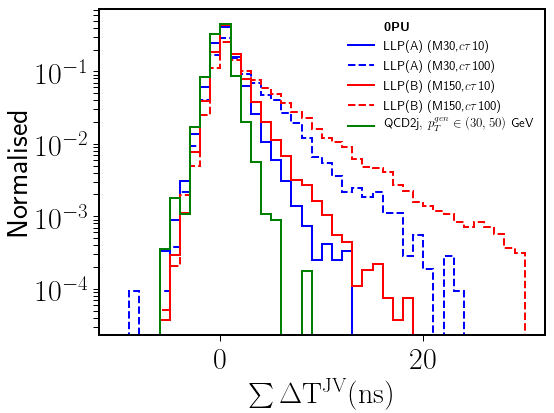}\qquad
    \includegraphics[width=0.45\textwidth]{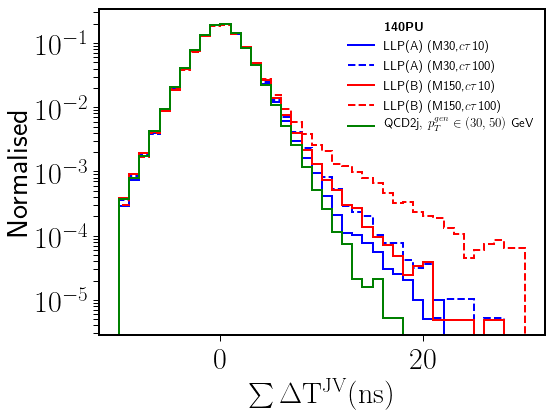}
    \caption{Distributions of sum of all ECAL crystals associated with a jet ($\Delta T_{sum}$) when timing of each crystal is calibrated with respect to the origin ({\it top}), primary vertex, PV ({\it center}), and jet vertex, JV ({\it bottom}) in the ideal situation with 0 PU ({\it left}) and beginning of HL-LHC with 140 PU ({\it right}) for two benchmarks each from LLP scenarios (A) and (B), and QCD dijet events in the bin with $p_T^{gen}\in\{30,50\}$\,GeV.}
    \label{fig:sum-time}
\end{figure}

\begin{figure}[t]
\centering
    \includegraphics[width=0.45\textwidth]{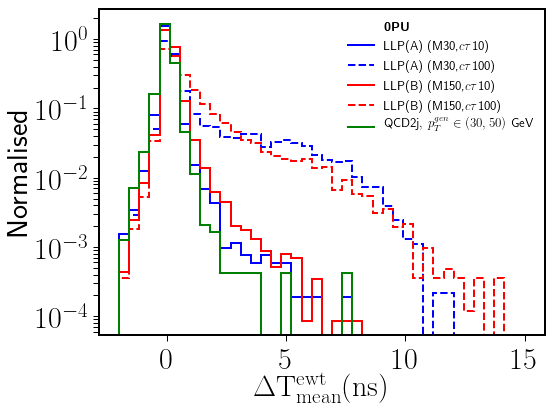}\qquad
    \includegraphics[width=0.45\textwidth]{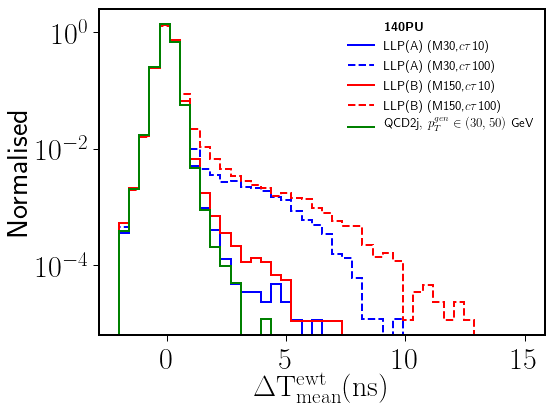}\\ 
    \includegraphics[width=0.45\textwidth]{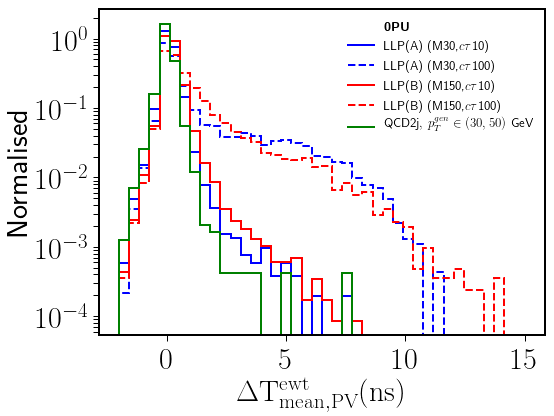}\qquad 
    \includegraphics[width=0.45\textwidth]{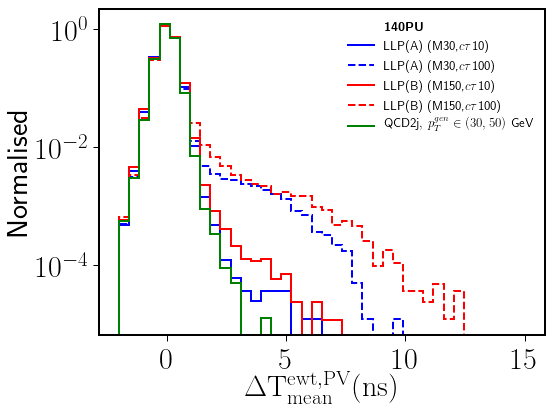}\\ 
    \includegraphics[width=0.45\textwidth]{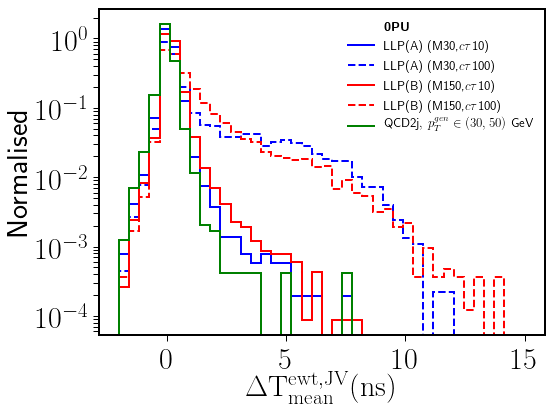}\qquad 
    \includegraphics[width=0.45\textwidth]{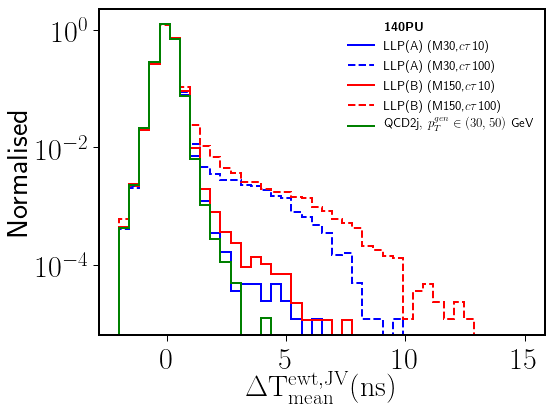}
    \caption{Distributions of mean energy-weighted timing ($\Delta T_{mean}^{Ewt}$) of a jet when timing of each crystal is calibrated with respect to the origin ({\it top}), primary vertex, PV ({\it center}), and jet vertex, JV ({\it bottom}) in the ideal situation with 0 PU ({\it left}) and beginning of HL-LHC with 140 PU ({\it right}) for two benchmarks each from LLP scenarios (A) and (B), and QCD dijet events in the bin with $p_T^{gen}\in\{30,50\}$\,GeV.}
    \label{fig:mean-ewt-time}
\end{figure} 

\begin{figure}[t]
\centering
    \includegraphics[width=0.45\textwidth]{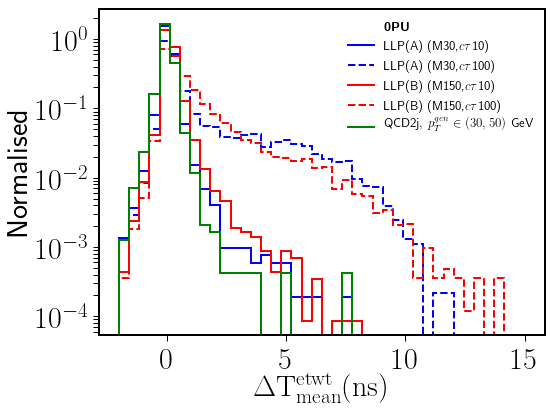}\qquad
    \includegraphics[width=0.45\textwidth]{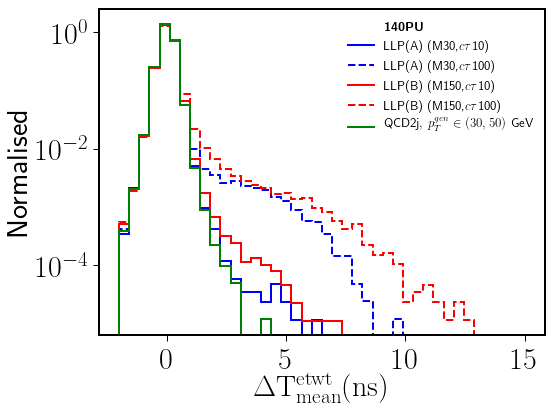}\\ 
    \includegraphics[width=0.45\textwidth]{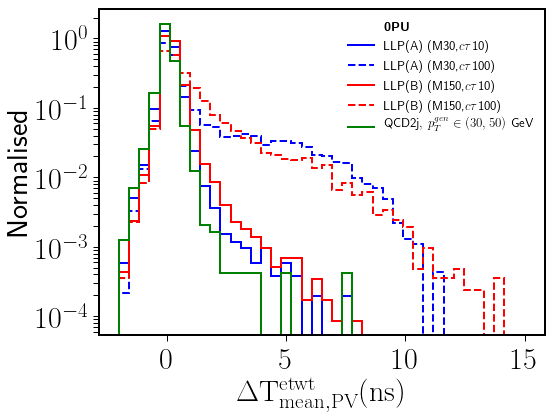}\qquad 
    \includegraphics[width=0.45\textwidth]{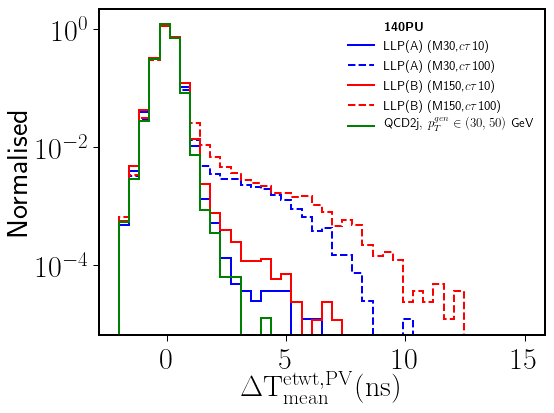}\\ 
    \includegraphics[width=0.45\textwidth]{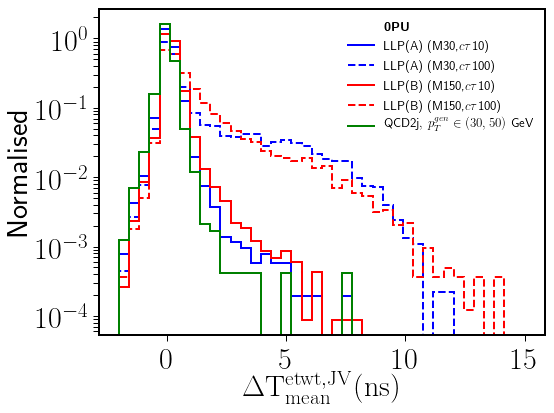}\qquad 
    \includegraphics[width=0.45\textwidth]{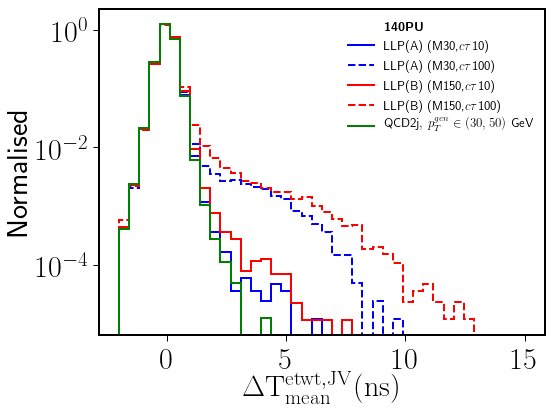}
    \caption{Distributions of mean transverse energy-weighted timing ($\Delta T_{mean}^{ETwt}$) of a jet when timing of each crystal is calibrated with respect to the origin ({\it top}), primary vertex, PV ({\it center}), and jet vertex, JV ({\it bottom}) in the ideal situation with 0 PU ({\it left}) and beginning of HL-LHC with 140 PU ({\it right}) for two benchmarks each from LLP scenarios (A) and (B), and QCD dijet events in the bin with $p_T^{gen}\in\{30,50\}$\,GeV.}
    \label{fig:mean-etwt-time}
\end{figure} 

In addition, we compute the mean timing of the jet using only five or ten crystals with the maximum time delay at ECAL or maximum value of time delay multiplied by the energy of the crystal, given the jet has at least five or ten towers associated with it. Again, the timing of each crystal is calibrated with respect to the origin. We describe these variables in the following list:
\begin{itemize}
\item $\mathbf{\Delta T_{mean}^{Max 5}}$: mean of the timing of the 5 ECAL crystals with largest timing in the jet, $\sum_{i=1}^5 \Delta T_i/5$, where $i$ runs over the 5 crystals having the highest $\Delta T$.
\item $\mathbf{(\Delta T\times E)_{mean}^{Max 5}}$: mean of the energy multiplied timing of the 5 ECAL crystals with largest energy multiplied timing in the jet, $\sum_{i=1}^5 \Delta T_i\times E_i/5$, where $i$ runs over the 5 crystals having the highest $\Delta T\times E$.
\item $\mathbf{\Delta T_{mean}^{Max 10}}$: mean of the timing of the 10 ECAL crystals with largest timing in the jet, $\sum_{i=1}^{10} \Delta T_i/10$, where $i$ runs over the 10 crystals having the highest $\Delta T$.
\item $\mathbf{(\Delta T\times E)_{mean}^{Max 10}}$: mean of the energy multiplied timing of the 10 ECAL crystals with largest energy multiplied timing in the jet, $\sum_{i=1}^{10} \Delta T_i\times E_i/10$, where $i$ runs over the 10 crystals having the highest $\Delta T\times E$.
\end{itemize}
If the jet has less than five or ten towers within it, $\Delta T_{mean}^{Max 5}$ and $\Delta T_{mean}^{Max 10}$ are assigned same values as $\Delta T_{mean}$, and $(\Delta T\times E)_{mean}^{Max 5}$ and $(\Delta T\times E)_{mean}^{Max 10}$ are assigned same values as $\Delta T_{mean}^{Ewt}$, respectively. The motivation behind constructing such variables is to reduce the contribution from the timing of ECAL energy depositions coming from PU vertices which reduce the jet timing, as we have seen earlier. Moreover, variables using the towers with highest $\Delta T\times E$ ensure that low-energy PU energy deposits with high timing do not contaminate these variables $-$ for displaced jets from LLPs, usually towers with high time delay contribute and for prompt jets from QCD dijet events, towers with high energy which usually have lower time delays contribute. 

\begin{figure}[t]
\centering 
    \includegraphics[width=0.4\textwidth]{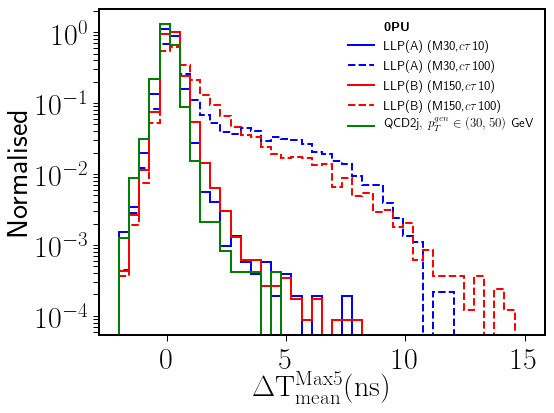}\qquad
    \includegraphics[width=0.4\textwidth]{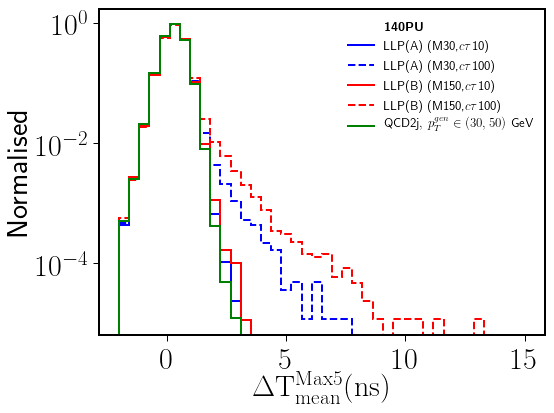}\\ 
    \includegraphics[width=0.4\textwidth]{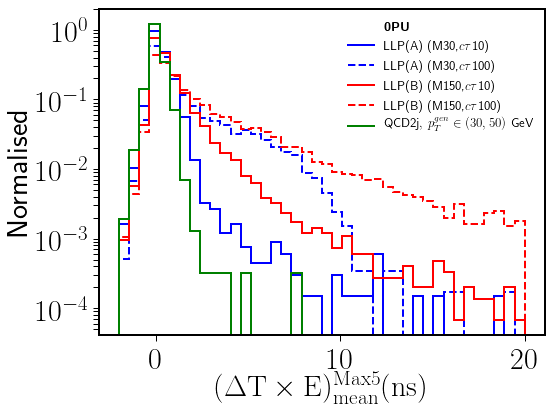}\qquad
    \includegraphics[width=0.4\textwidth]{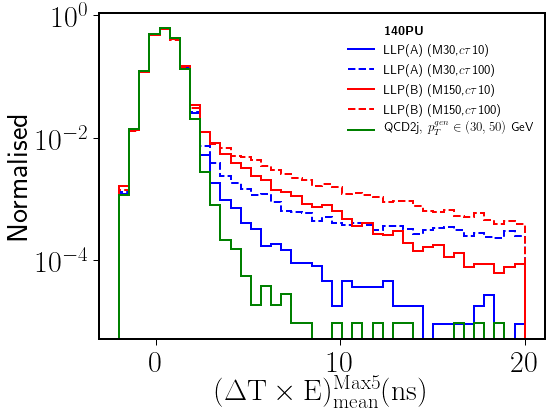}\\
    \includegraphics[width=0.4\textwidth]{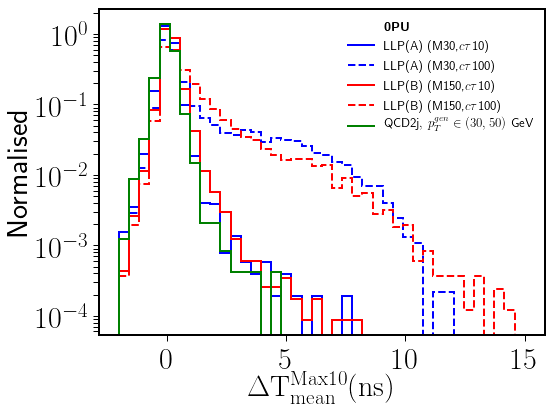}\qquad
    \includegraphics[width=0.4\textwidth]{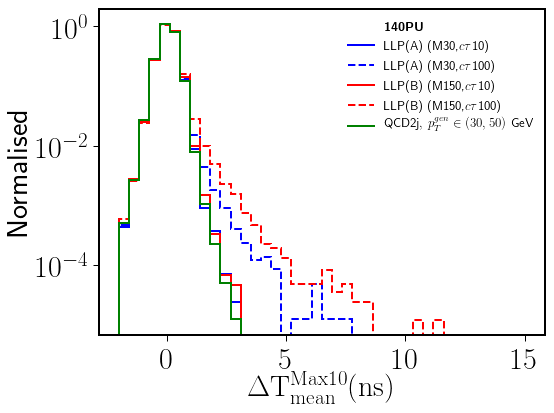}\\ 
    \includegraphics[width=0.4\textwidth]{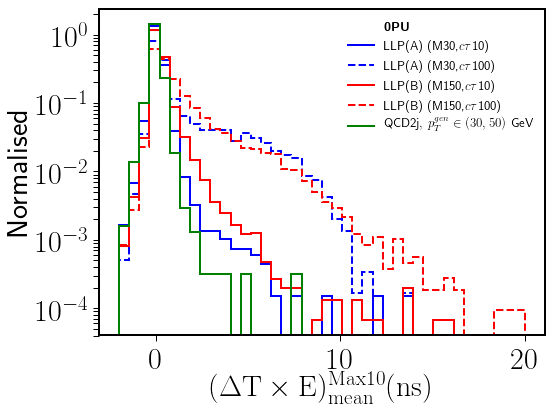}\qquad 
    \includegraphics[width=0.4\textwidth]{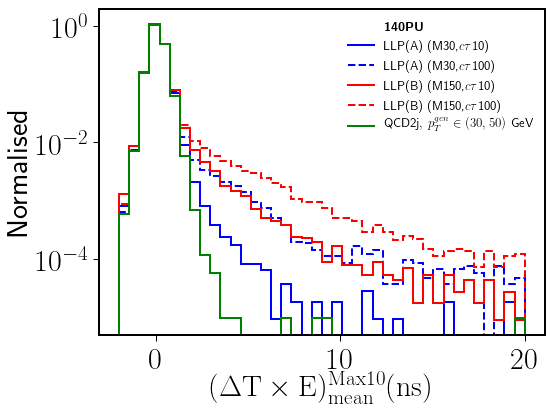}
\caption{Distributions of mean (mean energy-weighted) timing of a jet computed only with the 5 and 10 crystals having the maximum time delay shown in {\it first} ({\it second}) and {\it third} ({\it fourth}) rows respectively. Timing of each crystal is calibrated with respect to the origin in the ideal situation with 0 PU ({\it left}) and beginning of HL-LHC with 140 PU ({\it right}) for two benchmarks each from LLP scenarios (A) and (B), and QCD dijet events in the bin with $p_T^{gen}\in\{30,50\}$\,GeV.
}
\label{fig:max-time}
\end{figure}

In Figs.\,\ref{fig:mean-time}, \ref{fig:median-time}, \ref{fig:rms-time}, \ref{fig:sum-time}, \ref{fig:mean-ewt-time}, \ref{fig:mean-etwt-time}, and \ref{fig:max-time}, we have shown distributions of all of the above defined timing variables for two LLP scenarios (A) and (B) with two benchmark points each $-$ having decay lengths 10\,cm and 100\,cm and a mass of 30\,GeV in scenario (A) and 100\,GeV in scenario (B), along with one QCD bin having $p_T^{gen} \in$\,\{30,50\}\,GeV. All these distributions have been shown for the ECAL timing resolution corresponding to 1000\,fb$^{-1}$ luminosity.
For each case, we show the distributions of the variables in both the 0 PU and 140 PU scenarios to understand the effect of PU on these variables and to select the timing variables which are more PU resistant in addition to having the good capability to differentiate between the LLP signal and the background. We list down our observations below:
\begin{itemize}
    \item We observe a longer tail in the distributions of all the timing variables for LLP benchmarks where decay length is 100\,cm, however, for LLPs with a shorter decay length of 10\,cm, the distributions are not much distinguishable from the QCD background for most of the variables, even without any PU.
    \item Except for the $\sum \Delta T$ variable, for all the other variables, the timing reduces on adding 140 PU, since these are mostly average measures, unlike the former, which increases with increasing PU and hence increasing jet constituents.
    \item Compared to all other variables, the addition of PU has a lesser effect on the RMS timing of the jet, which might be due to its robustness against the presence of both negative and positive values in the data. The time difference can be either positive or negative depending on the spread of the PU vertices in the time and $z$ direction.
    \item We observe that timing variables with energy-weighted timing are more PU resistant than the rest of the variables. We find good discrimination between QCD prompt jets and displaced jets from LLPs for energy-weighted mean timing ($\Delta T_{mean}^{Ewt}$) even after adding 140 PU. Weighting the timing with the energy of the tower reduces contamination from low energy PU. 
    \item For shorter decay lengths, timing calculated using 5 and 10 towers having highest values of $\Delta T\times E$ in a jet is more PU resistant and 
    provides a good distinction between background and signal for both shorter and larger decay lengths even in the 140 PU environment of HL-LHC. 
    \item We can also see that the effect of calibration on the timing of ECAL towers from PV and JV has a negligible impact on jet timing compared to the situation where the timing of each ECAL tower is calibrated with respect to the origin.
\end{itemize}

After carefully studying the various timing variables $-$ how their distributions look for displaced jets and prompt jets, in the absence and presence of PU, we now proceed further to use these to construct triggers that will be useful for selecting events with LLPs decaying to jets efficiently, maintaining reasonable background rejection at L1 of HL-LHC CMS. The correlation matrices of these variables for two LLP benchmarks each from scenario (A) and (B) along with QCD dijet events ($p_T^{gen}\in \{30,50\}$\,GeV) has been shown in Fig.\,\ref{fig:corr} of Appendix\,\ref{app:corr}.

\subsection{Triggering using timing variables}
\label{ssec:trigger}

We have short-listed three timing variables out of the above-mentioned ones, which are the most PU resistant and provide a good distinction between the background and the signal $-$ $\Delta T_{mean}^{Ewt}$, $\Delta T_{RMS}$, and $(\Delta T\times E)_{mean}^{Max 5}$.
As we can see from the timing distributions of QCD jets after the addition of 140 PU, they have a comparatively longer tail in the timing distributions compared to the 0 PU scenario. We have discussed this earlier in Section\,\ref{ssec:qcd-time} that large jet timing in the QCD distribution can be due to various factors, like $\eta$-$\phi$ position as well as $p_T$ of the jet constituents, timing resolution of the ECAL crystals, the temporal and longitudinal spread of vertices and SM long-lived particles.
Besides these, the timing of the jet, which is solely calculated using ECAL towers, can be affected if the number of ECAL towers is relatively less and most of the energy of the jet is deposited in HCAL. In such cases, only a few ECAL towers are available to calculate the timing of the jet, which can be statistically insufficient to determine the timing of the jet accurately.

\begin{figure}[hbt!]
\centering
\includegraphics[width=\textwidth]{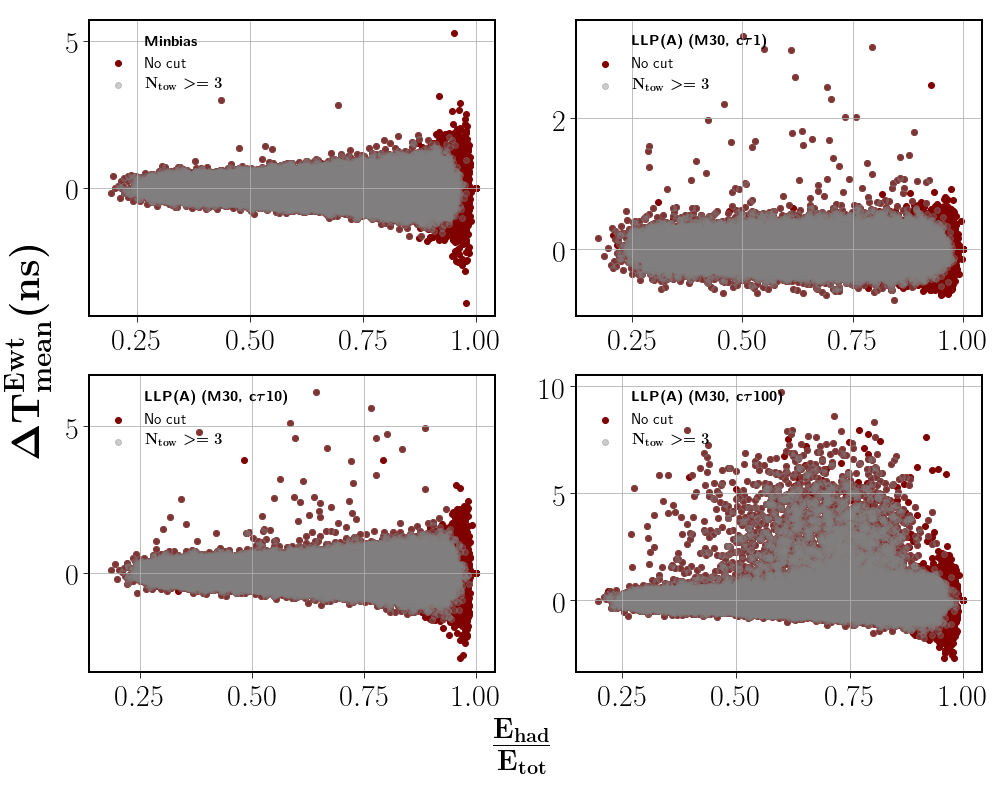}
\caption{Correlation between the energy-weighted mean timing of a jet $(\Delta T_{mean}^{Ewt})$ and the hadronic energy fraction of the jet ($\frac{E_{had}}{E_{tot}}$) without and with a cut on the number of ECAL towers, $N_{tow}\geq 3$ for jets from the minimum bias events ({\it top left}), and from decay of a 30\,GeV LLP from scenario (A) having $c\tau=$\,1\,cm ({\it top right}), 10\,cm ({\it bottom left}), and 100\,cm ({\it bottom right}).}
\label{fig:itwj_jethad}
\end{figure}

Fig.\,\ref{fig:itwj_jethad} shows the correlation between the $\Delta T_{mean}^{Ewt}$ time of a jet and the fraction of jet energy deposited in the HCAL ($\frac{E_{had}}{E_{tot}}$) for jets from minimum bias, where we use the $pp\rightarrow\nu\bar{\nu}$ process merged with 140 PU as discussed in Section\,\ref{ssec:L1trigger} along with a 30\,GeV LLP from scenario (A) with three different decay lengths $-$ 1\,cm, 10\,cm, and 100\,cm.
Our first observation is that the spread in the timing increases with increasing $\frac{E_{had}}{E_{tot}}$, which implies that these jets having higher values of $\Delta T_{mean}^{Ewt}$ have lesser number of associated ECAL energy deposits. For the higher decay length LLP benchmark, apart from this correlation, we have many jets with high timing even when the ECAL energy deposit is 25-50\% ($\frac{E_{had}}{E_{tot}}$ lies between 0.75-0.5).
We can see from Fig.\,\ref{fig:itwj_jethad}, putting a minimum cut on the number of ECAL towers, say $N_{tow}\geq 3$ significantly reduces the number of QCD prompt jets having higher timing without affecting jets coming from the signal processes. Displaced jets from LLP with the decay length of 100\,cm having high timing remains even after the $N_{tow}\geq 3$ cut. Therefore, we can use this cut to reduce QCD jets with high timing values, without affecting the signal much.

\begin{figure}[t]
\centering
\includegraphics[width=0.45\textwidth]{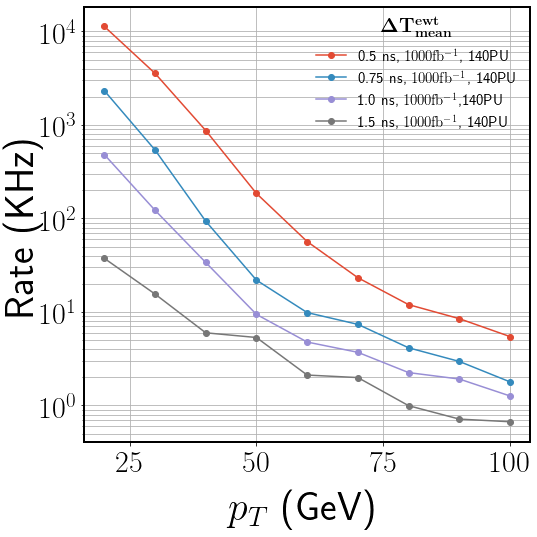}\qquad
\includegraphics[width=0.45\textwidth]{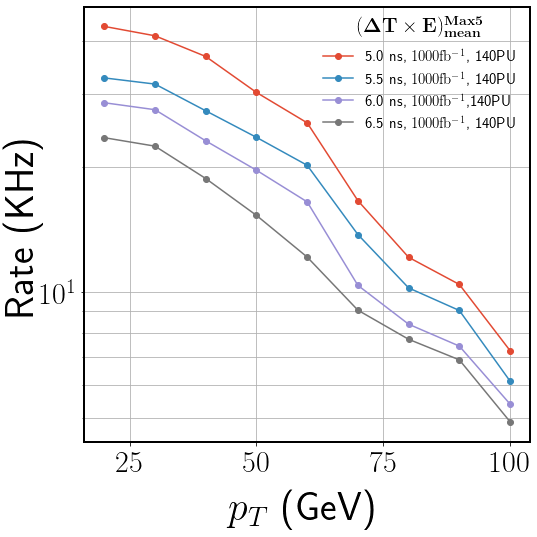}\\
\includegraphics[width=0.45\textwidth]{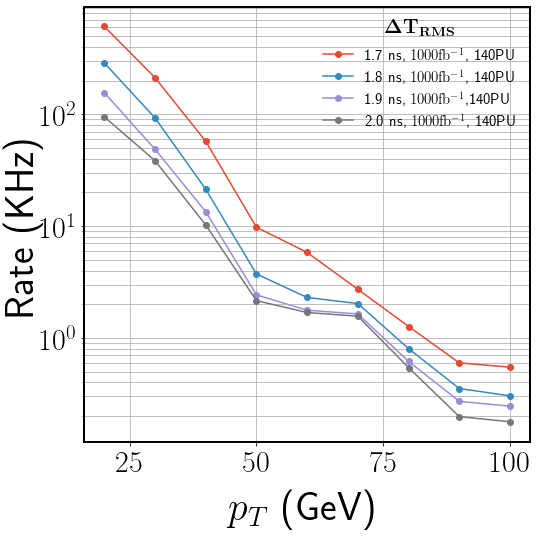}
\caption{Rate of background events as a function of jet $p_T$ with varying cuts on the $\rm {\Delta T_{mean}^{Ewt}}$ (\textit{top left}), $\rm {\Delta T_{mean}^{Max 5}}$ (\textit{top right}) and  $\rm {\Delta T_{RMS}}$ (\textit{bottom}) timing of the jet for ECAL timing resolution corresponding to 1000\,fb${-1}$ in the 140 PU scenario.}
\label{fig:rate_mean}
\end{figure}

A crucial part of constructing triggers is to ensure that the trigger rate is within the acceptable bandwidth, i.e., the number of events passing the trigger selection cuts per second is not impractically large. Therefore, it is essential to keep a check on the QCD dijet background rates from our triggers. We have generated our QCD background in different $p_T^{gen}$ bins, which we need to combine with the minimum bias sample, without any overlap in phase-space, which can lead to overestimation of rates. For this purpose, we have used the ``stitching'' procedure as described in Ref.\,\cite{Ehataht:2021rkh}, which we discuss and validate in Appendix\,\ref{app:stitch}.
In this procedure, each event is assigned a rate depending on the weight calculated for that event using Eq.\,\ref{eq:stitch}, explained in Appendix\,\ref{app:stitch}. The total rate of our trigger is computed by summing over the individually assigned rates of each background event, passing our trigger criteria.

To remind the readers, we are reconstructing jets using anti-$k_T$ clustering algorithm with jet cone radius $R=0.3$.
We have applied an energy and $\eta$ dependent timing resolution corresponding to 1000\,fb$^{-1}$ on the ECAL towers before calculating the jet timing. We have chosen the resolution corresponding to 1000\,fb$^{-1}$ since that is the integrated luminosity that is expected to be collected in the 140 PU scenario, which we have mostly focused on till now. The resolution corresponding to 1000\,fb$^{-1}$ lies somewhere between the best and the worst resolutions of the ECAL timing information at the beginning and end of HL-LHC runs.
Fig.\,\ref{fig:rate_mean} shows the rate as a function of jet $p_T$ with varying cuts on the three different jet timing measures identified from the previous section $-$ $\Delta T_{mean}^{Ewt}$ ({\it top left}), $(\Delta T\times E)_{mean}^{Max 5}$ ({\it top right}), and $\Delta T_{RMS}$ ({\it bottom}).
We demand the presence of at least 3 and 4 ECAL towers to calculate $\Delta T_{mean}^{Ewt}$ and $\Delta T_{RMS}$, respectively, in order to get rid of high jet timings coming from jets with lower ECAL energy depositions compared to HCAL. 
As we can see from Fig.\,\ref{fig:rate_mean}, for 40\,GeV jets, we can restrict background rate to less than $\approx$ 30\,kHz by putting $\Delta T_{mean}^{Ewt}>1$\,ns, or $(\Delta T\times E)_{mean}^{Max 5}>5.5$\,ns, and even to less than $\approx$ 20\,kHz by putting $\Delta T_{RMS}>1.9$\,ns.

\begin{table}[t]
\centering
\begin{tabular}{|c|c|c|c||}
\hline
Timing variable & Time (ns) & Number of ECAL towers & $p_T $ (GeV) \\ 
\hline\hline
 $\rm {\Delta T_{mean}^{Ewt}}$ & $>$ 1.1 & $\geq$ 3 & \multirow{3}{*}{$>35$} \\ 
\cline{1-3}
$\rm {(\Delta T\times E)_{mean}^{Max 5}}$ & $>$ 5.5 & $-$ &  \\ 
\cline{1-3}
$\rm {\Delta T_{RMS}}$ & $>$ 1.9 & $\geq$ 4 &  \\ 
\hline
\hline
\end{tabular} 
\caption{Selection cuts for the three dedicated timing triggers with thresholds for the three jet timing variables, number of ECAL towers associated with a jet and the jet $p_T$ which keeps the background rate $\approx$ 30\,kHz.}
\label{tab:eff}
\end{table}
 
These rate plots motivate the choice of appropriate $p_T$ and timing variable cuts which maintain reasonable background rates. The L1 trigger bandwidth is expected to be around 750\,kHz at HL-LHC. We believe that a rate of 30\,kHz might be reasonable enough to fix our trigger cuts. We then perform a cut based analysis by applying these cuts on the signal to calculate the signal efficiencies of these triggers for the fixed background rate of 30\,kHz. 
Our triggers are single jet triggers above some $p_T$ threshold with some minimum number of ECAL towers, and satisfy the jet timing cut. For each of the three variables, $\Delta T_{mean}^{Ewt}$, $\Delta T_{RMS}$, and $(\Delta T\times E)_{mean}^{Max 5}$, we have separate cuts such that the rate is not more than 30\,kHz.
We have enumerated the cut thresholds for each variable in Table\,\ref{tab:eff}. 
We calculate the efficiency of our timing triggers for all three LLP scenarios $-$ for LLP scenario (A), we calculate efficiency for LLPs in the mass range of \{10,50\}\,GeV and for LLP scenarios (B) and (C), we compute the efficiency for LLPs with masses in the range \{100,500\}\,GeV. For all scenarios, we vary the decay length in the range \{1,500\}\,cm. 

\begin{figure}[hbt!]
\centering
    \includegraphics[width=0.45\textwidth]{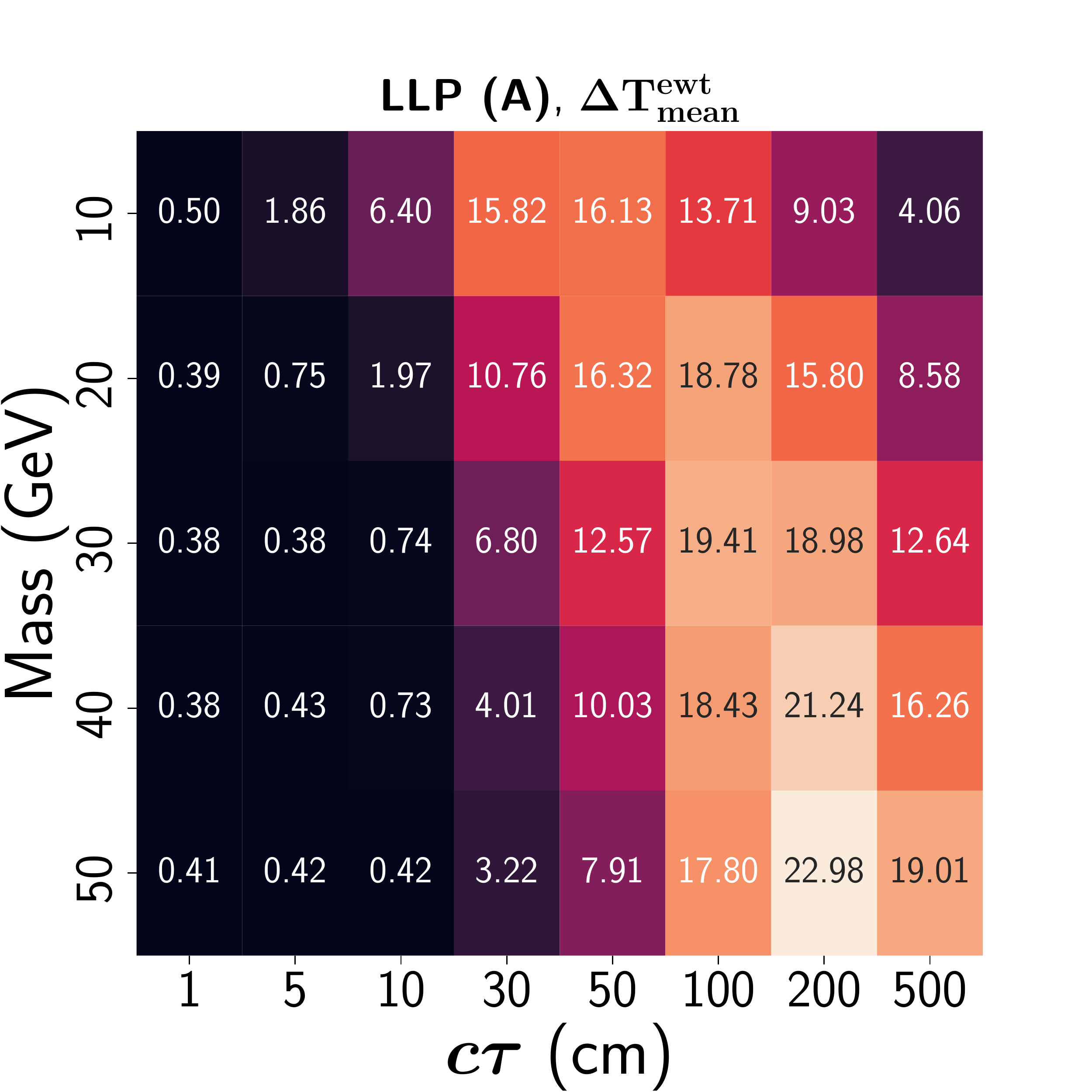}\qquad
    \includegraphics[width=0.45\textwidth]{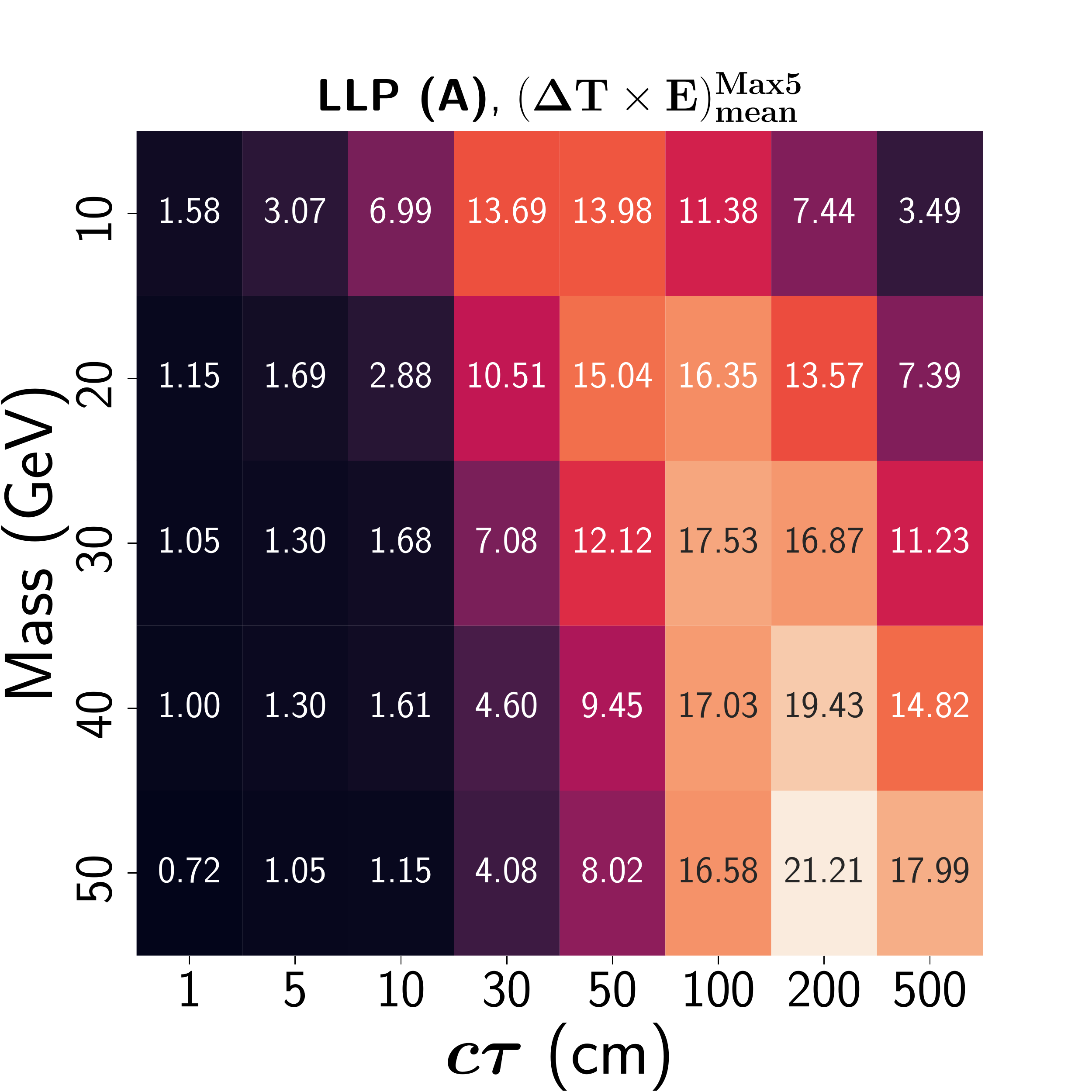}\\ 
    \includegraphics[width=0.45\textwidth]{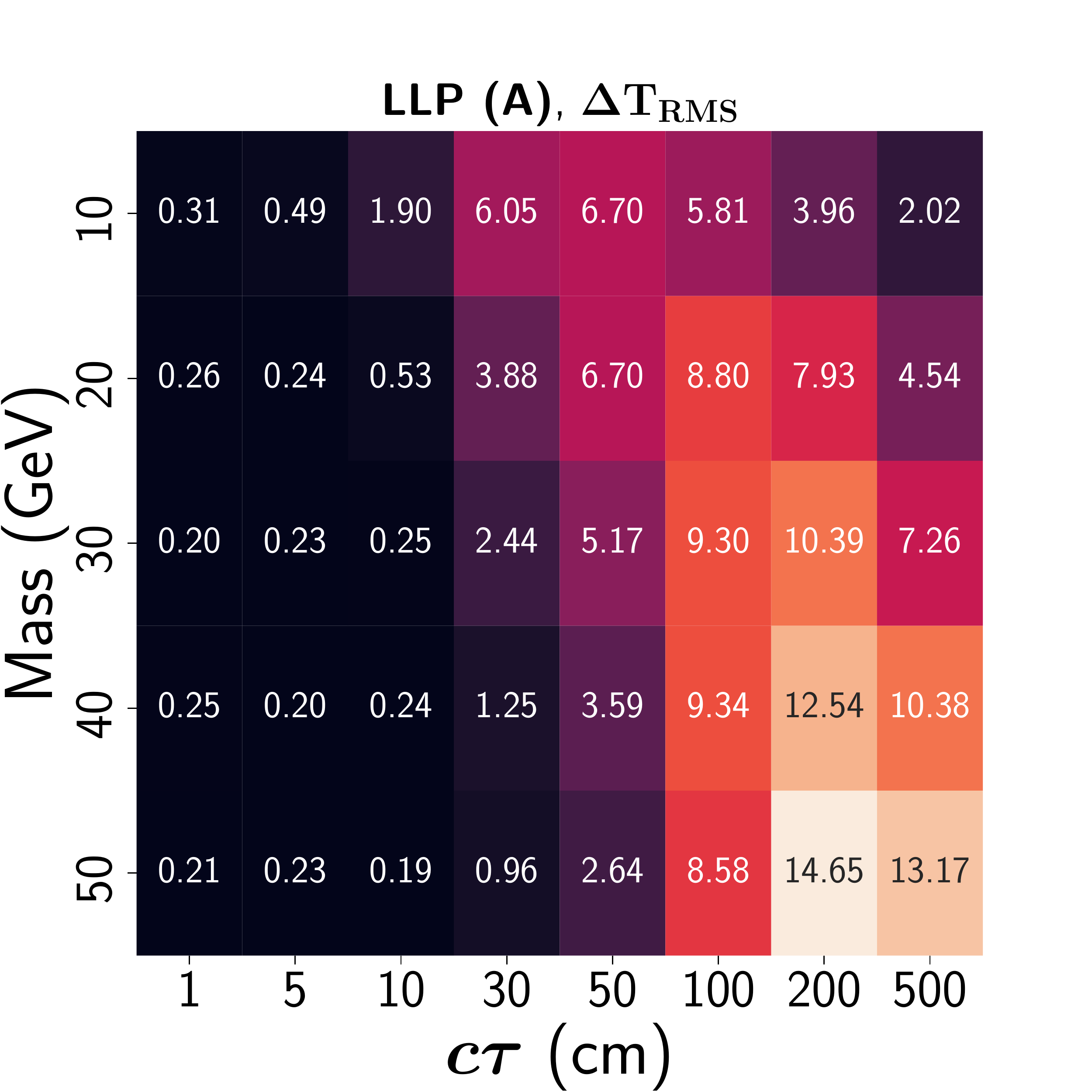}     
\caption{Efficiency of selecting events for LLP scenarios (A) for various values of LLP mass and decay length using $\Delta T_{mean}^{Ewt}$ ({\it top left}), $(\Delta T\times E)_{mean}^{max5}$ ({top right}), and $\Delta T_{RMS}$ ({\it bottom}). Efficiency is calculated by applying cuts on the jet $p_T$, respective timing variables, and number of ECAL towers according to Table\,\ref{tab:eff} to keep the background rate $\approx$ 30\,kHz.}
\label{fig:rate_eff_LLPA}
\end{figure}

\begin{figure}[hbt!]
\centering
    \includegraphics[width=0.45\textwidth]{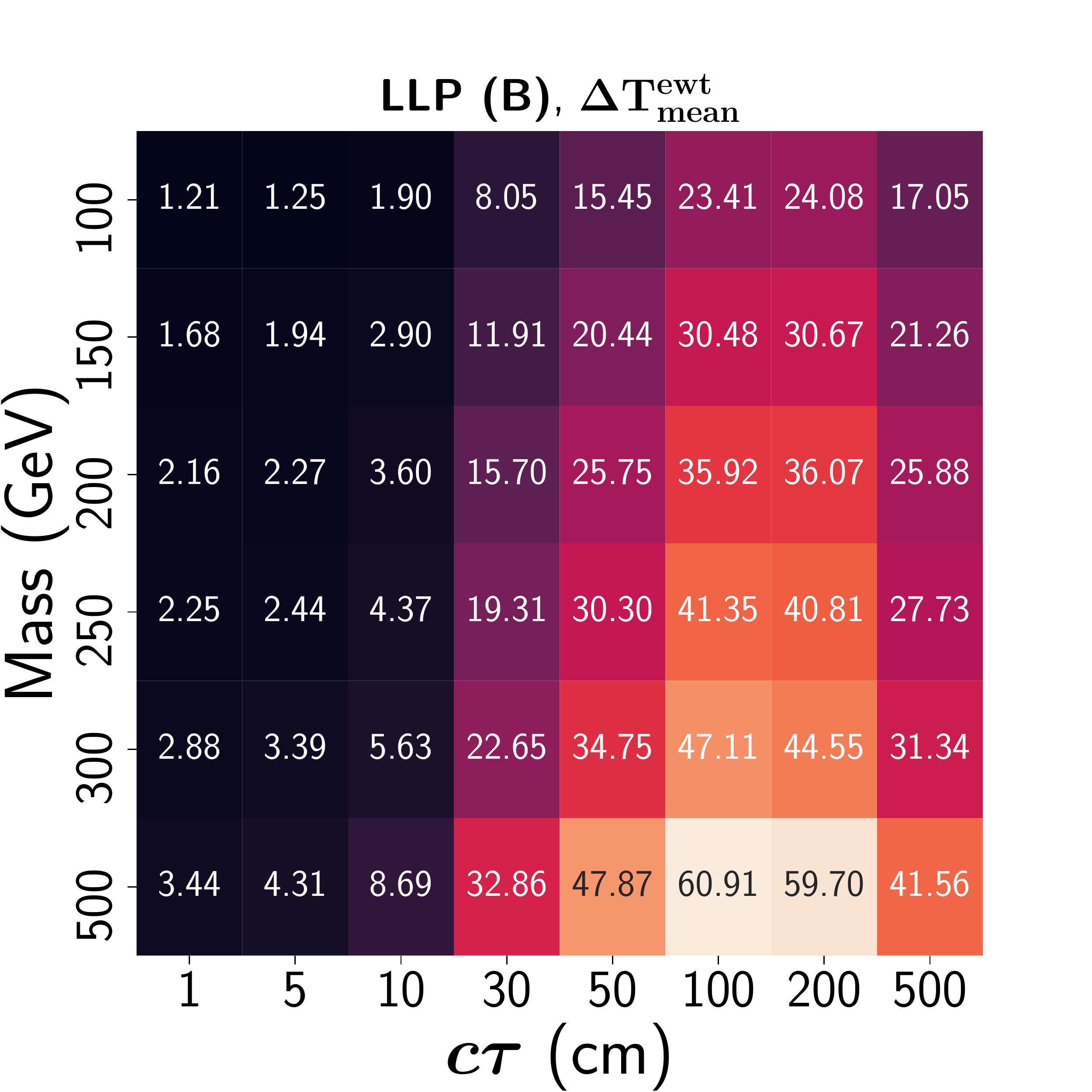}\qquad
    \includegraphics[width=0.45\textwidth]{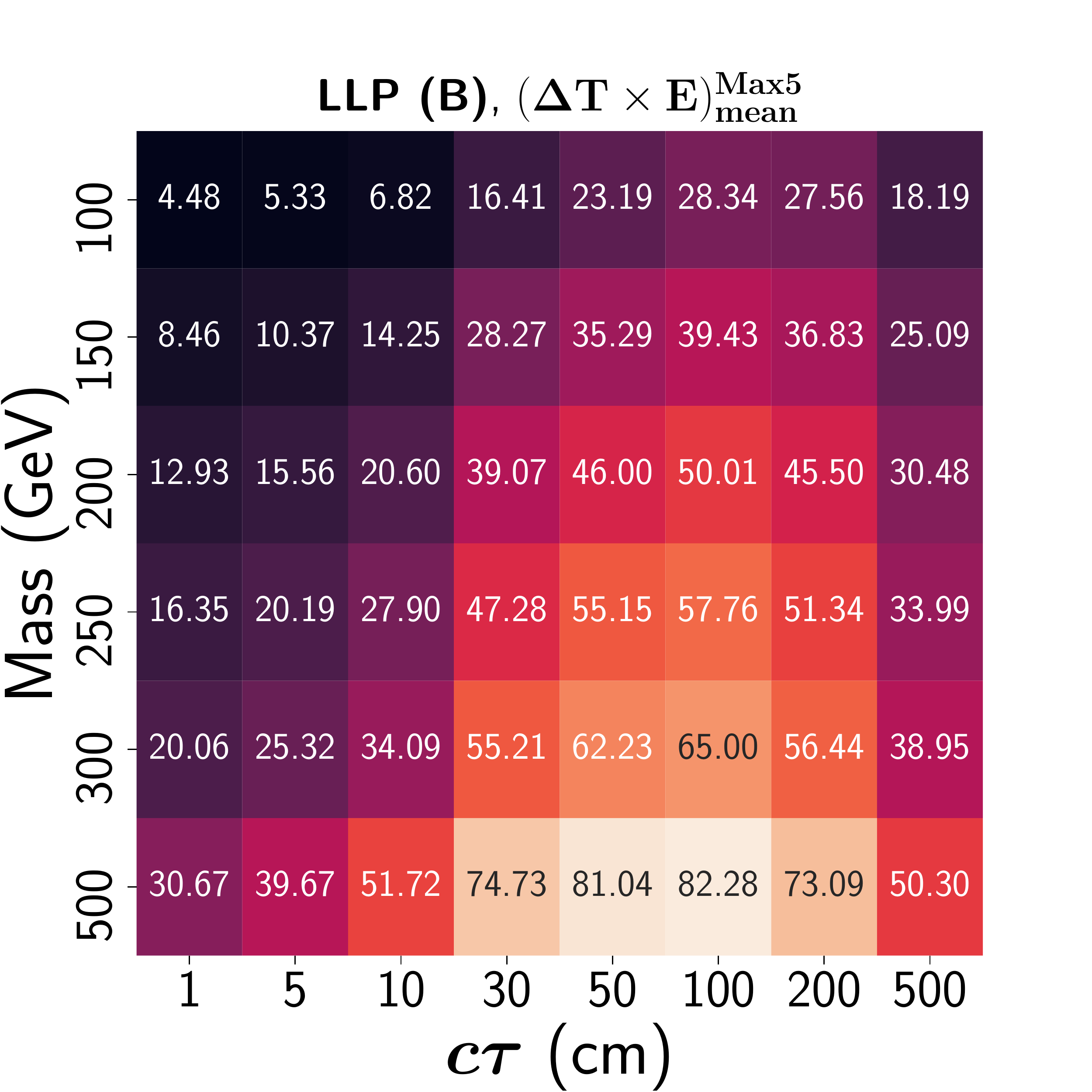}\\ 
    \includegraphics[width=0.45\textwidth]{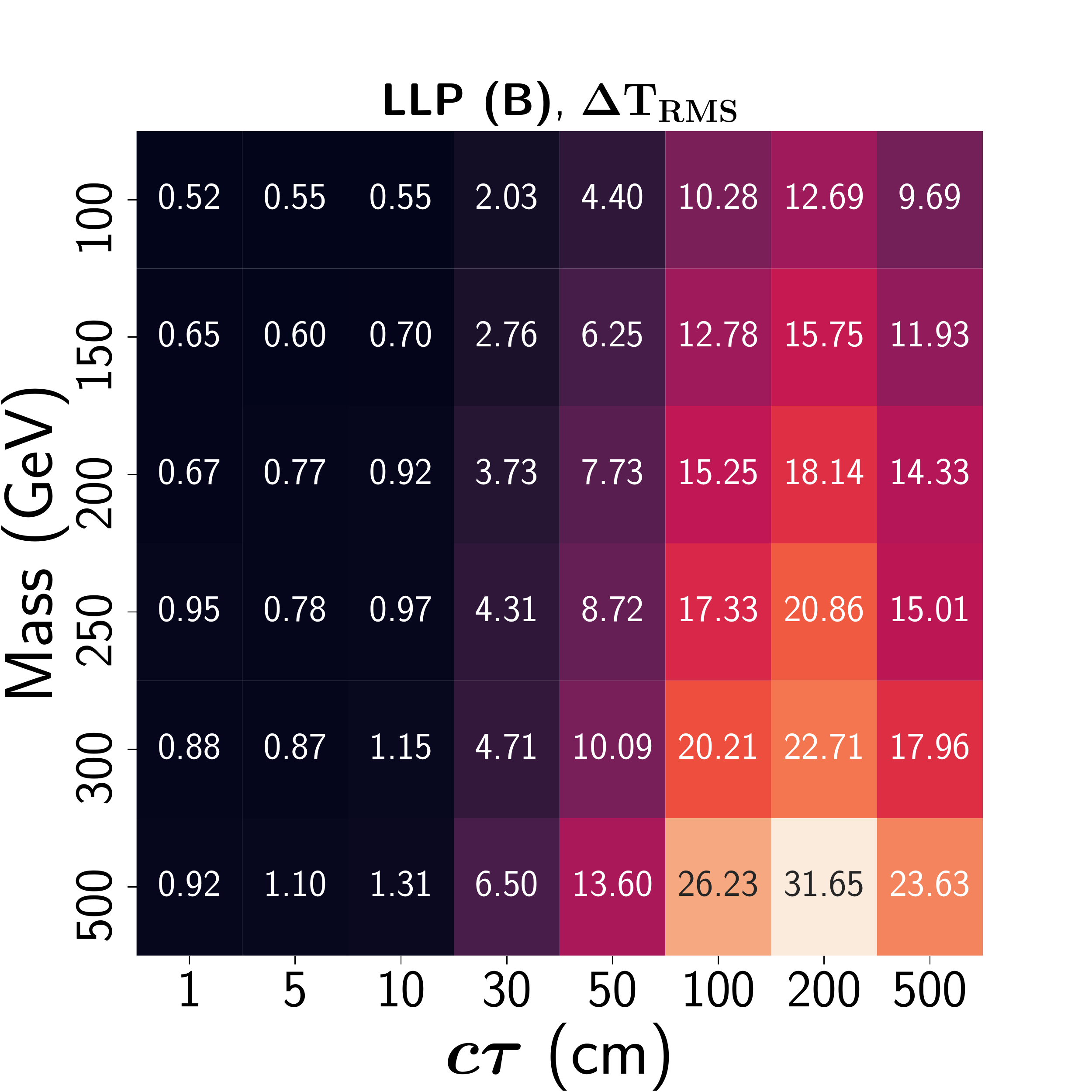}     
\caption{Efficiency of selecting events for LLP scenarios (B) for various values of LLP mass and decay length using $\Delta T_{mean}^{Ewt}$ ({\it top left}), $(\Delta T\times E)_{mean}^{max5}$ ({top right}), and $\Delta T_{RMS}$ ({\it bottom}). Efficiency is calculated by applying cuts on the jet $p_T$, respective timing variables, and number of ECAL towers according to Table\,\ref{tab:eff} to keep the background rate $\approx$ 30\,kHz.}
\label{fig:rate_eff_LLPB}
\end{figure}

\begin{figure}[hbt!]
\centering
    \includegraphics[width=0.45\textwidth]{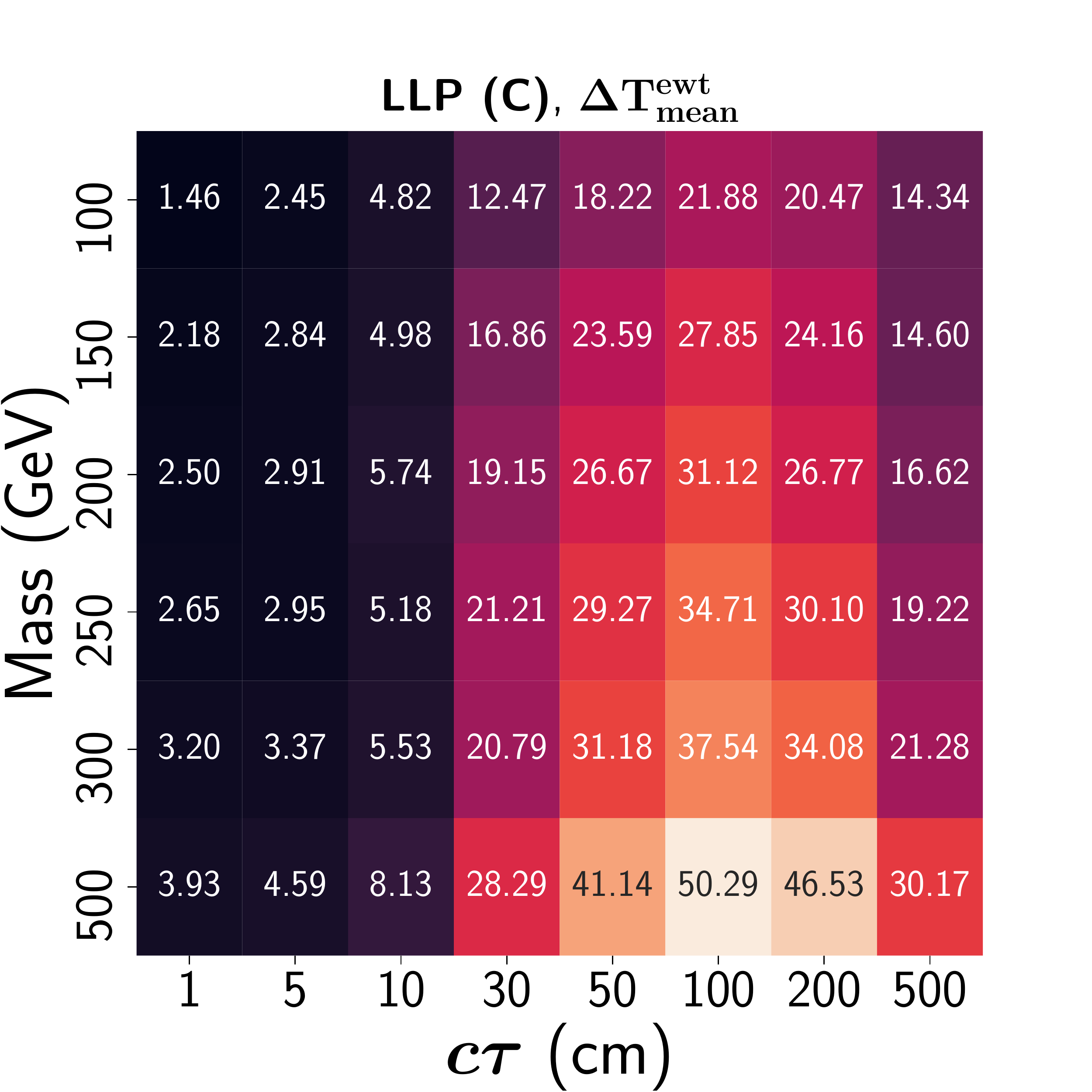}\qquad
    \includegraphics[width=0.45\textwidth]{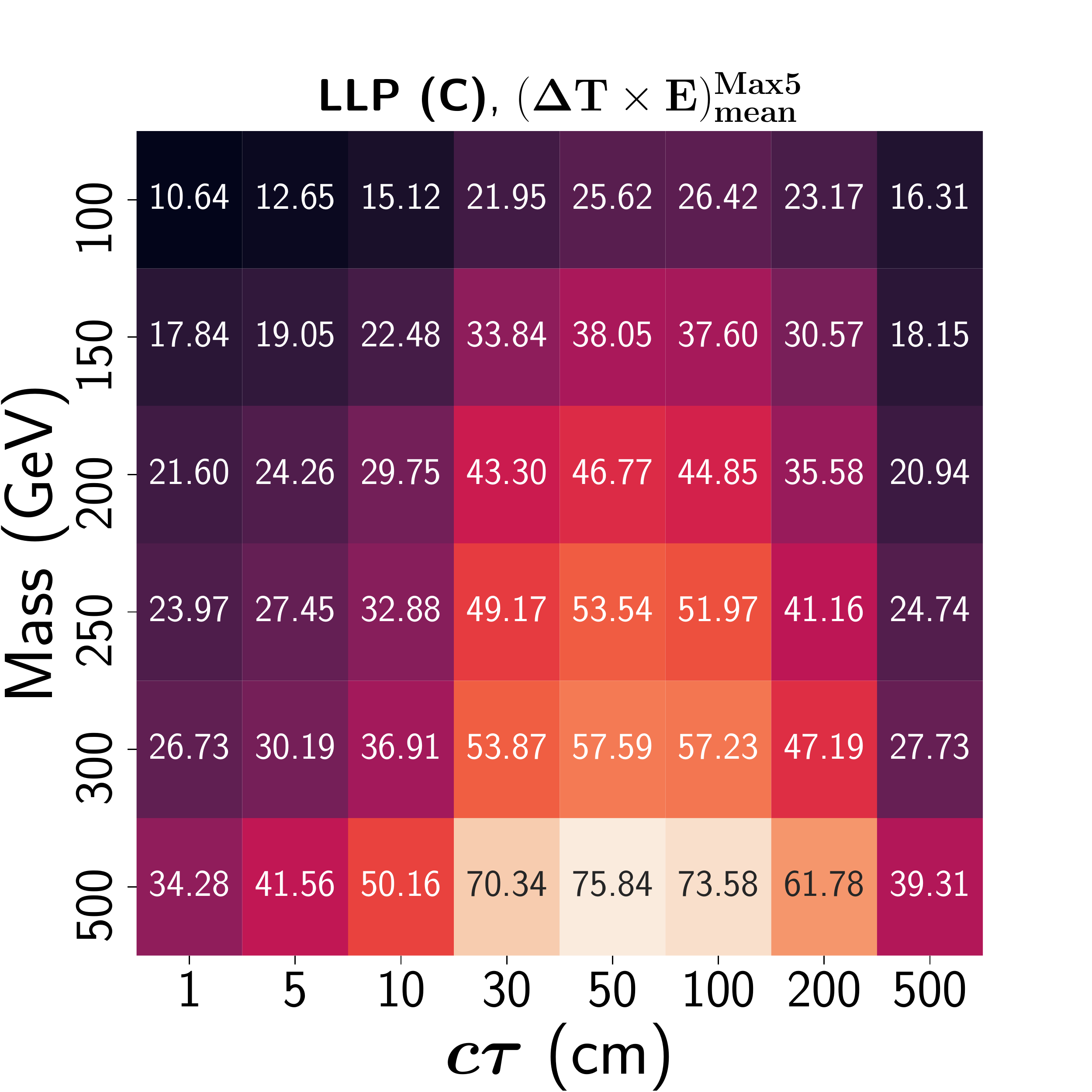}\\ 
    \includegraphics[width=0.45\textwidth]{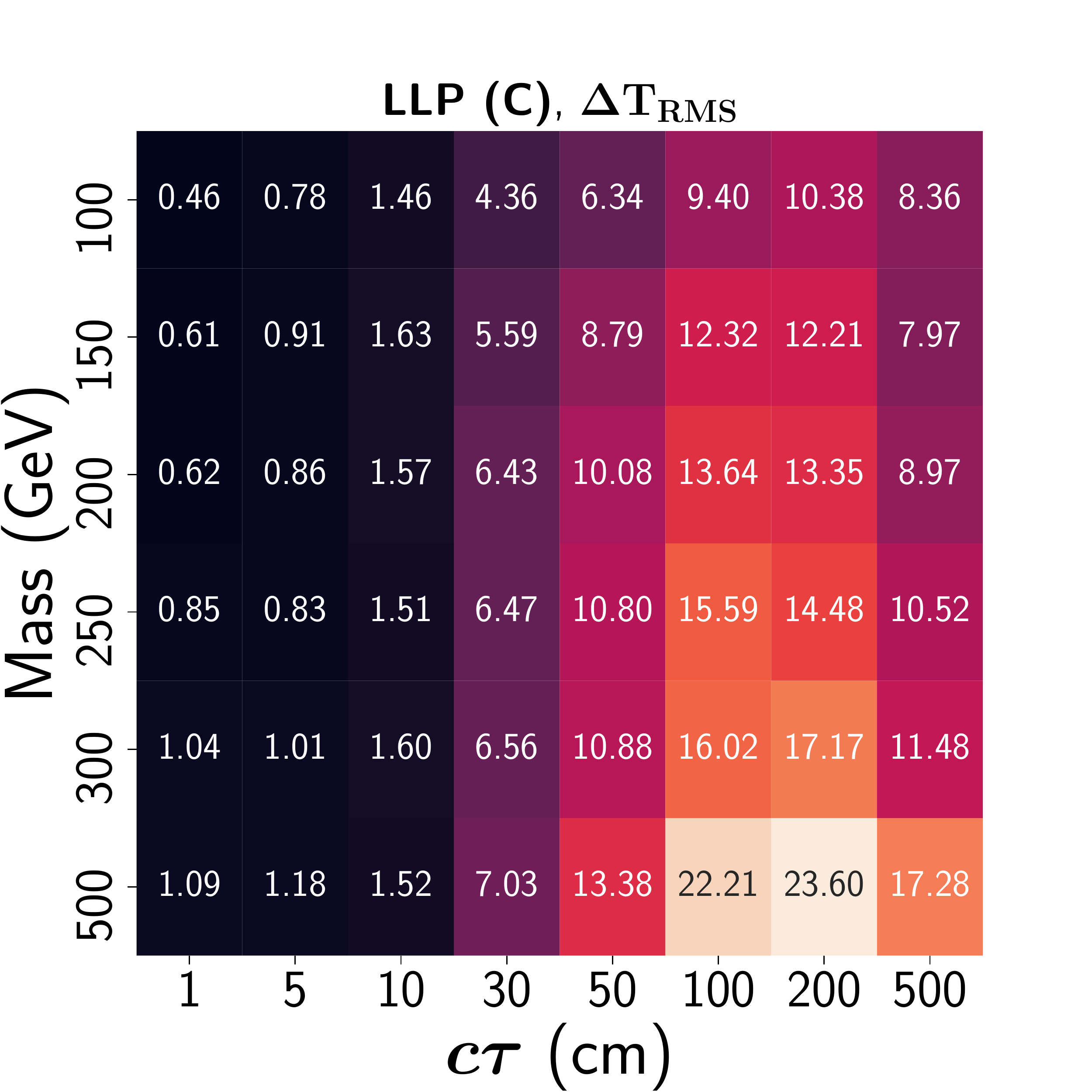}     
\caption{Efficiency of selecting events for LLP scenarios (C) for various values of LLP mass and decay length using $\Delta T_{mean}^{Ewt}$ ({\it top left}), $(\Delta T\times E)_{mean}^{max5}$ ({top right}), and $\Delta T_{RMS}$ ({\it bottom}). Efficiency is calculated by applying cuts on the jet $p_T$, respective timing variables, and number of ECAL towers according to Table\,\ref{tab:eff} to keep the background rate $\approx$ 30\,kHz.}
\label{fig:rate_eff_LLPC}
\end{figure}

We present the signal efficiencies in the form of grids in the plane of mass and decay length of the LLP for the three LLP scenarios, which can be directly used in other theoretical studies of LLPs.
Figs.\,\ref{fig:rate_eff_LLPA}, \ref{fig:rate_eff_LLPB}, and \ref{fig:rate_eff_LLPC} show the efficiency grids of the three LLP scenarios considered in this work with varying masses and decay lengths of the LLPs, each with the three sets of triggers, whose cuts are tabulated in Table\,\ref{tab:eff}.
We observe that efficiency increases with an increase in the mass of the LLPs, which is due to the fact that massive LLPs have lesser boost, and hence cause more time delay of their decay products. With increasing decay length, the efficiency first increases due to increasing time delays and then decreases for much larger $c\tau$ values because 
a significant number of LLPs decay after the ECAL without depositing a sufficient amount of energy in ECAL. We can see this for the LLP benchmark point with 500\,cm decay length, where the efficiency decreases in all the three LLP scenarios.
We discuss our results below:
\begin{itemize}
    \item \textbf{Scenario (A):} We observe that for LLP scenario (A), where the LLPs are produced from Higgs boson decay, energy-weighted mean timing ($\Delta T_{mean}^{Ewt}$) performs the best among the chosen three timing variables. An efficiency of around 16\% can be achieved for 10\,GeV LLP with 50\,cm decay length, 19\% can be achieved for 30\,GeV LLP with 100\,cm decay length, and for 50\,GeV LLP, an efficiency of around 23\% can be achieved for 200\,cm decay length, at a rate of around 30\,kHz.
    Comparing these efficiencies to the initial fraction of decays within the ECAL, as shown in Fig.\,\ref{fig:decay_frac}, we find that for 30\,GeV LLPs even though around 100\% of the decays happen within the ECAL for $c\tau=10$\,cm, only 0.74\% survives the timing cuts, since most of these have very low displacements, and hence lower jet timing. For 30\,GeV LLP with $c\tau=500$\,cm, 24\% of events have at least one LLP decaying within the ECAL, and 19.41\% pass the timing cuts.
    
    At HL-LHC, with an integrated luminosity of 1000\,fb$^{-1}$, if we translate these numbers to the sensitivity on the branching fraction of Higgs boson to decay to such LLPs, we get the following upper limits on Br($h\rightarrow XX$) assuming 100\% decay of the LLP to jets:
    
    \begin{itemize}
        \item Br($h\rightarrow XX$)$\lesssim 6.2\times 10^{-6}$ for $M_X=10$\,GeV, $c\tau=50$\,cm
        \item Br($h\rightarrow XX$)$\lesssim 5.1\times 10^{-6}$ for $M_X=30$\,GeV, $c\tau=100$\,cm
        \item Br($h\rightarrow XX$)$\lesssim 4.3\times 10^{-6}$ for $M_X=50$\,GeV, $c\tau=200$\,cm
    \end{itemize}
    
    \item \textbf{Scenario (B):} For pair produced LLPs in scenario (B), $(\Delta T\times E)_{mean}^{Max 5}$ performs better than the other two variables for LLP masses greater than 100\,GeV, especially for smaller decay lengths. With this variable, efficiencies of around 39\%, 58\%, and 82\% can be achieved for LLPs of mass 150\,GeV, 250\,GeV, and 500\,GeV, respectively, decaying to two jets for 100\,cm decay length in this scenario. 
    From Fig.\,\ref{fig:decay_frac}, we find that 100\% of events with 150\,GeV LLPs and $c\tau=10$\,cm have at least a single LLP decaying before the ECAL, out of which 14.25\% pass the trigger based on $(\Delta T\times E)_{mean}^{Max 5}$ variable, and for $c\tau=100$\,cm, these numbers are 80\% and 39\% respectively. With increasing mass and $c\tau$, we can retain more events after L1 from the parton level decay efficiencies within the ECAL.
    
    
     At HL-LHC, with an integrated luminosity of 1000\,fb$^{-1}$, we get the following upper limits on the production cross sections of such LLPs, assuming 100\% decay of the LLP to two jets:
    
    \begin{itemize}
        \item $\sigma\lesssim 0.13$\,fb for $M_X=150$\,GeV, $c\tau=100$\,cm
        \item $\sigma\lesssim 0.09$\,fb for $M_X=250$\,GeV, $c\tau=100$\,cm
        \item $\sigma\lesssim 0.06$\,fb for $M_X=500$\,GeV, $c\tau=100$\,cm
    \end{itemize}
    \item \textbf{Scenario (C):} Similar to scenario (B), $(\Delta T\times E)_{mean}^{Max 5}$ performs better than the other two variables in this case as well, and we get efficiencies of around 38\%, 54\%, and 76\% for LLPs of mass 150\,GeV, 250\,GeV, and 500\,GeV, respectively, decaying to three jets for 50\,cm decay length.
    
    At HL-LHC, with an integrated luminosity of 1000\,fb$^{-1}$, we get the following upper limits on the production cross sections of such LLPs, assuming 100\% decay of the LLP to three jets:
    
    \begin{itemize}
        \item $\sigma\lesssim 0.13$\,fb for $M_X=150$\,GeV, $c\tau=50$\,cm
        \item $\sigma\lesssim 0.09$\,fb for $M_X=250$\,GeV, $c\tau=50$\,cm
        \item $\sigma\lesssim 0.07$\,fb for $M_X=500$\,GeV, $c\tau=50$\,cm
    \end{itemize}
\end{itemize}

In each of the three scenarios, we have presented the level-1 sensitivity corresponding to 50 events passing our triggers. The actual discovery or exclusion limits will depend on several factors, like the identification of displaced vertex at the HLT and offline analyses, and suppression of SM backgrounds, which is beyond the scope of the present study. 

\section{Possible improvements and discussions}
\label{sec:improve}

Hitherto, we have discussed various timing variables of a jet and then identified three of them, depending on their ability to distinguish between the displaced and prompt jets and robustness against PU, and construct three different triggers for displaced jets using these variables.
In this section, we discuss various methods which can further help in improving the trigger performance that we discussed in the previous section. Moreover, we have done our analysis using a particular set of scenario $-$ ECAL timing resolution, cone-size of jets, and amount of PU, and we explore in this section whether our timing variables are robust and how much our results vary with these varying scenarios. Lastly, we also have some brief discussions on the prospect of triggering events passing our L1 trigger in the HLT level of the trigger as well.

\subsection{Impact of better resolution in the initial runs of HL-LHC}
\label{ssec:ini-runs}

\begin{figure}[hbt!]
\centering
\includegraphics[width=0.45\textwidth]{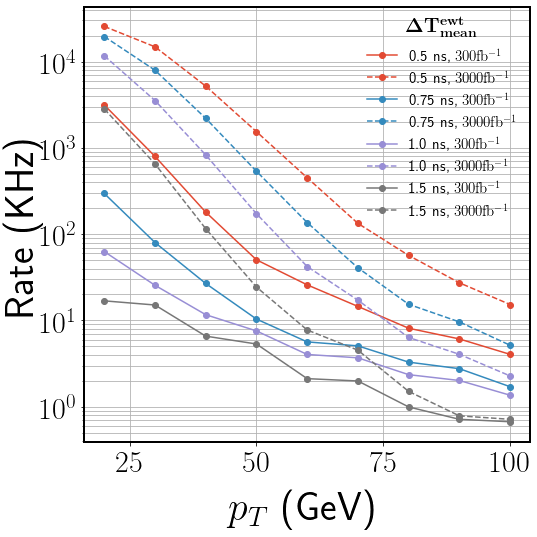}\qquad
\includegraphics[width=0.45\textwidth]{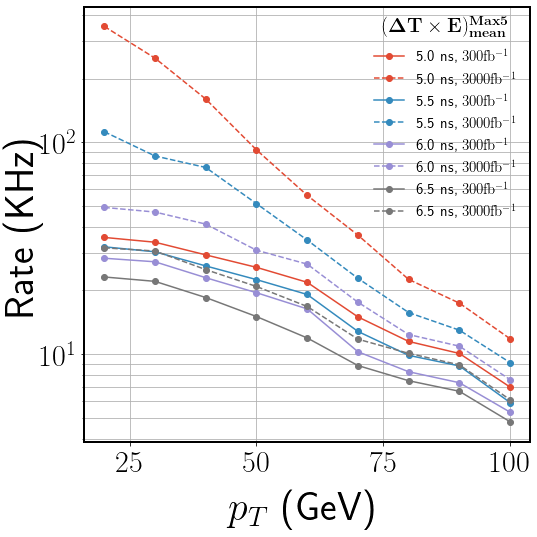}\\
\includegraphics[width=0.45\textwidth]{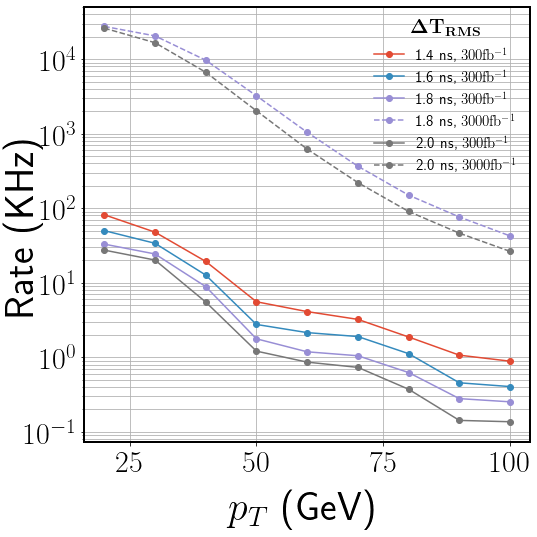}
\caption{Rate of background events as a function of jet $p_T$ with varying cuts on the $\rm {\Delta T_{mean}^{Ewt}}$ (\textit{top left}), $\rm {\Delta T_{mean}^{Max 5}}$ (\textit{top right}) and  $\rm {\Delta T_{RMS}}$ (\textit{bottom}) timing of the jet for ECAL timing resolutions corresponding to 300\,fb$^{-1}$ and 3000\,fb$^{-1}$ in the 140 PU scenario.}
\label{fig:mean_3003000}
\end{figure}

As we discussed in Section\,\ref{ssec:effect-of-reso}, timing resolution will play a very critical role in the timing-based trigger performance. As mentioned in the ECAL TDR\,\cite{CERN-LHCC-2017-011}, the timing resolution of ECAL detector will worsen over time with an increase in the collected integrated luminosity. Initial runs of CMS at HL-LHC will have better timing resolution for ECAL towers of the same energy as compared to the final runs at the end of HL-LHC. 
Till now, we have considered the resolution corresponding to 1000\,fb$^{-1}$ of collected luminosity, which is moderate, compared to the best resolution at 300\,fb$^{-1}$ integrated luminosity at the beginning of HL-LHC and a much worse resolution when around 3000\,fb$^{-1}$ of data is collected by the end of HL-LHC.
To understand the effect of the resolution, we now calculate the background rate as a function of the $p_T$ of the jet with varying cuts on the three jet timing variables, $\Delta T_{mean}^{Ewt}$, $(\Delta T\times E)_{mean}^{Max 5}$, and $\Delta T_{RMS}$, after applying timing resolution corresponding to an integrated luminosity of 300\,fb$^{-1}$ and 3000\,fb$^{-1}$, and is shown in Fig.\,\ref{fig:mean_3003000}.  

Looking back at Fig.\,\ref{fig:mean_ewt_diffRes}, we have seen that the timing distribution corresponding to 300\,fb$^{-1}$ has a much narrower peak than the resolution at 3000\,fb$^{-1}$. Therefore, in the latter case, we expect that we have to devise much stricter cuts on jet variables in order to keep the rate under control. We observe this is for all the three timing variables in Fig.\,\ref{fig:mean_3003000}, mostly in the $\Delta T_{RMS}$ rates.
From Fig.\,\ref{fig:mean_3003000}, we find that we can restrict the background rate at around 30\,kHz by selecting jets with $p_T>40$\,GeV and $\Delta T_{mean}^{Ewt}>0.75$\,ns, whereas for these same cuts the rate increases to around 2000\,kHz.
In order to keep the rate under control during final runs with 3000\,fb$^{-1}$, jet $p_T$ and timing need to be constrained more aggressively, which will deteriorate the efficiency of selecting LLP events. To achieve a rate below 30\,kHz, we have to use selection cuts of $p_T>50$\,GeV and $\Delta T_{mean}^{Ewt}>1.5$\,ns.

\begin{table}[hbt!]
\centering
\begin{tabular}{|c|c|c|c|c||}
\hline
LLP& Mass (GeV) & $\rm {\Delta T_{mean}^{Ewt}}$ & $\rm {\Delta T_{mean}^{Max 5}}$ & $\rm {\Delta T_{RMS}}$ \\ 
Scenario & c$\tau$ (cm) & (\%) & (\%) & (\%)\\
\hline
\hline
\multirow{2}{*}{LLP (A)} & 30, 10  & 0.88 & 1.79 & 0.48\\ 
\cline{2-5}
 &  30, 100 & 21.46 & 18.18 & 13.10\\ 
\hline 
\hline
\multirow{2}{*}{LLP (B)} & 150, 10  & 4.44 & 15.50 &1.54\\ 
\cline{2-5} 
&  150, 100  & 35.16 & 40.85 & 18.47\\ 
\hline 
\hline
\multirow{2}{*}{LLP (C)} & 150,  10 & 6.43 & 23.53 &3.18\\ 
\cline{2-5} 
&  150,  100 & 31.41 & 38.70 & 17.69\\ 
\hline
\hline
\end{tabular} 
\caption{Trigger efficiency for three benchmark masses from the three LLP scenarios with $c\tau=$10\,cm and 100\,cm calculated by putting suitable thresholds on the three timing variables to keep the background rate restricted at $\approx$ 30\,kHz when the ECAL timing resolution corresponding to 300\,fb$^{-1}$ luminosity is applied.}
\label{tab:300LLP}
\end{table}

However, in the initial HL-LHC runs, due to better resolution, we can apply more relaxed cuts maintaining low background rates, seen in Fig.\,\ref{fig:mean_3003000}. Let us now check whether these reduced cuts can improve the signal efficiency.
We use the same $p_T$ and $N_{tow}$ cuts as used by us in the 1000\,fb$^{-1}$ case mentioned in Table\,\ref{tab:eff}, with the timing cuts reduced for each variable to $\Delta T_{mean}^{Ewt}>0.85$\,ns, $(\Delta T\times E)_{mean}^{max5}>5.0$\,ns, and $\Delta T_{RMS}>1.4$\,ns.
We have calculated the efficiency of the timing trigger at 300\,fb$^{-1}$ using these cuts for three LLP scenarios for a few benchmark points, as listed in Table\,\ref{tab:300LLP}. 
There is improvement in the signal efficiency for each benchmark point in all the three LLP scenarios for triggers constructed using $\Delta T_{mean}^{Ewt}$, $(\Delta T\times E)_{mean}^{max5}$, and $\Delta T_{RMS}$. The improvement is not as large to compensate for the reduced collected luminosity compared to 1000\,fb$^{-1}$ (around a factor of 3), however, we can conclude here that the initial runs of CMS at HL-LHC will be slightly more sensitive to trigger LLP events, while keeping the rate under control.

\subsection{Inclusion of displaced tracks at L1}
\label{ssec:disp_tracks}

\begin{figure}[hbt!]
\centering
\includegraphics[width=0.42\textwidth]{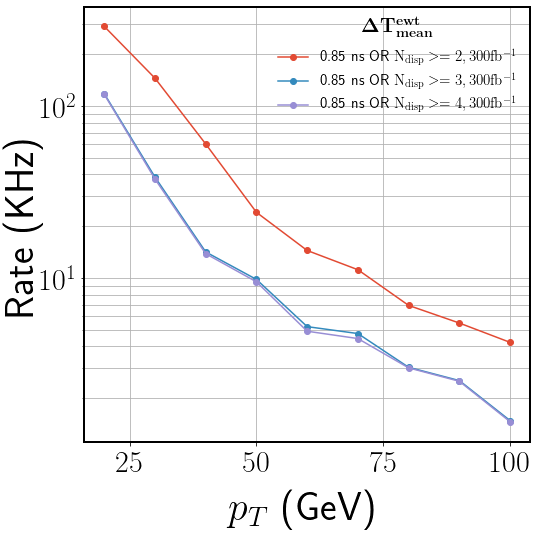}\qquad
\includegraphics[width=0.42\textwidth]{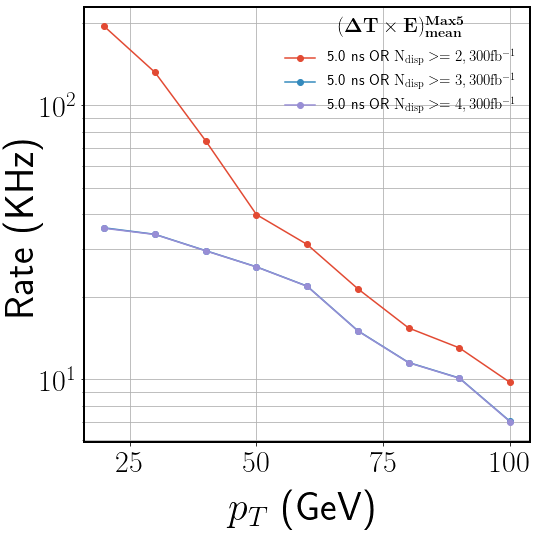}\\
\includegraphics[width=0.42\textwidth]{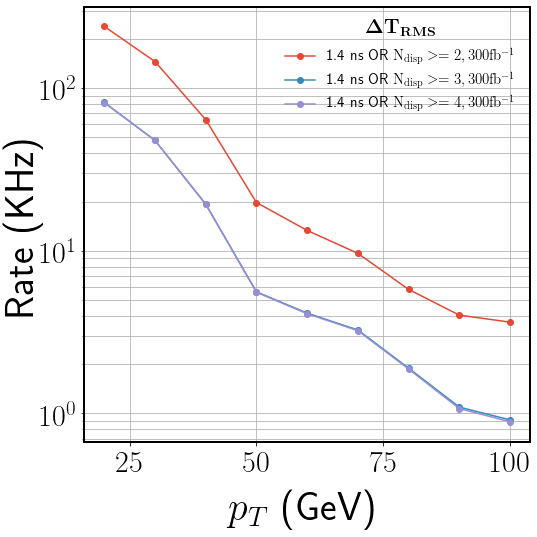}
\caption{Rate of background events as a function of jet $p_T$ with the combination of varying cuts on the $\rm {\Delta T_{mean}^{Ewt}}$ (\textit{top left}), $\rm {\Delta T_{mean}^{Max 5}}$ (\textit{top right}) and  $\rm {\Delta T_{RMS}}$ (\textit{bottom}) timing of the jet for ECAL timing resolution corresponding to integrated luminosity of 300\,fb$^{-1}$ and number of displaced L1 tracks inside a jet within $\Delta R=0.3$ at ECAL, in the 140 PU scenario.}
\label{fig:mean_300disp}
\end{figure}

With the inclusion of extended tracking at L1 for HL-LHC, tracks of displaced charged particles coming from the decay of LLPs can be reconstructed at L1. Availability of displaced tracks at L1 along with ECAL timing information will give an extra handle to trigger on LLP events. L1 triggers based on timing information of displaced objects start becoming sensitive in selecting LLPs with higher decay lengths and can miss LLPs with smaller decay lengths. Inclusion of displaced tracks information along with timing information to construct dedicated LLP triggers can significantly increase the signal efficiency, especially for LLP scenarios where LLP has a relatively shorter lifetime whose decay products will have very small time delay as measured in the ECAL. We have combined information about the multiplicity of displaced tracks inside a jet with the jet timing to construct a new hybrid trigger where a jet will be selected if it satisfies the timing threshold or if the jet has a certain number of displaced tracks inside $\Delta$ R = 0.3 of the jet axis. Displaced tracks are selected at L1 with efficiency as cited in the CMS Phase-II L1 TDR\,\cite{CERN-LHCC-2020-004}, and we also put a cut of $d_0>1$\,cm to reduce QCD background. In Fig.\,\ref{fig:mean_300disp}, we have shown the background rate as a function of jet $p_T$ for resolution corresponding to 300\,fb$^{-1}$ with varying cuts on the three timing variables, $\Delta T_{mean}^{Ewt}$, $(\Delta T\times E)_{mean}^{max5}$, and $\Delta T_{RMS}$, along with the number of displaced L1 tracks inside the jet.  

\begin{table}[hbt!]
\centering
\begin{tabular}{|c|c|c|c|c||}
\hline
LLP& Mass (GeV) & $\rm {\Delta T_{mean}^{Ewt}}$ & $\rm {\Delta T_{mean}^{Max 5}}$ & $\rm {\Delta T_{RMS}}$ \\ 
Scenario & c$\tau$ (cm) & (\%) & (\%) & (\%)\\
\hline
\hline
\multirow{2}{*}{LLP (A)} & 30, 10  & 4.60 & 5.41 & 4.14\\ 
\cline{2-5}
 &  30, 100 & 22.14 & 18.84 & 13.82\\ 
\hline 
\hline
\multirow{2}{*}{LLP (B)} & 150, 10  & 24.79 & 32.32 & 22.31\\ 
\cline{2-5} 
&  150, 100  & 39.45 & 44.42 & 23.79 \\ 
\hline 
\hline
\multirow{2}{*}{LLP (C)} & 150,  10 & 39.59 & 47.29 & 37.22\\ 
\cline{2-5} 
&  150,  100 & 39.64 & 44.52 & 26.69 \\ 
\hline
\hline
\end{tabular} 
\caption{Trigger efficiency for three benchmark masses from the three LLP scenarios with $c\tau=$10\,cm and 100\,cm calculated by putting suitable thresholds on the three timing variables as well as the number of displaced L1 tracks associated with the jet to keep the background rate restricted at $\approx$ 30\,kHz when the ECAL timing resolution corresponding to 300\,fb$^{-1}$ luminosity is applied.}
\label{tab:300dispLLP}
\end{table}

The {\it top left} panel of Fig.\,\ref{fig:mean_300disp} shows that the rates increase to about 60\,kHz when we select events having at least one $p_T>40$\,GeV jets with either at least 2 displaced L1 tracks associated or with $\Delta T_{mean}^{Ewt}>0.85$\,ns, same as in the previous section where we discuss the impact of resolution at 300\,fb$^{-1}$. 
We can further constrain the background rates to below 20\,kHz by requiring at least 3 displaced L1 tracks inside the jet, keeping the cuts on the timing variables the same, and increasing the minimum number of displaced tracks to 4 does not reduce the rate much.
We have calculated the signal efficiency for two benchmark points in all the three LLP scenarios by fixing the jet timing cut in each of the three timing variables to be the same as taken in the previous section and mentioned in Fig.\,\ref{fig:mean_300disp} or impose the condition of displaced track multiplicity to be greater than or equal to 3 in a $p_T>35$\,GeV jet. For these set of cuts, the background rate is restricted at $\approx$ 30\,kHz, and the signal efficiencies are shown in Table\,\ref{tab:300dispLLP}.
Comparing the values of signal efficiencies obtained in Table\,\ref{tab:300dispLLP} with Table\,\ref{tab:300LLP}, we can see that there is a huge improvement in the signal efficiency for benchmark points with $c\tau=10$\,cm across all three scenarios for trigger constructed using all the three timing variables. 

\subsection{Narrower jets with R=0.2}
\label{ssec:R0.2}

\begin{figure}[hbt!]
\centering
\includegraphics[width=0.4\textwidth]{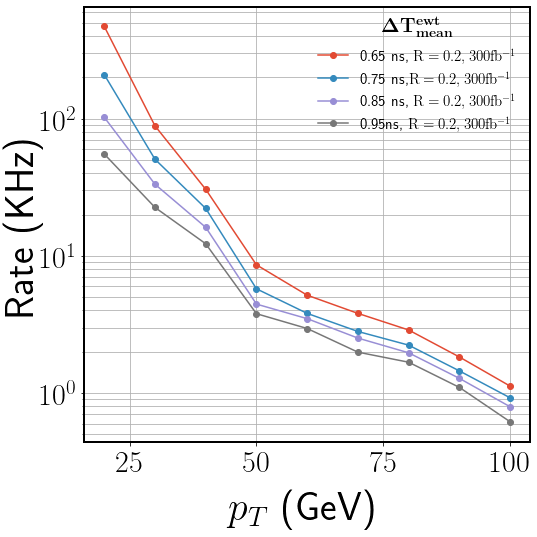}\qquad
\includegraphics[width=0.4\textwidth]{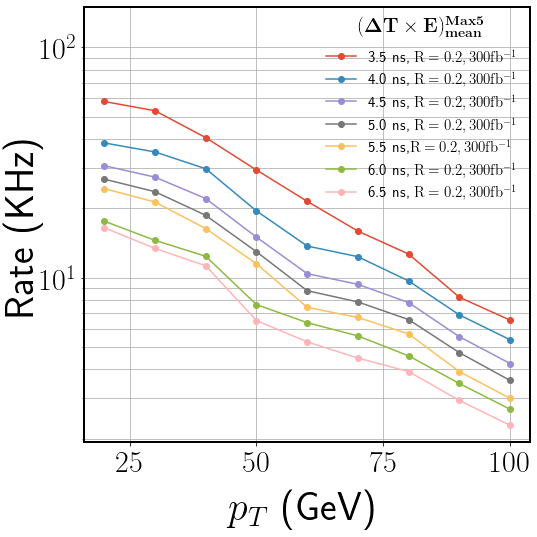}\\
\includegraphics[width=0.4\textwidth]{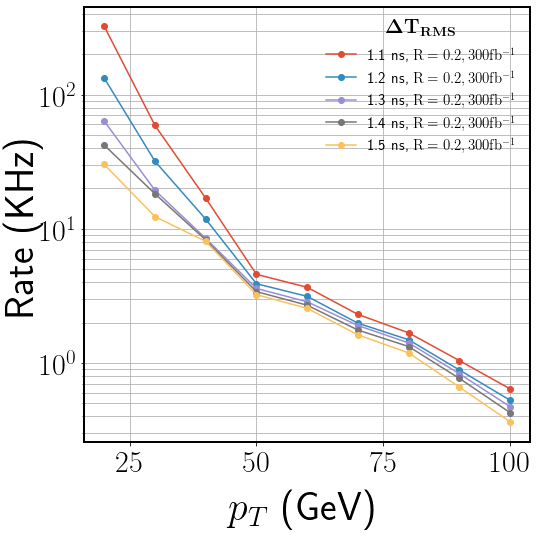}
\caption{Rate of background events as a function of jet $p_T$ with varying cuts on the $\rm {\Delta T_{mean}^{Ewt}}$ (\textit{top left}), $\rm {\Delta T_{mean}^{Max 5}}$ (\textit{top right}) and  $\rm {\Delta T_{RMS}}$ (\textit{bottom}) timing of $R=0.2$ jets for ECAL timing resolution corresponding to integrated luminosity of 300\,fb$^{-1}$ in the 140 PU scenario.}
\label{fig:mean_300R02}
\end{figure}

As we discussed in Section\,\ref{ssec:effect-of-PU}, jets with narrow cone size have lesser PU contribution, and QCD jets are most affected as jet cone size is reduced while LLP jets being narrow can be comfortably contained in a smaller area. In previous sections, we have used $R=0.3$ jets which reduced PU contamination inside jets to a considerable amount, and contamination of jet timing by energy deposits coming from PU was also reduced compared to $R=0.4$. We now study the effect of reducing the cone size further down to $R=0.2$ and how this affects the signal efficiency and background rates. 
We recluster jets using anti-$k_T$ jet clustering algorithm with a reduced jet cone size of $R=0.2$ and impose the same conditions as applied in the last section for $R=0.3$ jets when the resolution corresponding to 300\,fb$^{-1}$ is used. Fig.\,\ref{fig:mean_300R02} shows the background rates as a function of the jet $p_T$ with varying cuts on the three timing variables. We find that in going from jets with a cone size of $R=0.3$ to $R=0.2$, the background rate decreases for the same $p_T$ cut. This can be understood from Fig.\,\ref{fig:cone_size_QCD_LLP_3050_30100_3010_PT} which shows that the chances of prompt QCD jets passing a certain $p_T$ threshold become difficult in going to smaller cone sizes, especially for lower $p_T$ jets, which also dominate the rate.

\begin{table}[hbt!]
\centering
\begin{tabular}{|c|c|c|c|c||}
\hline
LLP& Mass (GeV) & $\rm {\Delta T_{mean}^{Ewt}}$ & $\rm {\Delta T_{mean}^{Max 5}}$ & $\rm {\Delta T_{RMS}}$ \\ 
Scenario & c$\tau$ (cm) & (\%) & (\%) & (\%)\\
\hline
\hline
\multirow{2}{*}{LLP (A)} & 30, 10  & 1.32 & 1.84 &0.77\\ 
\cline{2-5}
 &  30, 100 & 18.43 & 9.97 &9.96\\ 
\hline 
\hline
\multirow{2}{*}{LLP (B)} & 150, 10  & 7.71 & 17.09 &2.47\\ 
\cline{2-5} 
&  150, 100  & 36.96 & 34.27 &16.20 \\ 
\hline 
\hline
\multirow{2}{*}{LLP (C)} & 150,  10 & 8.73 & 24.83 &4.41\\ 
\cline{2-5} 
&  150,  100 & 30.13 & 29.33 &14.53 \\ 
\hline
\hline
\end{tabular} 
\caption{Trigger efficiency for three benchmark masses from the three LLP scenarios with $c\tau=$10\,cm and 100\,cm calculated by putting suitable thresholds on the three timing variables to keep the background rate restricted at $\approx$ 30\,kHz when the ECAL timing resolution corresponding to 300\,fb$^{-1}$ luminosity is applied and we consider $R=0.2$ jets.}
\label{tab:300R02LLP}
\end{table} 

Since the rates decrease slightly, we can reduce the cuts of the timing variables keeping the rates around 30\,kHz, which can help in increasing the signal efficiencies. Table\,\ref{tab:300R02LLP} shows the computed efficiencies when there is at least one jet in the event with $p_T>35$\,GeV and $\Delta T_{mean}^{Ewt}>0.75$\,ns, or $(\Delta T\times E)_{mean}^{Max5}>4.1$\,ns, or $(\Delta T_{RMS}>1.1$\,ns. We find from Table\,\ref{tab:300R02LLP} that the efficiencies increase slightly for the shorter decay length of 10\,cm, which is more prominent for scenarios (B) and (C) where the LLP has direct pair-production in the collider. For the 100\,cm decay length, the efficiencies slightly decrease when we consider $R=0.2$ jets instead of $R=0.3$ jets.

\subsection{The ultimate HL-LHC scenario with 200 PU}
\label{ssec:200PU}

\begin{figure}[hbt!]
\centering
\includegraphics[width=0.4\textwidth]{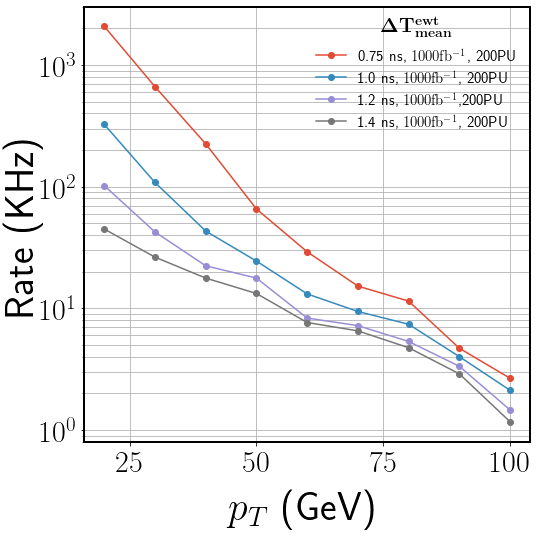}\qquad
\includegraphics[width=0.4\textwidth]{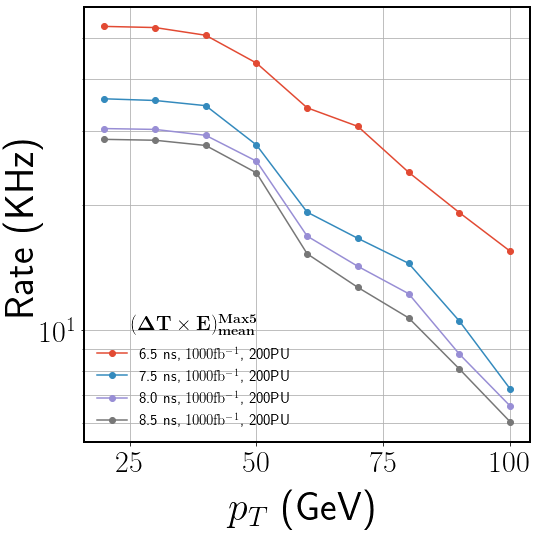}\\
\includegraphics[width=0.4\textwidth]{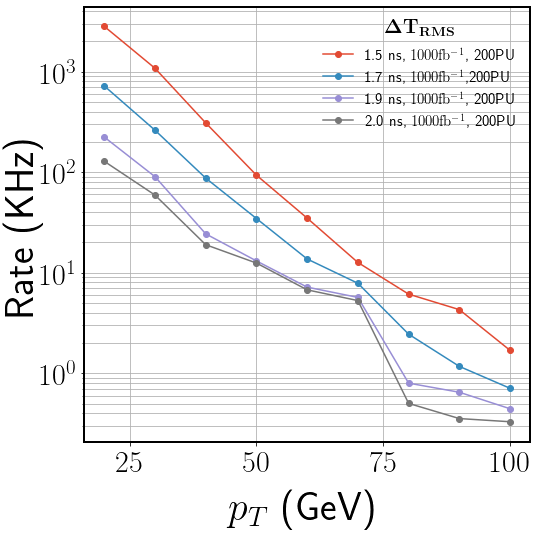}
\caption{Rate of background events as a function of jet $p_T$ with varying cuts on the $\rm {\Delta T_{mean}^{Ewt}}$ (\textit{top left}), $\rm {\Delta T_{mean}^{Max 5}}$ (\textit{top right}) and  $\rm {\Delta T_{RMS}}$ (\textit{bottom}) timing of $R=0.3$ jets for ECAL timing resolution corresponding to integrated luminosity of 1000\,fb$^{-1}$ in the 200 PU scenario.}
\label{fig:rate_200PU}
\end{figure}

For the ultimate scenario at HL-LHC, the number of PU interactions will reach up to 200 vertices per bunch crossing. With the increase in the number of PU events, the background rate will also increase. To keep the background under permissible range, we have to revisit the thresholds for the timing variables used in the previously discussed triggers. We calculate the rate for QCD background for the three timing variables as discussed above after smearing the timing of the ECAL towers with the resolution corresponding to an optimistic scenario where timing resolution corresponds to the one when integrated luminosity reach around 1000\,fb$^{-1}$, and the background now merged with an average of 200 PU vertices per hard collision instead of 140. In Fig.\,\ref{fig:rate_200PU} we have shown the rate as a function of $p_T$ for the three timing variables at 200 PU,
and comparing it with Fig.\,\ref{fig:rate_mean} we find that there is almost a two-fold increase in the rate when PU interactions increase from 140 to 200 for the $\Delta T_{mean}^{Ewt}$ variable, if the cuts are kept the same. To keep the rate under 30\,kHz, we have to put tighter constraints on the values of the timing variables $-$ $\Delta T_{mean}^{Ewt}>1.2$\,ns with $N_{tow} \geq$ 4 , $(\Delta T\times E)_{mean}^{Max5}>8$\,ns, and $\Delta T_{RMS}>2$\,ns (compared to 1.1\,ns with $N_{tow} \geq$ 3, 5.5\,ns, and 1.9\,ns at 140 PU), respectively.

\begin{table}[hbt!]
\centering
\begin{tabular}{|c|c|c|c|c||}
\hline
LLP& Mass (GeV) & $\rm {\Delta T_{mean}^{Ewt}}$ & $\rm {\Delta T_{mean}^{Max 5}}$ & $\rm {\Delta T_{RMS}}$ \\ 
Scenario & c$\tau$ (cm) & (\%) & (\%) & (\%)\\
\hline
\hline
\multirow{2}{*}{LLP (A)} & 30, 10  & 0.50 & 1.03 &0.24\\ 
\cline{2-5}
 &  30, 100 & 18.04 & 16.58 &7.76\\ 
\hline 
\hline
\multirow{2}{*}{LLP (B)} & 150, 10  & 2.12 & 8.37 &0.42\\ 
\cline{2-5} 
&  150, 100  & 27.26 & 34.76 &11.23 \\ 
\hline 
\hline
\multirow{2}{*}{LLP (C)} & 150,  10 & 4.52 & 17.66 &1.06\\ 
\cline{2-5} 
&  150,  100 & 25.51 & 34.16 &10.21 \\ 
\hline
\hline
\end{tabular} 
\caption{Trigger efficiency for three benchmark masses from the three LLP scenarios with $c\tau=$10\,cm and 100\,cm calculated by putting suitable thresholds on the three timing variables to keep the background rate restricted at $\approx$ 30\,kHz when the ECAL timing resolution corresponding to 1000\,fb$^{-1}$ luminosity is applied in the 200 PU scenario.}
\label{tab:200PU}
\end{table} 

With this tighter set of cuts, we now compute the signal efficiencies of the three LLP scenarios for two benchmark points in each scenario, shown in Table\,\ref{tab:200PU}. Due to the increased thresholds, signal efficiency decreases with the increase in PU. At 200 PU, $\rm {\Delta T_{mean}^{Max 5}}$ performs at par with $\rm {\Delta T_{mean}^{Ewt}}$ for LLP(A) benchmark with 100\,cm decay length while for benchmark with 10\,cm decay length, there is two-fold increase in the signal efficiency when compared with $\rm {\Delta T_{mean}^{Ewt}}$. Performance of $\rm {\Delta T_{mean}^{Max 5}}$ is considerably better than the other two variables for LLP scenarios (B) and (C) for both benchmark points.

\begin{figure}[hbt!]
\centering
\includegraphics[width=0.4\textwidth]{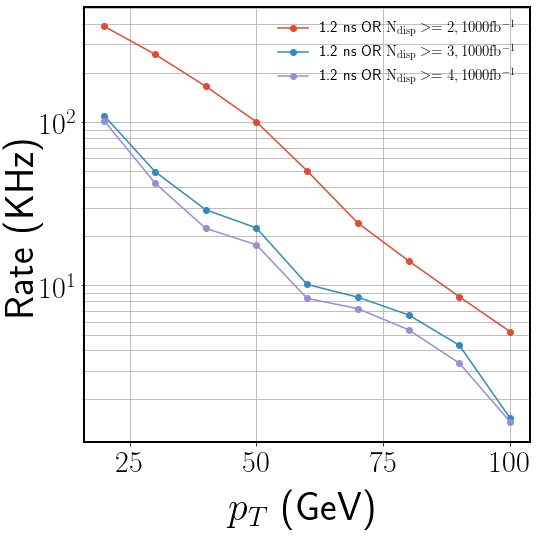}\qquad
\includegraphics[width=0.4\textwidth]{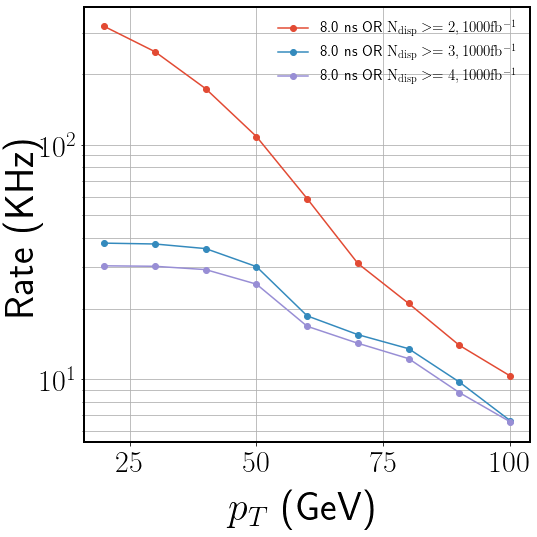}\\
\includegraphics[width=0.4\textwidth]{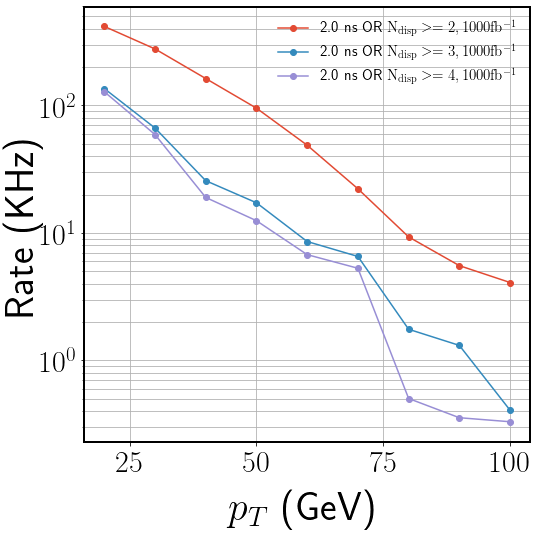}
\caption{Rate of background events as a function of jet $p_T$ with the combination of varying cuts on the $\rm {\Delta T_{mean}^{Ewt}}$ (\textit{top left}), $\rm {\Delta T_{mean}^{Max 5}}$ (\textit{top right}) and  $\rm {\Delta T_{RMS}}$ (\textit{bottom}) timing of the jet for ECAL timing resolution corresponding to integrated luminosity of 1000\,fb$^{-1}$ and number of displaced L1 tracks inside a jet within $\Delta R=0.3$ at ECAL, in the 200 PU scenario.}
\label{fig:disp_200PU}
\end{figure}

To improve the signal efficiency for LLP benchmarks with shorter decay lengths, we can perform a similar analysis as done for the 140 PU scenario in Section.\,\ref{ssec:disp_tracks} and combine the displaced tracks information with the timing information of the jet. Fig.\,\ref{fig:disp_200PU} shows the background rate as function of $p_T$ and $N_{disp}$ for three timing variables. As we can see from Fig.\,\ref{fig:disp_200PU}, the background rate for 200 PU can be restricted to 30\,kHz by requiring at least 4 displaced tracks inside the jet along with the respective timing, and $N_{tow}$ cuts as explained in the previous discussion. Signal efficiency for three LLP scenarios for two benchmark points each is calculated with an updated set of requirements and is tabulated in Table\,\ref{tab:200PU_disp}. Performance of  $\rm {\Delta T_{mean}^{Ewt}}$ improves by a factor of three for LLP scenario (A), while for LLP scenarios (B) and (C), signal efficiency improves by a factor of 6 for LLP benchmark with 10\,cm decay length. For $\rm {\Delta T_{mean}^{Max 5}}$, there is a two-fold improvement in the performance for LLP with 10\,cm decay length across all scenarios.

\begin{table}[hbt!]
\centering
\begin{tabular}{|c|c|c|c|c||}
\hline
LLP& Mass (GeV) & $\rm {\Delta T_{mean}^{Ewt}}$ & $\rm {\Delta T_{mean}^{Max 5}}$ & $\rm {\Delta T_{RMS}}$ \\ 
Scenario & c$\tau$ (cm) & (\%) & (\%) & (\%)\\
\hline
\hline
\multirow{2}{*}{LLP (A)} & 30, 10  & 1.50 & 1.99 &1.24\\ 
\cline{2-5}
 &  30, 100 & 18.19 & 16.72 &7.92\\ 
\hline 
\hline
\multirow{2}{*}{LLP (B)} & 150, 10  & 11.95 & 17.14 &10.51\\ 
\cline{2-5} 
&  150, 100  & 29.25 & 36.40 &13.61 \\ 
\hline 
\hline
\multirow{2}{*}{LLP (C)} & 150,  10 & 25.22 & 33.16 &22.88\\ 
\cline{2-5} 
&  150,  100 & 30.20 & 37.97 &15.41 \\ 
\hline
\hline
\end{tabular} 
\caption{Trigger efficiency for three benchmark masses from the three LLP scenarios with $c\tau=$10\,cm and 100\,cm calculated by putting suitable thresholds on the three timing variables as well as the number of displaced L1 tracks associated with the jet to keep the background rate restricted at $\approx$ 30\,kHz when the ECAL timing resolution corresponding to 1000\,fb$^{-1}$ luminosity is applied.}
\label{tab:200PU_disp}
\end{table} 


\subsection{Next level of trigger system: HLT}
\label{ssec:HLT}

High level trigger (HLT) is the second stage of the online trigger system after L1, where interesting events passing L1 are further sorted and selected after applying appropriate trigger criteria.
The high available latency period at HLT enables the reconstruction of physics objects with extended reach and improved accuracy. More complex physics objects can also be reconstructed at HLT after combining information from various sub-detectors. In the following two sections, we briefly discuss the extended reach of displaced track reconstruction, and the prospect of MTD at HLT for the search of LLPs decaying to displaced jets.

\subsubsection{Displaced tracking at HLT}
\label{sssec:disp-track-HLT}

\begin{figure}[hbt!]
\centering
\includegraphics[width=0.5\textwidth]{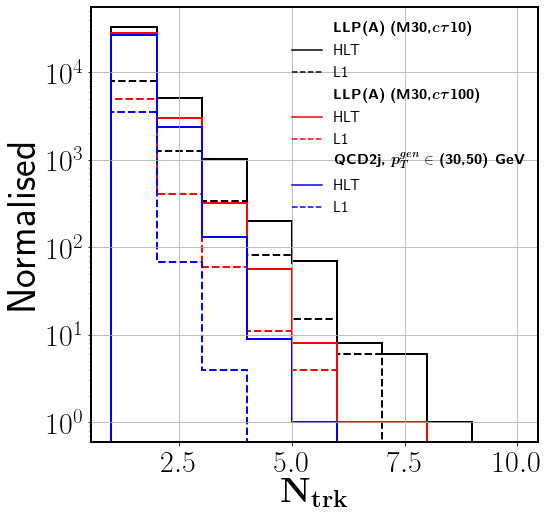}
\caption{Displaced track multiplicity inside a jet within $\Delta R = 0.3$ for jets from two LLP benchmarks from scenario (A) with $M_X=30$\,GeV and $c\tau=10$\,cm and 100\,cm, and QCD dijet process ($p_T^{gen}\in\{30,50\}$\,GeV) at L1 and HLT.}
\label{fig:ntrk_hlt}
\end{figure} 

Extended tracking at L1 for HL-LHC enables the  reconstruction of displaced tracks, but the coverage is limited. Only tracks with transverse impact parameter $|d_0| < $ 8 cm can be reconstructed with track reconstruction efficiency dwindling with increasing displacement. However, tracks with higher transverse displacement can be reconstructed at HLT with better efficiency. Higher tracking efficiency for tracks with large displacement means displaced jets can be tagged efficiently and accurately at HLT. Fig.\,\ref{fig:ntrk_hlt} shows the distribution of displaced track multiplicity inside the jet for some LLP benchmarks from scenario (A) and QCD dijet events at L1 and HLT. For displaced tracks reconstruction efficiency at HLT, we have used the same efficiency achievable for displaced tracks as a function of their transverse distance from the beamline ($L_{xy}$) at Run-3 of Phase-I LHC for CMS, taken from Ref.\,\cite{disptrackHLT}. As we can see from Fig.\,\ref{fig:ntrk_hlt}, the number of displaced tracks reconstructed and associated with a jet at HLT is relatively more in number when compared to L1, which might improve the chances of events identified as containing displaced jets at L1 to pass the HLT as well. 

\subsubsection{MIP timing detector at HLT}
\label{sssec:MTD-HLT}

As we discussed in Section\,\ref{sec:intro}, timing information from MTD might not be available at L1 due to bandwidth constraints. However, there might be plans to include it as some ``External Triggers'' at L1. Nevertheless, this information can be accommodated at HLT. At MTD, timing resolution of up to 30\,ps can be achieved for all tracks with $p_T >$ 0.7\,GeV in barrel ($|\eta| < $ 1.5) and  $p > $ 0.7\,GeV in endcap (1.5 $ < |\eta| < $ 3.0). Timing information from MTD can be used to construct similar timing variables as constructed using ECAL timing, and dedicated LLP triggers can be constructed using these timing variables for HLT. Better resolution and wider coverage of MTD might ensure that LLP events passing dedicated L1 triggers will also be efficiently picked up by the HLT for the offline analyses. In Ref.\,\cite{Bhattacherjee:2020nno}, we have discussed the possibility of using MTD to trigger on LLP events using BDT based L1 triggers as well as combine the information of L1 prompt tracks to improve their performance. However, since the availability of MTD information at L1 is uncertain, the same framework can be used to construct dedicated LLP triggers for HLT. 

\section{Summary and conclusion}
\label{sec:concl}
With several ongoing dedicated analyses with exciting future prospects, an ample amount of theory motivation supports the need for LLP searches to be performed with more vigour to look for physics beyond the standard model. 
LLPs can leave different types of exotic signatures in the detector based on their final state, as discussed in the text. In this paper, we have mainly focused on the scenarios where LLP decays to jets in the final state.
We have studied two well-motivated scenarios where LLPs are produced through the decay of the 125\,GeV Higgs boson and where they are directly pair-produced in the context of CMS detector at HL-LHC.  
With several future upgrades in the pipeline, the need to develop dedicated triggers for displaced searches becomes imminent. We have studied how ECAL based timing triggers at level-1 can be effectively used for LLP searches at HL-LHC in the present work. 
In Section\,\ref{sec:intro}, we have specified our goal in terms of several questions. We summarise our findings below:

\begin{itemize}
    \item {\bf Displaced jets and PUPPI:} We have performed a preliminary analysis to study how the PUPPI algorithm, usually optimised for prompt jets, performs for displaced jets. We find that it is difficult to differentiate jets from the decay of lighter LLPs and jets coming from minimum bias events, and therefore, they might be rejected at L1. 
    For heavier LLPs, for example, in LLP scenarios (B) and (C), PUPPI triggers will perform comparatively better, and we can trigger some events containing displaced jets, since energy deposited by LLP in the calorimeters for such scenarios will be sufficient enough.
    However, even for heavier LLPs, the signal efficiency will degrade with an increase in decay length.
    
    
    \item {\bf Timing variables $-$ factors affecting them and most optimal variables:} We have constructed and studied more than 20 timing variables and their performance for jets from QCD dijet events and some LLP benchmarks, both without and with the 140 PU environment of HL-LHC. We have found that smaller cone sizes can contain most of the energy deposition of displaced jets, and help in reducing the PU contribution, and therefore, we consider jets clustered using anti-$k_T$ with $R=0.3$. Resolution of the ECAL barrel timing also plays a crucial role in determining the jet timing, and it degrades with increasing collected luminosity. Another major factor affecting jet timing, is the spread of vertices in the temporal and longitudinal directions. The presence of SM long-lived hadrons, like $K_S$ and $\Lambda$, inside the jet and their decay products, can also increase the jet timing. From the timing distributions of the many timing variables, we have identified three variables that have good separation power between the LLP signal and QCD prompt jets background as well as are fairly robust against PU $-$ $\Delta T_{mean}^{Ewt}$, $(\Delta T\times E)_{mean}^{max5}$, and $(\Delta T)_{RMS}$. 
    
    \item {\bf Accurate background rate calculation using ``stitching'': } We have combined rates of QCD dijet events from different $p_T^{gen}$ bins, and the minimum bias events using the ``stitching'' procedure to ensure we do not double count any region of phase space. From the background rates for varying cuts on the jet $p_T$ and one from the three timing variables, we fix the thresholds for our cuts for a feasible rate, like 30\,kHz, and then apply these cuts on our LLP signal benchmarks. To the best of our knowledge, such an estimation of background rates for various timing based triggers in different PU scenarios considering proper resolutions corresponding to the degrading ECAL timing resolutions with increasing luminosity is not available elsewhere in the literature.
    
    \item {\bf Performance of different triggers based on the optimal variables:} While $\Delta T_{mean}^{ewt}$ has better performance for LLPs in scenario (A) for lighter LLPs with large decay length, $(\Delta T\times E)_{mean}^{Max 5}$ performs better than the other variables for LLPs in all three scenarios for lighter as well heavier LLPs for both small and large decay lengths. Also, in scenario (A), for lower decay lengths $(\Delta T\times E)_{mean}^{max5}$ based timing trigger has better performance.
    Keeping the background rate restricted at approximately 30\,kHz with 140 PU at HL-LHC with ECAL timing resolution corresponding to an integrated luminosity of around 1000\,fb$^{-1}$, we can achieve signal efficiency of around 19\% for $M30, c\tau100$ from $\Delta T_{mean}^{Ewt}$ based trigger and efficiency of around 1.7\% for $M30, c\tau10$ from $(\Delta T\times E)_{mean}^{max5}$ based trigger for LLP scenario (A). For heavier pair produced LLPs, we can achieve signal efficiency of around 39\% and 38\% for benchmark $M150, c\tau100$ in scenario (B) and (C) respectively. For benchmarks with shorter decay lengths, signal efficiency of around 14\% and 22\% can be achieved for $M150, c\tau10$ in scenarios (B) and (C), respectively, with the $(\Delta T\times E)_{mean}^{max5}$ based trigger.
    
    \item {\bf Possibilities for improvement, and the 200 PU scenario:}
    The initial HL-LHC runs will have better ECAL timing resolution at 300\,fb$^{-1}$ of collected luminosity, which helps us to reduce the thresholds without affecting the rate, and hence improve the signal efficiency.
    Using narrower jet cone size can also slightly improve the performance for LLP benchmarks with lower decay lengths, but signal efficiency for LLPs with larger decay lengths have a slight reduction. For the ultimate scenario at HL-LHC, an increase in PU interaction from 140 to 200 will increase the QCD background rate by around a factor of 2 for the $\Delta T_{mean}^{Ewt}$ based trigger, which implies that the thresholds of the cuts have to be increased.
    For ECAL timing resolution corresponding to 1000\,fb$^{-1}$ and 200 PU, we can achieve signal efficiency of around 18\% with the $\Delta T_{mean}^{ewt}$ based trigger for LLP benchmark $M30, c\tau100$ in LLP scenario (A). We get signal efficiency of around 35\%, 34\% for benchmarks $M150, c\tau100$, and $M150, c\tau10$ in LLP scenarios (B) and (C) respectively with the $(\Delta T\times E)_{mean}^{max5}$ based trigger. 
    
    \item {\bf Inclusion of displaced L1 tracking and its complementarity with timing:} The inclusion of displaced L1 tracks along with timing information to construct displaced jet triggers will significantly increase the signal efficiency for LLPs, especially the ones with shorter decay lengths. Signal efficiency of approximately 5\%, 25\% and 40\% can be achieved for benchmarks $M30, c\tau10$, $M150, c\tau10$, and $M150, c\tau10$ in LLP scenarios (A), (B), and (C) respectively when the $\Delta T_{mean}^{Ewt}$ trigger is combined inclusively with demanding at least 3 displaced tracks inside a jet. Displaced tracks based triggers are more sensitive to lower decay lengths as compared to delayed jet triggers, and hence, improve the sensitivity in a complementary region.
    Signal efficiency for shorter decay lengths at 200PU can also be improved with the inclusion of displaced tracks information along with timing information as demonstrated for 140 PU scenario.
    
\end{itemize}

In conclusion, the feasibility of ECAL timing to construct various dedicated triggers at L1 of HL-LHC has been studied in detail in this work, considering various aspects, like the high PU environment of HL-LHC, degrading resolution, etc, and it is sensitive to LLPs having a range of masses and lifetimes in various scenarios. The background rate plots and signal efficiencies presented in the form of grids in the mass and decay length plane for the three LLP scenarios can be directly used in other theoretical studies of LLPs.

\section*{Acknowledgement}
\label{sec:ackn}
TG acknowledges the financial support from the DST Woman Scientist-A Research Grant Ref. No. SR/WOS-A/PM-42/2018 and the facilities provided at Indian Institute of Science, Bengaluru. 
BB, RS, and PS would like to thank Swagata Mukherjee, Camellia Bose, and Rahool Kumar Barman for useful discussions.

\appendix

\section*{Appendix}

\section{PUPPI weights}
\label{app:puppiwt}

\begin{figure}[hbt!]
\centering
\includegraphics[width=0.48\textwidth]{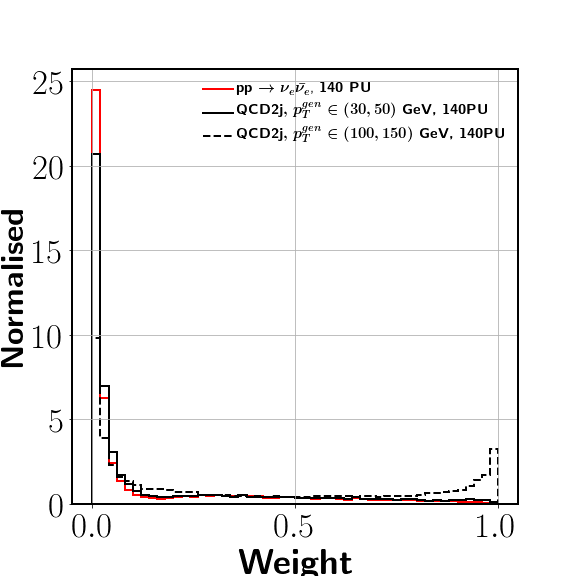}~
\includegraphics[width=0.48\textwidth]{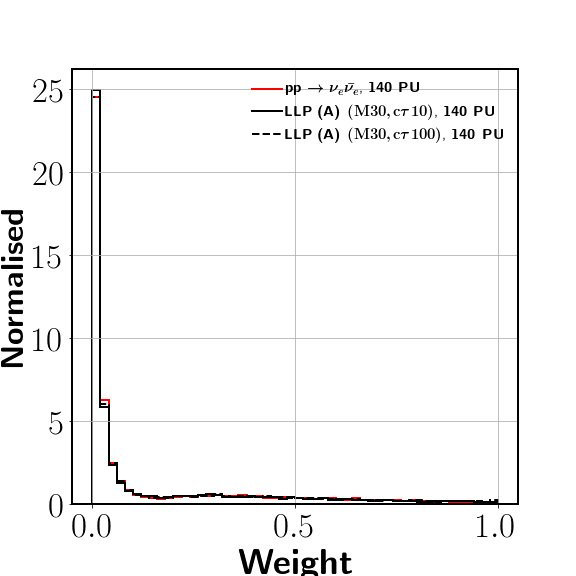}\\
\includegraphics[width=0.48\textwidth]{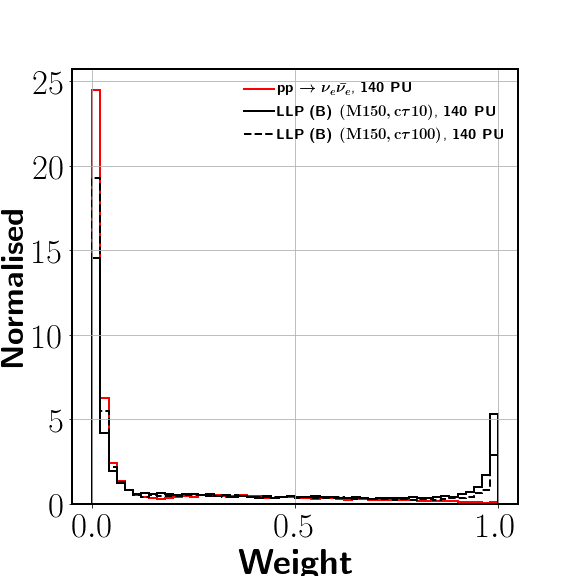}
\caption{Distribution of weight of each neutral particle calculated using the procedure followed in Ref.\,\cite{CERN-LHCC-2020-004} for QCD dijet process with $p_T^{gen}\in\{30,50\}$\,GeV and \{100,150\}\,GeV (\textit{top left}), and LLP of mass 30\,GeV from scenario (A) (\textit{top right}) and 150\,GeV from scenario (B) (\textit{bottom}) with $c\tau=10$\,cm and 100\,cm each, all merged with an average of 140 PU vertices. Distribution for minimum bias events is also shown for comparison in each case.}
\label{fig:puppi_weights}
\end{figure} 

Weight is calculated for each neutral particle using following formulae: 

\begin{equation}
w_i = \frac{1}{1+e^{-x_{tot}}}
\end{equation}
\begin{equation}
x_{tot} = x_\alpha + x_{p_T} - x_{PU}
\end{equation}
\begin{equation}
x_{\alpha} = \rm{min}(\rm{max}(c_\alpha (\alpha-\alpha_0), -x_\alpha^{max}),+x_\alpha^{max})
\end{equation}
\begin{equation}
x_{p_T} =  c_{p_T}(p_T - p_T^0)
\end{equation}
\begin{equation}
x_{PU} =  \rm{log}(N_{PU}/200)+c_0
\end{equation}

Here, $x_{\alpha}^{max}$, $c_{p_T}$, $c_0$ and $c_{\alpha}$ are tunable parameters (which can be read from a look out table as given in Ref.\,\cite{CERN-LHCC-2020-004}) while value of $\alpha_0$ and $p_T^0$ represents mean values from PU. $N_{PU}$ is the number of tracks not associated to the identified PV. Finally, $p_T$ of the neutral particle is multiplied by the weight as calculated above. Charged PF tracks coming from the PV are combined together with re-weighted neutral particles to participate in the jet clustering. Fig.\,\ref{fig:puppi_weights} shows the PUPPI weights for neutral particles calculated using the above formula for QCD dijet and LLP decaying to jets processes.

\section{Cone size}
\label{app:conesize}

\begin{figure}[hbt!]
\centering
\includegraphics[width=\textwidth]{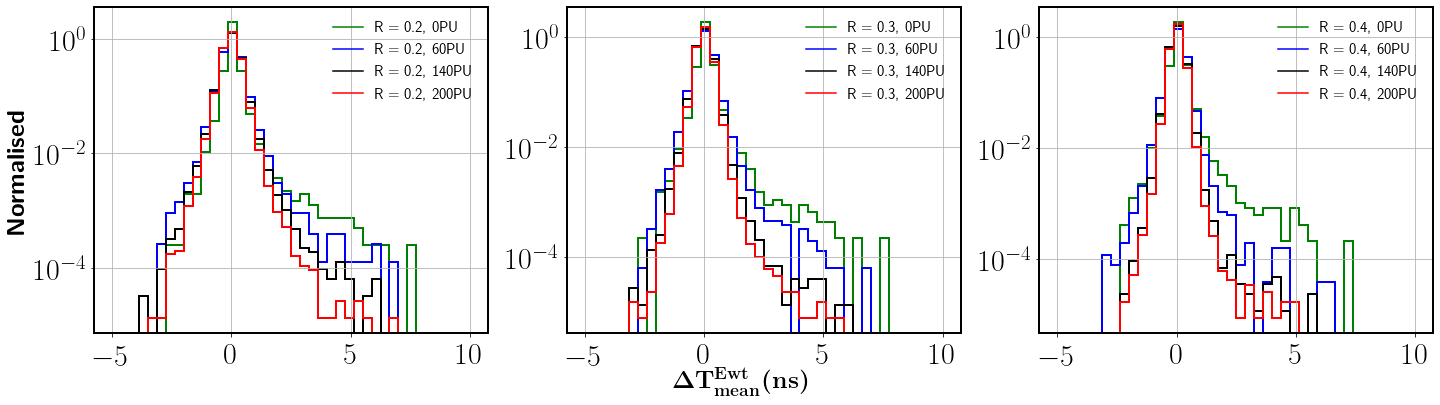}\\
\includegraphics[width=\textwidth]{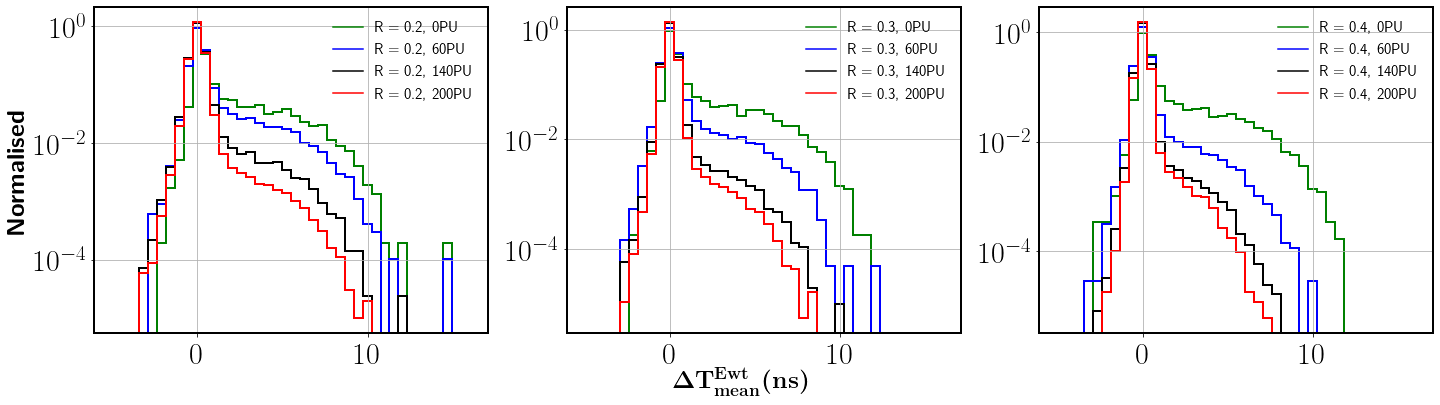}
\caption{The timing of the jets ($\Delta T_{mean}^{Ewt}$) for LLP scenario (A) with benchmark point ($M_X$= 30\,GeV, $c\tau$ = 10\,cm) (\textit{top}) and ($M_X$= 30\,GeV, $c\tau$ = 100\,cm) (\textit{bottom}) with jet cone radius $R =$\,0.2 (\textit{left}), 0.3 (\textit{center}), and 0.4 ({\it right}) for 0 PU, 60 PU, 140 PU, and 200 PU scenarios.}
\label{fig:cone}
\end{figure}

\clearpage
 
\section{Correlation matrices of timing variables}
\label{app:corr}

\begin{figure}[hbt!]
\centering
\includegraphics[width=0.8\textwidth]{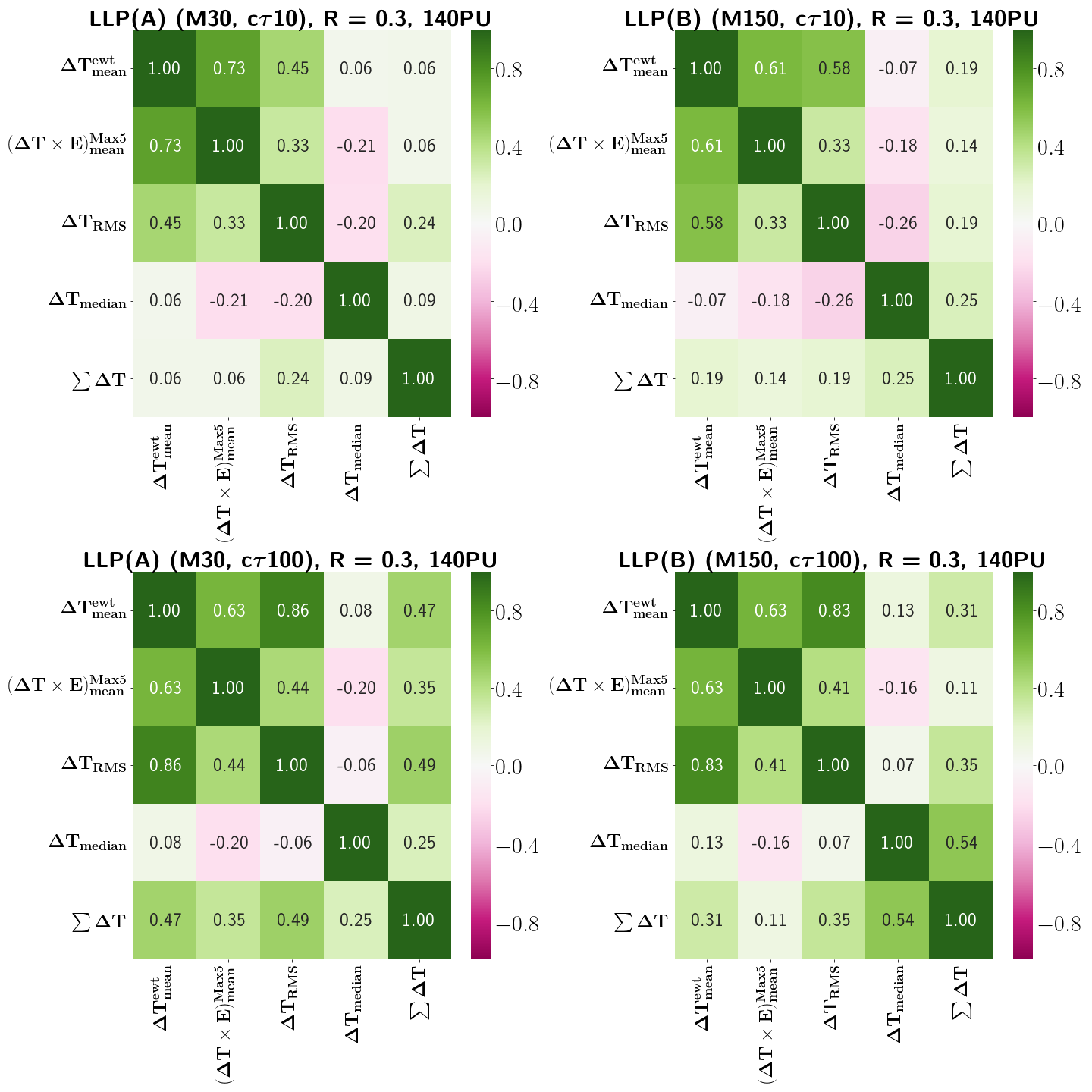}\\
\includegraphics[width=0.4\textwidth]{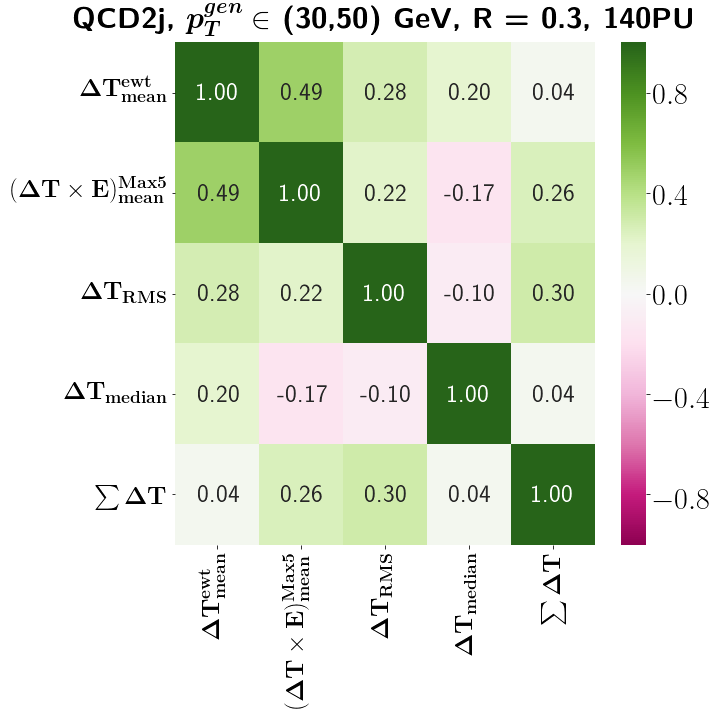}
\caption{Correlation between the various timing variables defined in Section\,\ref{ssec:timing-vars} for four benchmark points of LLP $-$ two from scenario (A) with $M_X=30$\,GeV, $c\tau=10$\,cm ({\it top left} and $c\tau=100$\,cm {\it center left}), and two from scenario (B) with $M_X=150$\,GeV, $c\tau=10$\,cm ({\it top right} and $c\tau=100$\,cm {\it center right}), and QCD dijet process ($p_T\in\{30,50\}$\,GeV) ({\it bottom}) using the resolution corresponding to 1000\,fb$^{-1}$ luminosity.}
\label{fig:corr}
\end{figure} 

\clearpage

\section{Calculation of QCD background rate using ``Stitching''}
\label{app:stitch}

\begin{figure}[hbt!]
\centering
\includegraphics[width=0.45\textwidth]{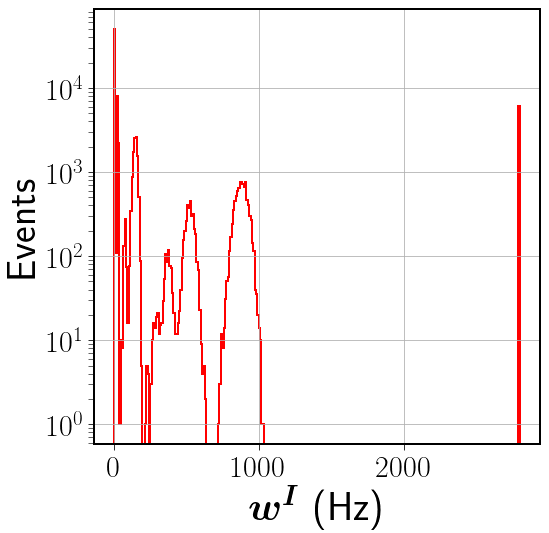}\qquad
\includegraphics[width=0.45\textwidth]{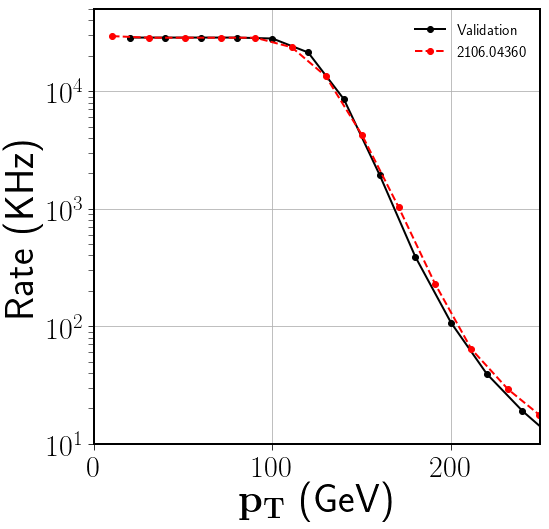}
\caption{Event weight in terms of rate for QCD events ({\it left}) and rate of the single jet trigger at HL-LHC using the ``stitching'' procedure, validated with Ref.\,\cite{Ehataht:2021rkh} ({\it right}).}
\label{fig:stitching}
\end{figure}

In order to calculate background event rate, we generate QCD samples in 7 $p_T^{gen}$ bins covering phase space \{30,50\}\,GeV, \{50,75\}\,GeV, \{75,100\}\,GeV, \{100,125\}\,GeV, \{125,150\}\,GeV, \{150,175\}\,GeV, \{175,200\}\,GeV and $>$200\,GeV. QCD samples in different $p_T^{gen}$ bins are stitched together along with minbias events in order to calculate the event rate accurately. Event weight in terms of rate is calculated using the following formula as given in Ref.\,\cite{Ehataht:2021rkh}:

\begin{equation}
w^{I} = \frac{F}{N_{incl}+\sum_j N_j \times \frac{n_j}{(N_{PU}+1)\times p_j}}
\label{eq:stitch}
\end{equation}
where, $F$ corresponds to the $pp$ collision frequency of approximately 28 MHz. $N_{incl}$ corresponds to total number of events containing $\rm{N_{PU}}$+1 minbias collisions where $N_{PU}$ is the mean number of pile-up events in a collision. $N_j$ is the number of events in a particular QCD sample for the $j^{th}$ $p_T^{gen}$ bin. $n_j$ refers to number of inelastic $pp$ interaction from PU or hard interaction falling in the $p_T^{gen}$ interval $j$. $p_j$ corresponds to the probability of single $pp$ inelastic collision to fall in the $j^{th}$ $p_T^{gen}$ bin. The probabilities $p_j$ for a particular $p_T^{gen}$ bin is calculated by taking the ratio of cross-section of the collision for that particular bin and cross-section when no condition is imposed on the $p_T^{gen}$. We have shown the weight calculated using monte-carlo stitching method in Fig.\,\ref{fig:stitching} (\textit{left}). We have also compared the HL-LHC single jet rate as calculated in Ref.\,\cite{Ehataht:2021rkh} to our calculations in Fig.\,\ref{fig:stitching} (\textit{right}) and rate matches fairly well.

\clearpage


\providecommand{\href}[2]{#2}\begingroup\raggedright\endgroup

\end{document}